\documentclass[twocolumn]{aastex63}

\usepackage{graphicx}
\usepackage{amsmath,amssymb}
\usepackage{color}
\usepackage{units}
\usepackage{hyperref}
\usepackage{gensymb}
\usepackage{wrapfig,lipsum,booktabs}

\newcommand\um{$\,\rm{\mu m}$}

\newcommand\Msun{M$_{\rm \odot}$}
\newcommand{\Lya}{Lyman-$\alpha$}
\newcommand{\Hb}{H$\beta$\,}
\newcommand{\Ha}{H$\alpha$\,}
\newcommand{\zs}{$z_{\rm spec}$\,}
\newcommand{\zp}{$z_{\rm phot}$\,}
\newcommand{\Oii}{[O\textsc{ii}]}
\newcommand{\Hii}{H\textsc{ii}}
\newcommand{\Oiii}{[O\textsc{iii}]}
\newcommand{\Ciii}{C\textsc{iii}]}
\newcommand{\Cii}{[C\textsc{ii}]}
\newcommand{\Muv}{M_{\text{UV}}}

\begin{document} 

\title{The Web Epoch of Reionization Lyman-$\alpha$ Survey (WERLS) I. MOSFIRE Spectroscopy of $\mathbf{z \sim 7-8}$ Lyman-$\alpha$ Emitters\footnote{The data presented herein were obtained at the W. M. Keck Observatory, which is operated as a scientific partnership among the California Institute of Technology, the University of California and the National Aeronautics and Space Administration. The Observatory was made possible by the generous financial support of the W. M. Keck Foundation.}}

\author[0000-0003-3881-1397]{Olivia R. Cooper}\altaffiliation{NSF Graduate Research Fellow}
\affiliation{The University of Texas at Austin, 2515 Speedway Boulevard Stop C1400, Austin, TX 78712, USA}

\author[0000-0002-0930-6466]{Caitlin M. Casey}
\affiliation{The University of Texas at Austin, 2515 Speedway Boulevard Stop C1400, Austin, TX 78712, USA}
\affiliation{Cosmic Dawn Center (DAWN), Denmark}

\author[0000-0003-3596-8794]{Hollis B. Akins}
\affiliation{The University of Texas at Austin, 2515 Speedway Boulevard Stop C1400, Austin, TX 78712, USA}

\author{Jake Magee}
\affiliation{The University of Texas at Austin, 2515 Speedway Boulevard Stop C1400, Austin, TX 78712, USA}

\author{Alfonso Melendez}
\affiliation{The University of Texas at Austin, 2515 Speedway Boulevard Stop C1400, Austin, TX 78712, USA}

\author{Mia Fong}
\affiliation{The University of Texas at Austin, 2515 Speedway Boulevard Stop C1400, Austin, TX 78712, USA}

\author[0000-0001-8169-7249]{Stephanie M. Urbano Stawinski}
\affiliation{Department of Physics and Astronomy, University of California, Irvine, 4129 Frederick Reines Hall, Irvine, CA 92697, USA}

\author[0000-0001-9187-3605]{Jeyhan S. Kartaltepe}
\affiliation{Laboratory for Multiwavelength Astrophysics, School of Physics and Astronomy, Rochester Institute of Technology, 84 Lomb Memorial Drive, Rochester, NY 14623, USA}

\author[0000-0001-8519-1130]{Steven L. Finkelstein}
\affiliation{The University of Texas at Austin, 2515 Speedway Boulevard Stop C1400, Austin, TX 78712, USA}

\author[0000-0003-2366-8858]{Rebecca L. Larson}
\affiliation{Laboratory for Multiwavelength Astrophysics, School of Physics and Astronomy, Rochester Institute of Technology, 84 Lomb Memorial Drive, Rochester, NY 14623, USA}

\author[0000-0003-1187-4240]{Intae Jung}
\affiliation{Space Telescope Science Institute, 3700 San Martin Drive Baltimore, MD 21218, USA}

\author{Ash Bista}
\affiliation{Laboratory for Multiwavelength Astrophysics, School of Physics and Astronomy, Rochester Institute of Technology, 84 Lomb Memorial Drive, Rochester, NY 14623, USA}

\author[0000-0002-6184-9097]{Jaclyn B. Champagne}
\affiliation{Steward Observatory, University of Arizona, 933 N. Cherry Ave, Tucson, AZ 85721, USA}

\author[0000-0003-2332-5505]{\'Oscar A. Ch\'avez Ortiz}
\affiliation{The University of Texas at Austin, 2515 Speedway Boulevard Stop C1400, Austin, TX 78712, USA}

\author[0000-0003-3038-8045]{Sadie Coffin}
\affiliation{Laboratory for Multiwavelength Astrophysics, School of Physics and Astronomy, Rochester Institute of Technology, 84 Lomb Memorial Drive, Rochester, NY 14623, USA}

\author[0000-0003-1371-6019]{M. C. Cooper}
\affiliation{Department of Physics \& Astronomy, University of California, Irvine, 4129 Reines Hall, Irvine, CA 92697, USA} 

\author[0000-0003-4761-2197]{Nicole Drakos}
\affiliation{Department of Astronomy and Astrophysics, University of California, Santa Cruz, 1156 High Street, Santa Cruz, CA 95064}

\author[0000-0002-9382-9832]{Andreas L. Faisst}
\affiliation{Caltech/IPAC, 1200 E. California Boulevard, Pasadena, CA, USA, 91125}

\author[0000-0002-3560-8599]{Maximilien Franco}
\affiliation{The University of Texas at Austin, 2515 Speedway Boulevard Stop C1400, Austin, TX 78712, USA}

\author[0000-0001-7201-5066]{Seiji Fujimoto}\altaffiliation{NASA Hubble Fellow}
\affiliation{The University of Texas at Austin, 2515 Speedway Boulevard Stop C1400, Austin, TX 78712, USA}

\author[0000-0001-9885-4589]{Steven Gillman}
\affiliation{Cosmic Dawn Center (DAWN), Denmark}
\affiliation{DTU-Space, Technical University of Denmark, Elektrovej 327, DK-2800 Kgs. Lyngby, Denmark}

\author[0000-0002-0236-919X]{Ghassem Gozaliasl}
\affiliation{Department of Physics, University of Helsinki, P.O. Box 64, FI-00014 Helsinki, Finland} 

\author[0000-0003-0129-2079]{Santosh Harish}
\affiliation{Laboratory for Multiwavelength Astrophysics, School of Physics and Astronomy, Rochester Institute of Technology, 84 Lomb Memorial Drive, Rochester, NY 14623, USA}

\author[0000-0001-6251-4988]{Taylor A. Hutchison}\altaffiliation{NASA Postdoctoral Fellow}
\affiliation{Astrophysics Science Division, NASA Goddard Space Flight Center, 8800 Greenbelt Rd, Greenbelt, MD 20771, USA }

\author[0000-0002-6610-2048]{Anton M. Koekemoer}
\affiliation{Space Telescope Science Institute, 3700 San Martin Dr., Baltimore, MD 21218, USA}

\author[0000-0002-5588-9156]{Vasily Kokorev}
\affiliation{Kapteyn Astronomical Institute, University of Groningen, P.O. Box 800, 9700 AV Groningen, The Netherlands}

\author[0000-0002-3535-4066]{Jitrapon Lertprasertpong}
\affiliation{Laboratory for Multiwavelength Astrophysics, School of Physics and Astronomy, Rochester Institute of Technology, 84 Lomb Memorial Drive, Rochester, NY 14623, USA}

\author[0000-0001-9773-7479]{Daizhong Liu}
\affiliation{Max-Planck-Institut für Extraterrestrische Physik (MPE), Giessenbachstr. 1, D-85748 Garching, Germany}

\author[0000-0002-7530-8857]{Arianna S. Long}\altaffiliation{NASA Hubble Fellow}
\affiliation{The University of Texas at Austin, 2515 Speedway Boulevard Stop C1400, Austin, TX 78712, USA}

\author[0000-0001-6251-4988]{Casey Papovich}
\affiliation{Department of Physics and Astronomy, Texas A\&M University, College Station, TX 77843-4242 USA}
\affiliation{George P. and Cynthia Woods Mitchell Institute for Fundamental Physics and Astronomy, Texas A\&M University, College Station, TX 77843-4242 USA}

\author[0000-0003-0427-8387]{R. Michael Rich}
\affiliation{Department of Physics and Astronomy, UCLA, PAB 430 Portola Plaza, Box 951547, Los Angeles, CA 90095, USA} 

\author[0000-0002-4271-0364]{Brant E. Robertson}
\affiliation{Department of Astronomy and Astrophysics, University of California, Santa Cruz, 1156 High Street, Santa Cruz, CA 95064} 

\author[0000-0003-4352-2063]{Margherita Talia}
\affiliation{University of Bologna—Department of Physics and Astronomy ``Augusto Righi'' (DIFA), Via Gobetti 93/2, I-40129 Bologna, Italy}
\affiliation{INAF, Osservatorio di Astrofisica e Scienza dello Spazio, Via Gobetti 93/3, I-40129, Bologna, Italy}

\author[0000-0002-8163-0172]{Brittany N. Vanderhoof}
\affiliation{Laboratory for Multiwavelength Astrophysics, School of Physics and Astronomy, Rochester Institute of Technology, 84 Lomb Memorial Drive, Rochester, NY 14623, USA}
\affiliation{Space Telescope Science Institute, 3700 San Martin Dr., Baltimore, MD 21218, USA}

\author[0000-0001-7160-3632]{Katherine E. Whitaker}
\affiliation{Department of Astronomy, University of Massachusetts Amherst, 710 N Pleasant Street, Amherst, MA 01003, USA}
\affiliation{Cosmic Dawn Center (DAWN), Denmark}

\author[0000-0002-7051-1100]{Jorge A. Zavala}
\affiliation{National Astronomical Observatory of Japan, 2-21-1 Osawa, Mitaka, Tokyo 181-8588, Japan}

\date{\today}

\begin{abstract}

We present the first results from the Web Epoch of Reionization \Lya\ Survey (WERLS), a spectroscopic survey of \Lya\ emission using Keck I/MOSFIRE and LRIS. WERLS targets bright ($J<26$) galaxy candidates with photometric redshifts of $5.5\lesssim z \lesssim 8$ selected from pre-\textit{JWST} imaging embedded in the Epoch of Reionization (EoR) within three \textit{JWST} deep fields: CEERS, PRIMER, and COSMOS-Web. Here, we report 11 $z\sim7-8$ \Lya\ emitters (LAEs; 3 secure and 8 tentative candidates) detected in the first five nights of WERLS MOSFIRE data. We estimate our observed LAE yield is $\sim13$\%, broadly consistent with expectations assuming some loss from redshift uncertainty, contamination from sky OH lines, and that the Universe is approximately half-ionized at this epoch, whereby observable \Lya\ emission is unlikely for galaxies embedded in a neutral intergalactic medium. Our targets are selected to be UV-bright, and span a range of absolute UV magnitudes with $-23.1 < \Muv < -19.8$. With two LAEs detected at $z=7.68$, we also consider the possibility of an ionized bubble at this redshift. Future synergistic Keck+\textit{JWST} efforts will provide a powerful tool for pinpointing beacons of reionization and mapping the large scale distribution of mass relative to the ionization state of the Universe.

\end{abstract}


\section{Introduction}
\label{sec:intro}

The first billion years of the Universe hosts its most pivotal transition --- from a neutral to ionized medium --- for which we have yet to determine primary drivers or a precise timeline. From the earliest work on this phase change \citep[e.g.][]{1970arons} to today, considerable progress has been made to constrain the processes and timing of this transition --- the Epoch of Reionization (EoR) --- through both theoretical and observational efforts. Observations of some of the first light sources as they ionized a then neutral intergalactic medium (IGM) have revealed that the reionization process most likely finished around $z\sim5.5-6$ \citep{2011zheng,2016kakiichi,2016castellano} and was halfway completed by $z \sim 7-8$ \citep{2013robertson,2014faisst,2016finkelstein}. Around this halfway point, a high neutral fraction of the IGM has been fairly constrained from somewhat sparse measurements of \Lya\ emitters (LAEs), where the conversion to a neutral fraction has a high systematic uncertainty \citep{2013treu,2019mason,2019hoag,2021bolan}. 

While observations have provided some constraints on the timeline of the EoR, the duration and patchiness of reionization, as well as its main driving sources, remain unclear. In the case where all galaxies have relatively high escape fractions of ionizing photons ($f_{\rm esc} \sim20\%$) --- therefore massive, UV-bright galaxies dominate reionization \citep{2020naidu} --- a late reionization is favored. However, some observations show that galaxies with high $f_{\rm esc}$ are rare \citep{2016izotov}, with small samples of local detections of cosmologically relevant escape fractions $f_{\rm esc}>5\%$ \citep[e.g.][]{2022flury}, and few observations with very high escape fractions $f_{\rm esc}\gtrsim20\%$ \citep[e.g.][]{2018izotov,2020saha,2021marques-chaves}. Simulations predict that $f_{\rm esc}$ depends on halo mass, with higher $f_{\rm esc}$ from fainter galaxies in lower-mass halos \citep{2015paardekooper,2016faisst,2023bremer}. This suggests the case where more numerous UV-faint galaxies dominate reionization, and favors an earlier start to reionization, that evolves smoothly in time \citep{2015mason,2019finkelstein}. Around the instantaneous redshift of reionization \citep[$z_{\rm reion}=7.68\pm 0.79$;][]{2020collaboration} --- which serves as a mean reionization redshift by assuming the process was instantaneous --- the former scenario predicts the IGM is $<1/3$ ionized, while the latter scenario (wherein faint galaxies dominate) predicts $>1/2$ of the IGM is ionized \citep{2019finkelstein}. Taking a census of massive galaxies at $z>7$ can help answer both questions regarding the duration and sources of reionization; in particular, did intrinsically bright galaxies or faint galaxies drive reionization?

In constraining this problem, we are faced with a relative shortage of intrinsically UV-bright $z > 7$ galaxies currently known and spectroscopically confirmed \citep[see ][for a summary of the 15 $z>7.2$ spectroscopic confirmations pre-\textit{JWST}]{2020ouchi}. The pre-\textit{JWST} sample of $z>7$ EoR galaxies has been gathered and confirmed via direct detection of candidate \Lya\ Break Galaxies (LBGs) in relatively small, pencil-beam fields with \textit{Hubble Space Telescope} \citep[e.g.][]{2010bunker,2010mclure,2010finkelstein,2012finkelstein,2012oesch,2012yan}, from sources with ``IRAC excess'' attributed to intense \Oiii+\Hb\ line emission at $z\sim8$ polluting the IRAC 4.5\,\um{} band \citep[e.g.][]{2015smit,2022roberts-borsani}, and from spectroscopic follow-up mainly targeting \Lya\ \citep[e.g.][]{2012ono,2012shibuya,2013finkelstein,2015zitrin,2015oesch,2016inoue,2016song,2017stark,2017hoag,2019jung,2019hu,2020jung,2022larson} or \Cii\ emission lines \citep[e.g.][]{2018smit,2022bouwens,2022schouws}. Now with \textit{JWST}, perhaps the most impressive early results illustrate its ease of spectroscopic detection for galaxies that were previously undetectable, as spectra have been gathered for relatively large samples of EoR galaxies from both NIRSpec \citep[e.g.][]{2023bunker,2023cameron,2023curtis-lake,2023arrabalharo,2023fujimoto} and the NIRCam Grism \citep[e.g.][]{2023oesch}.

Several of these spectroscopically confirmed early EoR galaxies (both from \textit{JWST} and pre-\textit{JWST}) exhibit \Lya\ in emission \citep[e.g.][]{2015oesch,2015zitrin,2018hoag,2018hashimoto,2018pentericci}, with detections of \Lya\ as high as \zs$=10.60$ \citep[GN-z11;][]{2023bunker}. However, detecting \Lya\ from galaxies embedded in the EoR --- especially at $z > 8$ when the Universe is thought to have been predominantly neutral --- involves both technical and physical challenges. From a technical standpoint, while redshift identification via rest-frame optical nebular emission lines is very efficient with NIRSpec, \Lya\ can still be elusive, even to \textit{JWST}. For example, while \Lya\ was detected in GN-z11 at \zs$=10.60$ \citep[anchored by multi-line confirmation in][]{2023bunker}, it was only seen in higher resolution spectra, and was undetected in PRISM observations. Additionally, \Lya\ detections with NIRSpec can be difficult due to slit losses, particularly important for \Lya\ emission which can extend beyond the small {0}\farcs{3} NIRSpec slitlets. Detecting \Lya\ is also challenging from a physical perspective, as \Lya\ photons from EoR galaxies should resonantly scatter by the mostly neutral IGM at a relatively low threshold for HI column density \citep[$\rm N_{HI} > 10^{17} \rm cm^{-2}$][]{2014dijkstra,2016stark}. This concern is partially alleviated by assuming an inhomogeneous reionization process. 

Indeed, cosmological simulations indicate that reionization was likely a patchy process \citep{2017furlanetto,2018daloisio}, producing ionized bubbles in the surrounding IGM growing from 5 -- 20 cMpc at $z > 8$ to 30 -- 100 cMpc at $z \sim 7$ (10 -- 40 arcmin). Constraints from spectra of quasars near the end of reionization support this picture of patchiness \cite[e.g.][]{2015becker}. Observational clues of this patchy reionization have also been noted in the distribution of LAEs within the EoR and the large-scale bubbles of ionization they may live in. Recent studies report two or more spectroscopically confirmed sources at the same redshift \citep[e.g.][]{2019jung,2020tilvi}, from which a bubble size is inferred based on estimated ionizing radiation encompassing galaxies within that overdensity. For example, \citet{2022larson} find a candidate LAE at $z=8.7$ with Keck/MOSFIRE near a known source at a similar redshift \citep{2015zitrin} in the Extended Groth Strip (EGS), and report the tentative, serendipitous result of an ionized bubble. Further analysis of fainter galaxies within the $z=8.7$ overdensity in EGS suggest this ionized bubble could be fairly large \citep[][Larson et al., in prep.]{2023whitler}. Other studies report apparent overdensities in the EoR --- potentially pointing to large ionized bubbles --- but are limited to uncertain photo-$z$'s to approximate an encompassed comoving volume \citep[e.g.][]{2021endsley}. These sparse observations loosely match theoretical expectations, with the tentative $z=6.8$ bubble \citep{2021endsley} at an estimated radius of $\sim23$\,cMpc, and the potential $z=8.7$ overdensity up to $\sim30$\,cMpc \citep{2022larson,2023whitler}. Recent simulations show that for a fixed ionization fraction, bubble size distributions vary with the dominant source of ionizing output, wherein dominance of low mass haloes produces lots of smaller bubbles and dominance of high mass haloes produces fewer, larger bubbles \citep{2022kannan}. The ionization history is also encoded in the patchiness of reionization; at $z\sim7.5-8$, ionized bubbles should be rare in the late-reionization scenario and more common in the early-reionization scenario \citep[e.g.][]{2019finkelstein}.

The Web Epoch of Reionization \Lya\ Survey (WERLS)\footnote{This survey was originally named \textit{Webb} Epoch of Reionization \Lya\ Survey, in reference to the telescope name,
but later renamed to emphasize the scientific goal of mapping
the cosmic web on large scales as well as to be inclusive and
supportive to members of the LGBTQIA+ community.} is designed to conduct this census of \Lya\ emission in a sample of photometrically-selected UV-bright ($M_{\text{UV}} \lesssim -20$) EoR galaxies in areas covered by \textit{JWST} imaging, on scales large enough to capture bubbles. WERLS is designed to expand the sample of spectroscopically confirmed EoR LAEs at $z\sim5.5-8$ as well as map ionized bubbles in the IGM on large scales. In this paper, we present the first semester of Keck I/MOSFIRE data from WERLS, including spectra for a subset of UV-bright EoR LAEs from $z\sim7-8$. We describe the sample and observations in \textsection 2, and in \textsection 3 we present analysis of the spectroscopic data. In \textsection 4 we detail photometric and spectroscopic characterization of the individual sources, in \textsection 5 we discuss the implications of our measurements, and we present a summary in \textsection 6. The full target list including spectroscopic results and redshift measurements for filler targets are presented in the Appendix. All magnitudes are quoted in the AB system \citep{1983oke}, we assume a Chabrier initial mass function (IMF) \citep{2003chabrier}, and we assume a \textit{Planck} cosmology throughout this paper, adopting $H_0 = 67.74 \rm{\,km\,s^{-1}\,Mpc^{-1}}$ and $\Omega_{\lambda} = 0.691$ \citep{2016collaboration}.

\section{Observations \& Sample}

WERLS is a 29-night NASA key strategic mission support program (PIs: Casey \& Kartaltepe) using two multi-object spectrometers on Keck I, the MultiObject Spectrometer for Infra-Red Exploration \citep[MOSFIRE;][]{2012mclean} and the Low Resolution Imaging Spectrometer \citep[LRIS;][]{1995oke,2010rockosi}. The primary objective of WERLS is to target $\sim800$ galaxy candidates embedded within the latter half of the EoR in order to conduct a census of \Lya\ and correlate the \Lya-inferred location of ionized structures in the IGM to galaxy density maps measured with \textit{JWST}/NIRCam.

By combining these \Lya\ detections with \textit{JWST} imaging (currently being obtained), we can then map the underlying galaxy density distribution in the same areas where we have mapped the inferred ionization state of the IGM. The WERLS program uses two instruments, LRIS, optimal for detecting \Lya\ at $z\lesssim7$, and MOSFIRE in $Y$-band, optimal for detecting \Lya\ at $z\sim7-8$. 

In this paper, we focus only on candidate $z\sim7-8$ LAEs detected using MOSFIRE data collected over the first 5 nights of WERLS in the 2022A semester. Initial results from LRIS observations in 2022A will be presented in a companion paper (Urbano Stawinski et al. in prep).

\subsection{Sample Selection}

Our spectroscopic targets have been selected specifically across three extragalactic fields that have approved deep near-infrared imaging from \textit{JWST}/NIRCam during its first year of observations: the COSMOS-Web Cycle 1 program \citep[0.54 deg$^2$, GO\#1727, PIs: Casey \& Kartaltepe,][]{2022casey}, the Cosmic Evolution Early Release Science Survey \citep[CEERS, 0.03\,deg$^2$, ERS\#1345, PI: S. Finkelstein,][]{2022finkelstein}\footnote{Available for download at \href{ceers.github.io/releases.html}{ceers.github.io/releases.html} and on MAST as High Level Science Products via \href{doi:10.17909/z7p0-8481}{doi:10.17909/z7p0-8481}}, and the Public Release IMaging for Extragalactic Research Cycle 1 program (PRIMER, 0.07\,deg$^2$, GO\#1837, PI: J. Dunlop) in the Cosmic Assembly Near-infrared Deep Extragalactic Legacy Survey (CANDELS) regions of the COSMOS and Ultra Deep Survey (UDS) fields. These fields together encompass 0.7\,deg$^2$ and constitute the largest (by area) extragalactic surveys planned in the first year of \textit{JWST} observations. Here we describe the target selection used for the MOSFIRE observations in each of the three fields.

Within the COSMOS-Web footprint, we first select WERLS targets via deep ground-based imaging from the COSMOS2020 catalog \citep[][]{2022weaver}. These COSMOS targets include EoR candidates selected via well-constrained photometric redshifts from analysis of all \textit{Spitzer} data \citep{2022euclidcollaboration} in addition to deep near-infrared and optical imaging from UltraVISTA DR4 \citep[][]{2012mccrackena} and Subaru/Hyper Suprime-Cam, which increases depth by $\sim1\,$mag relative to previous observations in the same field \citep[][]{2016laigle}. We perform photometric selection using the \texttt{Farmer} photometry \citep[][]{2019weaver} and use photometric redshifts fit using \textsc{LePHARE} \citep[][]{2011arnouts,2006ilbert,1999arnouts}. The total WERLS sample (selected for both LRIS and MOSFIRE) has been selected within the COSMOS-Web footprint as $J < 26$ continuum sources with $z_{\rm phot} > 6$ with $\geq95$\% of their redshift probability density distribution (PDF) above $z = 5.5$, a conservative lower redshift bound to the end of reionization \citep[][]{2015becker}. WERLS targets are selected to be UV-bright; 90\% of the sample have $M_{\text{UV}} \lesssim -20$ if confirmed at their photometric redshifts; this is roughly equal to the characteristic magnitude $M_{*}$ of the luminosity function at these redshifts \citep{2019finkelstein}.

Additional targets within COSMOS as well as target selection in the Extended Groth Strip (EGS) and Ultra Deep Survey (UDS) fields have been selected from deep \textit{Hubble Space Telescope} (\textit{HST}) and \textit{Spitzer} imaging from the CANDELS fields \citep[][]{2011grogin,2011koekemoer,2015ashby} with $J < 26$ and with the same redshift criteria of $z_{\rm phot} > 6$ with 95\% of their redshift PDF above $z = 5.5$. This photometric selection for CANDELS sources utilizes \textsc{SourceExtractor} photometry \citep[][]{1996bertin} and \textsc{EAzY} redshift PDFs \citep[][]{2008brammer}. Given the increased depth of CANDELS near-infrared observations compared to ground-based observations, we also include slightly fainter EoR candidates ($J<27.5$) but include them as fillers rather than primary targets. In these CANDELS regions we used the photometric catalogs and photometric redshift results from \citet{2022finkelstein}.  We selected $z > 6$ galaxies using a modified version of the \citet{2022finkelstein} selection criteria (which had been optimized for $z > 8.5$), requiring $z\_\rm{best} > 6$, and 95\% of the redshift PDF above $z=5.5$.

Lower-redshift filler targets were included in the sample to increase the efficiency of observations, selected by magnitude and photometric redshift from the COSMOS2020 and CANDELS photometric catalogs. We select three categories of filler targets based on different redshift ranges optimized for other prominent emission lines: \Ha-emitters at $0.5<z<0.7$, \Oii-emitters at $1.6<z<2.0$, and \Ciii-emitters at $4.1<z<4.9$. Spectroscopic information for the 166 filler targets (an average of $\sim18$ fillers per slitmask, just over half of the targets per mask) are presented in the appendix, along with any new spectroscopic redshift measurements. Stars (used for alignment and for flux calibration) were also placed on each mask; these are taken from Gaia DR3 \citep{2023gaiacollaboration} and registered to the same reference astrometry as our source catalogs.

\begin{figure*}
    \centering
    \includegraphics[angle=0,trim=0in 0in 0in 0in, clip, width=1.\textwidth]{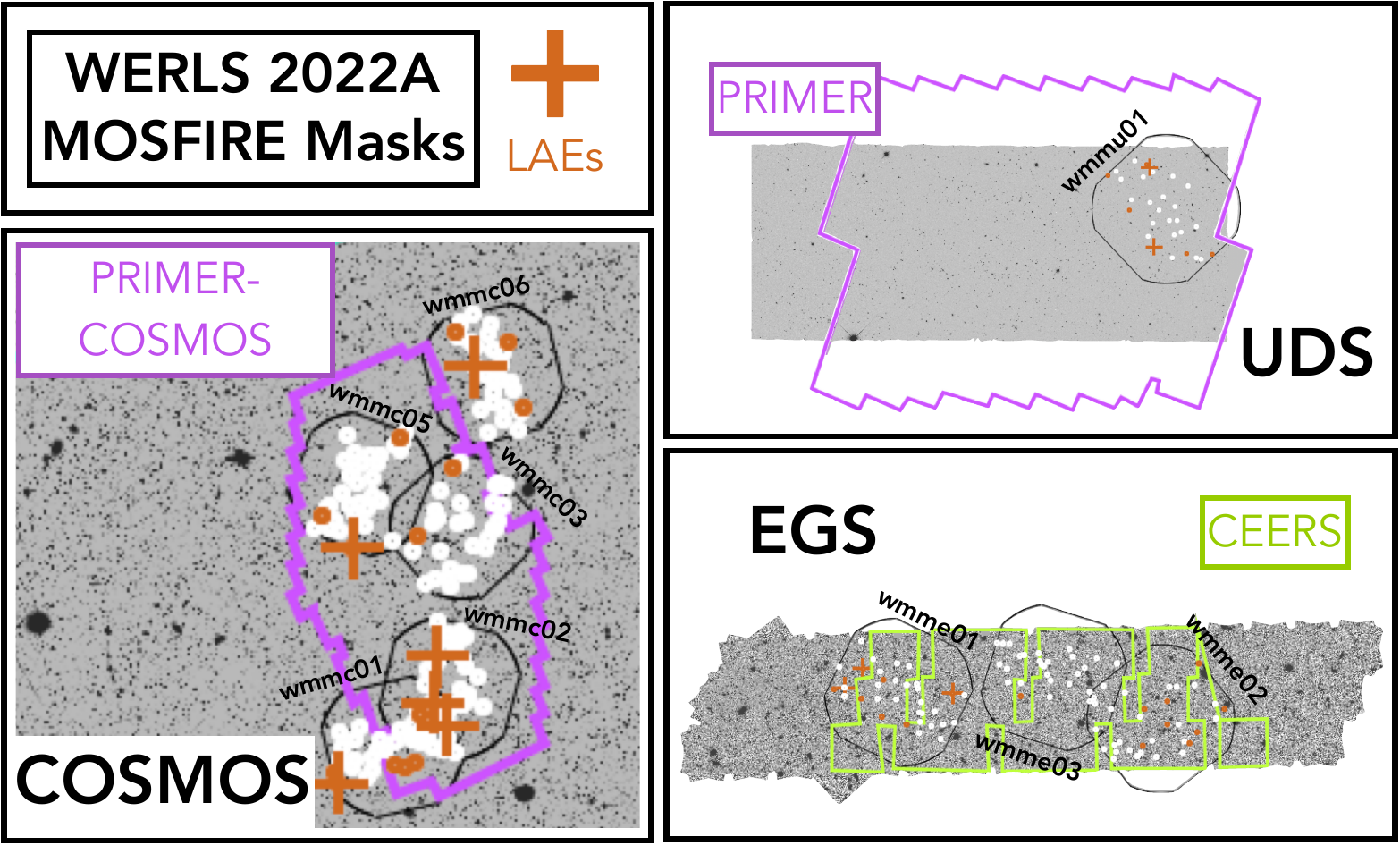}
    \caption{Positions of our WERLS 2022A MOSFIRE observations with the field of view of each MOSFIRE mask in black boxes (labeled by mask name) and all targets in each mask marked as white points. MOSFIRE Primary targets are noted as orange points, and the EoR LAEs reported in this paper are marked with orange pluses. In each field --- COSMOS, UDS, and EGS --- the \textit{JWST} coverage is overlaid with PRIMER in purple and CEERS in chartreuse. COSMOS-Web coverage extends beyond the entire image of COSMOS shown here. The figures are projected on the CANDELS \textit{HST}/F160W images for EGS and UDS, and the COSMOS2020 chimean image for COSMOS.}
    \label{fig:fields}
\end{figure*}

From this COSMOS2020 and CANDELS-based target list, we then designed optimized slitmask configurations in the MOSFIRE Automatic GUI-based Mask Application (\textsc{MAGMA}\footnote{https://www2.keck.hawaii.edu/inst/mosfire/magma.html}) for our Keck/MOSFIRE observations. The fields and targets observed relative to the \textit{JWST} deep fields in which they lie are shown in Figure \ref{fig:fields}. Mask pointings were selected to maximize the number of EoR targets on slits, and in general, targets were prioritized by brightness ($J_{\rm mag}$ as measured in UltraVISTA $J$-band or \textit{HST} WFC3/F125W, drawn from the survey from which a given source was selected) and photometric redshift corresponding to emission lines falling within the MOSFIRE $Y$-band wavelength coverage. Each mask also had at least one star placed on a slit to monitor seeing conditions and potential pointing drift throughout observations. 

The broader WERLS Primary target sample (\Lya\ targets selected photometrically at $z_{\rm phot} \gtrsim 5.5$) is optimized for observability with both LRIS and MOSFIRE; targets are then sorted into subsamples based on their redshift PDFs:

\begin{enumerate}
    \item Primary MOSFIRE targets: $\gtrsim50$\% of their photometric redshift PDF within $7.0<z<8.2$, corresponding to the MOSFIRE $Y$-band wavelength coverage for \Lya\ emission
    \item Primary LRIS targets: $\gtrsim50$\% of their photometric redshift PDF at $z<7$, corresponding to the LRIS wavelength coverage for \Lya\ emission
    \item Primary targets for both MOSFIRE+LRIS: broader redshift PDFs split roughly evenly between the wavelength ranges of both instruments, with $\sim$25\% of their photometric redshift PDFs within $7.0<z<8.2$
\end{enumerate}

Selection for these three categories resulted in 114 WERLS Primary \Lya\ targets across our nine 2022A MOSFIRE pointings. From these 114 WERLS Primary targets, 33 were MOSFIRE Primary targets, 54 were both MOSFIRE+LRIS Primary targets, and 27 were LRIS Primary targets. Given the breadth of the photometric redshift PDFs, we expect some LRIS Primary targets might be detected in \Lya\ emission within the MOSFIRE wavelength range, and vice versa; so, lower-redshift EoR candidates are still added to MOSFIRE masks as high priority fillers, though they do not drive the choice of pointing, which was based on the higher-$z$ MOSFIRE Primary subsample. Further, depending on specific LRIS and MOSFIRE mask design, a subset of the WERLS Primary targets were observed with both instruments in order to capture a wider wavelength range for possible \Lya\ detection. 

Primary target source density was highest within the CANDELS fields due to their depth, and as a result, COSMOS pointings for 2022A were clustered in the PRIMER-COSMOS area. The effective area covered by the nine slitmasks is $\sim0.05\,$deg$^2$ across the three fields, with the majority of the covered area in COSMOS. In total for the nine MOSFIRE masks observed in 2022A, 276 galaxy candidates (114 WERLS Primary and 162 fillers) and 15 stars were placed on slits. 

\subsection{Keck/MOSFIRE Observations}

\begin{deluxetable*}{ccclccccccc}
\setlength{\tabcolsep}{0.03in}
\tablecaption{Summary of 2022A MOSFIRE Observations \label{tab:obs}} 
\tablehead{
\colhead{Mask Name} & \colhead{RA} &\colhead{Dec} & \colhead{Date(s) Observed\tablenotemark{a}} & \colhead{$N_{\rm targs}$\tablenotemark{b}} & \colhead{$N_{\rm fillers}$\tablenotemark{c}} & \colhead{$N_\text{seq}$} & \colhead{Slit Width} & \colhead{Seeing\tablenotemark{d}} & \colhead{Airmass} & \colhead{$5\sigma$ Depth\tablenotemark{e}}\\
\colhead{} & \colhead{} & \colhead{} & \colhead{(UTC)} & \colhead{}  & \colhead{}  & \colhead{(\# ABBA)} & \colhead{(arcsec)} & \colhead{(arcsec)} & \colhead{} & \colhead{(mag)}}
\startdata
{wmmu01} & {02:17:04.68} & {--05:10:30.00} & {Feb 12, 13, 14} & {5 (23)} & {9} & {14} & {0.7--1.0} & {0.7--1.2} & {1.4} & {20.4}\\
{wmmc06} & {10:00:16.13} & {+02:29:18.60} & {Apr 17} & {2 (6)} & {19} & {15} & {1.0} & {0.9--1.3} & {1.0} & {20.0}\\
{wmmc03} & {10:00:22.34} & {+02:21:57.60} & {Feb 13, 14 \& Mar 14} & {9 (13)} & {10} & {33} & {0.7--1.0} & {0.9--1.3} & {1.7} & {20.7}\\
{wmmc02} & {10:00:24.60} & {+02:14:19.68} & {Feb 12, 13} & {6 (13)} & {13} & {20} & {0.7--1.0} & {0.7--1.3} & {1.6} & {20.5}\\
{wmmc01} & {10:00:35.47} & {+02:11:31.92} & {Feb 12} & {4 (14)} & {13} & {20} & {0.7} & {0.6--0.9} & {1.6} & {21.4}\\
{wmmc05} & {10:00:41.57} & {+02:24:03.96} & {Feb 14 \& Mar 14} & {3 (8)} & {18} & {32} & {0.7--1.0} & {0.7--1.3} & {1.1} & {21.3}\\
{wmme02} & {14:19:24.05} & {+52:48:20.88} & {Apr 17} & {1 (20)} & {11} & {17.5} & {1.0} & {0.8--1.3} & {1.2} & {19.9}\\
{wmme03} & {14:19:35.83} & {+52:53:24.36} & {Mar 14} & {1 (8)} & {25} & {8} & {0.7} & {0.6--0.9} & {1.2} & {20.0}\\
{wmme01} & {14:20:10.44} & {+52:58:28.92} & {Feb 12, 13, 14 \& Mar 14} & {3 (12)} & {21} & {20} & {0.7--1.0} & {0.7--1.2} & {1.2} & {20.3}\\
\enddata
\tablenotetext{}{
$^{a}$\footnotesize All dates listed are from the year 2022.\\
$^{b}$\footnotesize Number of MOSFIRE Primary targets on mask (excludes filler targets); in parentheses we give the total number of WERLS EoR targets (some of which are fainter than the primary sample).\\
$^{c}$\footnotesize Number of filler targets on mask.\\
$^{d}$ Seeing measured from the full-width half maximum (FWHM) estimated from continuum object (bright star) placed on each science mask.\\
$^{e}$ Limiting $5\sigma$ magnitude measured via $Y$-band magnitude and MOSFIRE SNR of bright star placed on each science mask.\\}

\end{deluxetable*}

Observations were taken with MOSFIRE \citep[][]{2012mclean} on the Keck I telescope using the $Y$-band spectroscopic filter to optimize for detection of \Lya\ redshifted to $7.0<z<8.2$. Observations were obtained over 5 nights in 2022A: 2022-Feb-12, 2022-Feb-13, 2022-Feb-14, 2022-Mar-14, and 2022-Apr-17 UTC. Individual science frames were taken with 180\,sec exposures, with a goal of $\sim4$\,hr of total exposure time per mask. We use a standard ABBA dither pattern with nod distance of $1\farcs25$ (unless contaminants landed on high priority slits at this distance, then $1\farcs5$ nods were taken). Exposures in $Y$-band were 180\,sec and taken in MCDS sampling mode with 16 reads; for each mask we aim for 20 ABBA sequences to achieve a nominal 4 hours total exposure time, but there are cases where we took more sequences if weather conditions were not ideal. 

Four of the nine slitmasks were not observed for the full 20 sequences mostly due to limited observability due to weather; observational details for each MOSFIRE slitmask are listed in Table \ref{tab:obs}. We adopted slit widths of 0\farcs7 (affording a spectral resolution of R $\sim$ 3500) or 1\farcs0 depending on the seeing during observations. The seeing varies through the nights from 0\farcs6 to 1\farcs3, as measured directly from our spectroscopic data using the monitoring star. Weather conditions varied across the 5 nights: 2022-Feb-12 was clear with good seeing throughout the night, 2022-Feb-13 was clear with variable seeing throughout the night, 2022-Feb-14 had wind, fog, and snow, leading to the dome closing twice and poor seeing, 2022-Mar-14 was clear with good seeing, and 2022-Apr-17 began with high humidity and delays in the dome opening followed by some cloud cover the rest of the night, leading to moderate seeing.

\subsection{MOSFIRE Spectroscopic Data Reduction}

The data were reduced using two reduction pipelines independently to ensure robust noise characterization in order to build confidence in our candidate faint line detections for these high redshift targets.

First, we use the \texttt{PypeIt} data reduction package \citep[][]{2020prochaskaa}, which is designed to be a general use spectroscopic pipeline and can be used for a range of instruments and facilities. We iteratively worked with the \texttt{PypeIt} team to determine the optimal data reduction configuration for Keck I/MOSFIRE parameters in \texttt{PypeIt} and reduce the data in ABBA sequence blocks, using the spectral trace of a bright star in one of the slits on each mask as a position reference. The output from \texttt{PypeIt} is reduced and co-added 2D spectra, from which we optimally extract 1D spectra at the centroid of the emission line using the technique of \citet[][]{1986horne}, with a 7 pixel (1\farcs26) spatial aperture, matched to the typical seeing FWHM level from our observations. In a few cases, candidate emission line detections were slightly offset by 1-2 pixels from the target position, however in all cases the candidate line was $<0.3''$ away from the source centroid, well within expectations given positional accuracy and possible \Lya\ emission offsets from the broadband imaging centroid. We also extract 1D spectra using a boxcar for comparison, but typically achieve a higher SNR using optimal extraction; we ultimately adopt the optimally extracted 1D spectra for our measurements.

Second, we use the public MOSFIRE data reduction pipeline (\textsc{MosfireDRP}\footnote{https://keck-datareductionpipelines.github.io/MosfireDRP/}) to reduce the raw data. The \textsc{MosfireDRP} pipeline provides a sky-subtracted, flat-fielded, and rectified 2D slit spectrum per slit object. The reduced spectra are wavelength-calibrated using telluric sky emission that is built specifically for the instrument. We extract 1D spectra from the combined 2D spectra via both optimal and boxcar extraction schemes as above, ultimately adopting the optimally extracted 1D spectra.

\begin{figure*}[ht!]
    \centering
    \includegraphics[angle=0,trim=0in 0in 0in 0in, clip, width=1.\textwidth]{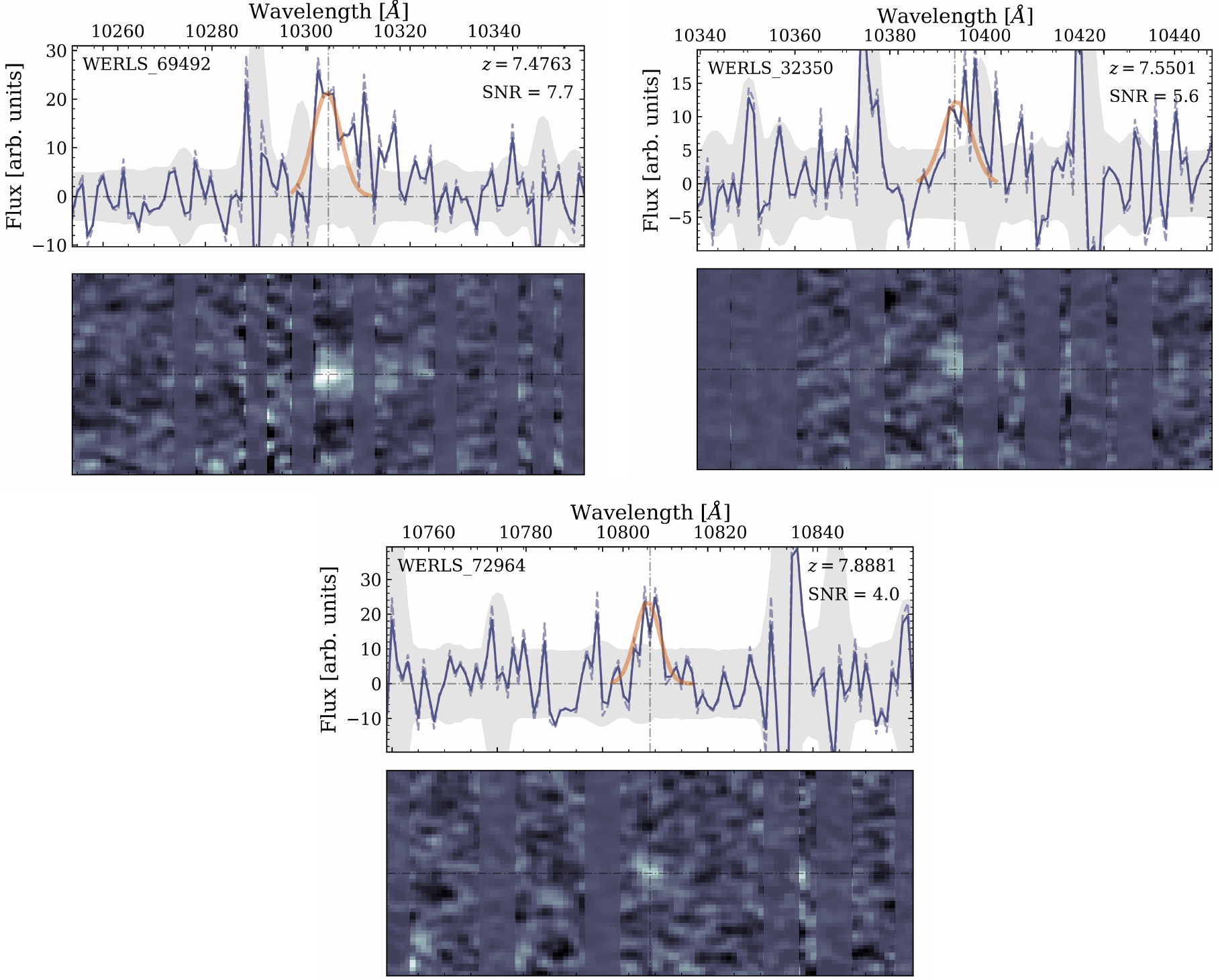}
    \caption{WERLS MOSFIRE $Y$-band spectra for the three secure LAE detections, with the 1D spectrum above and corresponding 2D spectrum (matched in wavelength space) below for each source. Each 1D spectrum shows the Gaussian-smoothed signal as a blue solid line, the unsmoothed signal as a blue dashed line, the error spectrum in solid gray, and the Gaussian fit to the \Lya\ emission line in orange. Each panel lists the source ID, the \Lya-derived spectroscopic redshift, and the SNR of the \Lya\ detection for each source. In the 2D spectra, sky lines are marked with blue bars, and the location of the line both spatially and spectrally is marked with the dashed crosshairs. These \Lya\ detections are categorized as secure because both their spectra and SED fits are robust (see \textsection 3.2 for more details and \textsection 4 for individual source notes).}
    \label{fig:securespec}
\end{figure*}

\begin{figure*}[ht!]
    \centering
    \includegraphics[angle=0,trim=0in 0in 0in 0in, clip, width=0.77\textwidth]{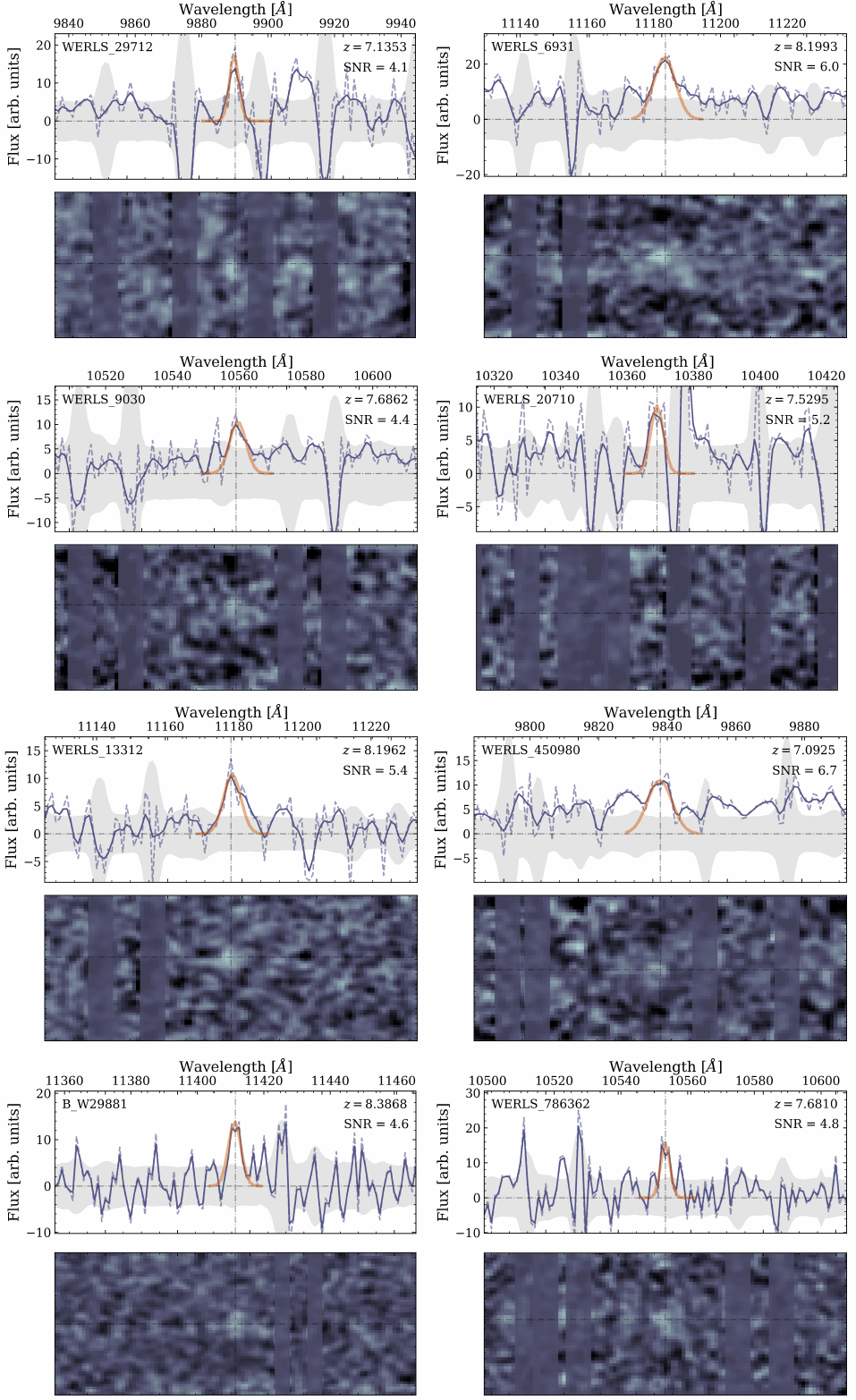}
    \caption{WERLS MOSFIRE $Y$-band spectra for the eight tentative LAE detections, with markers styled as in Figure \ref{fig:securespec}. These \Lya\ detections are categorized as tentative because their spectra and/or line identifications are less robust than the secure detections (see \textsection 3.2 for more details and \textsection 4 for individual source notes).}
    \label{fig:tentativespec}
\end{figure*}

While some teams have reported that there is non-negligible slit drift in the spatial direction (up to $\sim1$pixel hr$^{-1}$) in their MOSFIRE observations that necessitated correction \citep[e.g.,][]{2015kriek, 2016song, 2019jung, 2020hutchison, 2022larson}, we checked slit alignments carefully during the course of our observations using our reference star and found no significant slit drift. Note that all of our masks were only observed for a maximum of four hours where many datasets that see significant drift coadd data taken over a longer duration. While drift is automatically accounted for in our \texttt{PypeIt} reductions, it is not accounted for in our \textsc{MosfireDRP} reductions, except when co-adding data from different nights where it was critical to account for a global offset in the final reduced product.

\section{Spectroscopic Analysis}

In this first WERLS paper, we focus only on the goal of taking a census of LAEs in the EoR, limited to the WERLS MOSFIRE 2022A dataset in-hand. Here, we present the MOSFIRE spectroscopic constraints for this initial sample.

\subsection{Lyman-$\alpha$ Emission Line Vetting}

Our candidate \Lya\ emission lines were all first identified via visual inspection of the 2D spectra initially from either the \textsc{MosfireDRP} or \texttt{PypeIt} reduction. Inclusion in this paper as a tentative or secure detection requires the candidate emission line to be present at $\geq4\sigma$ in both independent reductions, which reinforces our confidence in the robustness of the detection. We check for a bright positive signal with spatial width well-matched to seeing and spectral width at least as broad as one spectral resolution element. Such a positive signal then has an integrated SNR greater than the average noise \citep[with a threshold of $\geq4\sigma$, consistent with other \Lya\ spectroscopic works, e.g.][]{2022jung}, with OH forest sky emission features masked out. We verify that candidate emission line detections also have negative signals; real astronomical signal should be accompanied by symmetric negative signals with SNR $\sqrt{2}$ lower than the positive signal, spatially offset above and below the positive signal at the expected separation based on the mask's adopted nod amplitude. Using these criteria, we inspect both data reduction products to ensure the feature is not an obvious artifact. All LAEs reported in this paper satisfied our criteria in both reductions; in some cases the SNR of the line varied slightly within the noise between reductions. 

To ensure the nature of the emission line candidates as (spectrally isolated) \Lya, we search for multiple emission lines in the $Y$-band spectra, which implies the original line is not \Lya\ and the source is at a lower redshift. A likely contaminant in this case is a source with \Oii$\lambda\lambda3727,3729$ emission; galaxies with \Oii\ emission in $Y$-band may be misidentified as EoR galaxies if the Balmer break is mistaken for the \Lya\ break in a photometric redshift solution. However, in this case, given the spectral resolution of our MOSFIRE data, this doublet should be resolvable (at $7-8\,\rm \AA$ at the expected \Oii-emitter redshift of $z\sim1.8$) and therefore distinguishable from \Lya\ for uncontaminated (with respect to sky lines) spectral regions broader than the doublet width. For all candidate lines, we also verify alignment on the slit mask by optimally extracting the 1D spectrum at the spatial position of the detection, and verify the spatial positions of the negatives match expectations given the nod amplitude for those observations.

To further assess credibility of the \Lya\ identification, we consider the available photometric constraints. Any excess emission in Spitzer/IRAC Channel 2 (4.5\um) over Spitzer/IRAC Channel 1 (3.6\um) provides increased credibility for our \Lya\ emission line candidates, as the presence of strong \Oiii+\Hb emission would cause an ``IRAC excess'' from $z\sim7-9$, encompassing our entire target redshift range. Additionally, by examining \textit{HST}/UltraVISTA cutouts of the target showing the slit overlay, we check for potential low-$z$ contaminants by ruling out any large, bright targets at or near the target position on the slit that could be serendipitous sources of emission lines.

\subsection{LAEs}

From this first round of emission line searching and vetting, we find 35 candidate EoR LAEs out of 114 primary targets, for which we compiled all available spectroscopic and photometric information. This compiled information was visually inspected and vetted by 22 of this paper's co-authors independently. Each of the 22 co-authors commented on the source and ranked its quality as ``Bogus'', ``Very Tentative'', ``Tentative'', or ``Secure'', which were then assigned numerical values 1-4. From this inspection, we settled on 11 total LAEs to present in this paper based on our aggregate confidence in their reliability. This sample of 11 was divided into two categories: secure ($N=3$) and tentative ($N=8$). We ultimately categorize sources using all available data with a holistic assessment, described as follows. Secure sources were strongly detected in both \textsc{MosfireDRP} and \texttt{PypeIt} reductions, have a clear \Lya\ break in their photometry, and have spectroscopic redshift solutions well-matched to the photometric redshift PDF(s). All secure sources also had average scores above 3.0/4.0 from the co-author vetting survey, wherein a maximum score of 4.0 reflects the scenario wherein all 22 co-authors voted the source as ``Secure''. Tentative sources were so classified because their \Lya\ lines were lower SNR, their photometric redshift may not align well with the identified line, or some other reason that casts the security of the identification in doubt. Justification of placement in the secure or tentative subsample is described on a per-source basis below.

The $Y$-band spectra of the secure sample is shown in Figure \ref{fig:securespec}, and the spectra of the tentative sample is shown in Figure \ref{fig:tentativespec}. Redshift solutions for our 11 LAEs are found by Gaussian fit (rather than an asymmetric Gaussian as our signals are faint) to the centroid of the emission line, with no velocity offset applied. Target information and redshifts are listed in Table \ref{tab:LAE}, and characteristics for each source are detailed further in \textsection4.

\begin{deluxetable*}{ccccccccccccc}
\setlength{\tabcolsep}{0.05in}
\tablewidth{1.\textwidth}
\tabletypesize{\footnotesize}
\tablecaption{LAEs and Redshift Information \label{tab:LAE}} 
\tablehead{
\colhead{ID} & \colhead{RA} &\colhead{Dec} & \colhead{$z_{\rm phot}$} & {$M_{\rm UV}$} & {$\beta$} & {SFR}  & {$M_\star$} & \colhead{Mask} & \colhead{$z_{\rm{Ly}\alpha}$} & \colhead{SNR}  & \colhead{Other refs.} \\
	\colhead{}  & \colhead{J2000}  & \colhead{J2000}  & \colhead{}  & \colhead{ABmag} & \colhead{}  & \colhead{\Msun\,yr$^{-1}$} & \colhead{log \Msun} & \colhead{} & \colhead{} & \colhead{} & \colhead{}\\
	\colhead{(1)}  & \colhead{(2)} & \colhead{(3)} & \colhead{(4)} & \colhead{(5)} & \colhead{(6)} & \colhead{(7)}  & \colhead{(8)} & \colhead{(9)} & \colhead{(10)} & \colhead{(11)} & \colhead{(12)} 
}
\startdata
WERLS\_72964 & 10:00:19.40 & +02:29:36.64 & $8.5_{-0.7}^{+0.2}$ & $-23.14_{-0.04}^{+0.04}$ & --- * &  --- * & --- * & wmmc06 & 7.8881 & 4.0 & \\ 
WERLS\_69492$\dagger$ & 14:20:12.08 & +53:00:26.82 & $7.8_{-0.4}^{+0.2}$ & $-21.85_{-0.04}^{+0.04}$ & $-1.79_{-0.13}^{+0.16}$ & $90_{-20}^{+40}$ & $9.7_{-0.2}^{+0.3}$ & wmme01 & 7.4763 & 7.7 & RB16, S17, J22, J23 \\ 
WERLS\_32350$\dagger$ & 14:19:59.77 & +52:56:31.09 & $8.3_{-1.3}^{+0.2}$ & $-21.86_{-0.11}^{+0.12}$ & $-2.1_{-0.3}^{+0.3}$ &  $40_{-20}^{+50}$ & $9.5_{-0.3}^{+0.5}$ & wmme01 & 7.5501 & 5.6 & J22, J23 \\
\hline
    WERLS\_29712 & 02:17:06.97 & --05:12:15.77 & $6.8_{-5.5}^{+0.1}$ & $-20.59_{-0.04}^{+0.05}$ & $-1.96_{-0.04}^{+0.05}$ & $0.1_{-0.1}^{+0.4}$ & $9.5_{-0.3}^{+0.3}$ & wmmu01 & 7.1353 & 4.1 & \\
    WERLS\_6931 & 02:17:07.82 & --05:08:35.09 & $6.8_{-5.3}^{+0.8}$ & $-20.23_{-0.06}^{+0.07}$ & $-2.6_{-0.06}^{+0.08}$ &  $4_{-1}^{+1}$ & $8.5_{-0.2}^{+0.3}$ & wmmu01 & 8.1993 & 6.0 & \\
    WERLS\_9030 & 10:00:24.79 & +02:12:28.66 & $7.0_{-5.6}^{+0.4}$ & $-21.95_{-0.05}^{+0.05}$ & $-2.5_{-0.2}^{+0.1}$ &  $24_{-18}^{+5}$ & $9.1_{-0.1}^{+0.5}$ & wmmc02 & 7.6862 & 4.4 & \\
    WERLS\_20710 & 10:00:26.71 & +02:15:47.20 & $6.5_{-4.8}^{+0.08}$ & $-20.64_{-0.05}^{+0.07}$ & $-1.96_{-0.11}^{+0.16}$ & $10_{-4}^{+7}$ & $9.2_{-0.1}^{+0.1}$ & wmmc02 & 7.5295 & 5.2 & \\
    WERLS\_13312 & 10:00:27.42 & +02:13:35.54 & $8.1_{-1.5}^{+0.1}$ & $-20.21_{-0.17}^{+0.19}$ & $-1.6_{-0.6}^{+1.1}$ & $10_{-10}^{+70}$ & $9.7_{-0.7}^{+1.0}$ & wmmc02 & 8.1962 & 5.4 & \\
    WERLS\_786362 & 10:00:42.72 & +02:20:58.85 & $7.41_{-0.12}^{+0.09}$ & $-22.1$** & $-3.15$** & ---** & $9.4$** & wmmc05 & 7.6810 & 4.8 & \\ 
    WERLS\_450980 & 10:00:45.58 & +02:09:43.34 & $7.3_{-0.8}^{+0.1}$ & $-22.07_{-0.02}^{+0.02}$ & $-2.10_{-0.04}^{+0.03}$ & $65_{-5}^{+5}$ & $9.4_{-0.1}^{+0.1}$ & wmmc01 & 7.0925 & 6.7 & \\ 
    WERLS\_29881 & 14:20:20.28 & +53:00:31.28 & $6.4_{-4.9}^{+0.4}$ & $-19.81_{-0.14}^{+0.18}$ & $-2.2_{-0.3}^{+0.5}$ & $5_{-4}^{+9}$ & $8.9_{-0.5}^{+0.7}$ & wmme01 & 8.3868 & 4.6 &  \\ 
\enddata
\tablecomments{Columns: (1) WERLS object ID, (2) Right ascension, (3) Declination, (4) peak \textsc{EAzY} photometric redshift and inner 68\% uncertainty, (5) UV magnitude measured from \texttt{Bagpipes} with inner 68\% uncertainty, (6) UV slope ($\beta$) measured from \texttt{Bagpipes} with inner 68\% uncertainty, (7) star formation rate (SFR) measured from \texttt{Bagpipes} with inner 68\% uncertainty, (8) stellar mass ($M_\star$) measured from \texttt{Bagpipes} with inner 68\% uncertainty, (9) WERLS MOSFIRE mask for object, (10) spectroscopic redshift measured from \Lya\ emission line, (11) \Lya\ emission line detection signal-to-noise ratio (SNR), and (12) other literature works that report \Lya\ emission for the source (RB16 = \citet{2016roberts-borsani}, S17 = \citet{2017stark}, J22 = \citet{2022jung}, J23 = \citet{2023jung}).\\ 
*WERLS\_72964 has blended IRAC photometry, therefore we do not report physical properties for the source given its contaminated infrared photometry. \\
**Estimated properties apply if WERLS\_786362 is indeed an EoR galaxy and not a brown dwarf; we are unable to constrain a physically plausible SFR and do not report uncertainties as properties are taken directly from a generated galaxy model rather than estimated from a best-fit model to the data (see \textsection4.2.4. for more details). \\
$\dagger$Sources with secure multi-line systemic redshifts from CEERS \textit{JWST}/NIRSpec data.}
\end{deluxetable*}

\section{Characterization of LAEs}

\subsection{Photometric Characterization}

Photometric measurements are drawn from the deepest available catalog, which for the majority of our sample, were the CANDELS catalogs \citep{2011grogin,2011koekemoer,2015ashby}. For one source in COSMOS that is not in the CANDELS catalog (WERLS\_450980), we use COSMOS2020 photometry \citep{2022weaver}. Only one source was in both the CANDELS and COSMOS2020 catalogs (WERLS\_786362), however, this source also had \textit{JWST} imaging available. 

We search publicly available imaging from \textit{JWST} \citep{2023rigby,2023menzel,2023mcelwain}; at the time of writing, \textit{JWST}/NIRCam imaging data \citep{2023rieke} existed for five sources: WERLS\_9030, WERLS\_20710, \& WERLS\_786362 in PRIMER-COSMOS, and WERLS\_29712 \& WERLS\_6931 in PRIMER-UDS. Reduction of the PRIMER data was carried out as in Franco et al. (in prep). The raw NIRCam imaging in PRIMER-COSMOS was reduced with \textit{JWST} Calibration Pipeline version Pipeline 1.10.0 \citep{2022bushouse}, with the Calibration Reference Data System (CRDS)\footnote{\href{jwst-crds.stsci.edu}{jwst-crds.stsci.edu}} pmap-1075, which corresponds to NIRCam instrument mapping imap-0252. For the imaging in PRIMER-UDS, we use the publicly available \texttt{grizli} reduction \citep{2023grizli}. From the reduced NIRCam imaging, we measure aperture photometry using \texttt{photutils} \citep{2023bradley} from images PSF homogenized to F444W \citep[using \texttt{pypher};][]{2016boucaud}.

\begin{figure*}[ht!]
    \centering
    \includegraphics[angle=0,trim=0in 0in 0in 0in, clip, width=1.\textwidth]{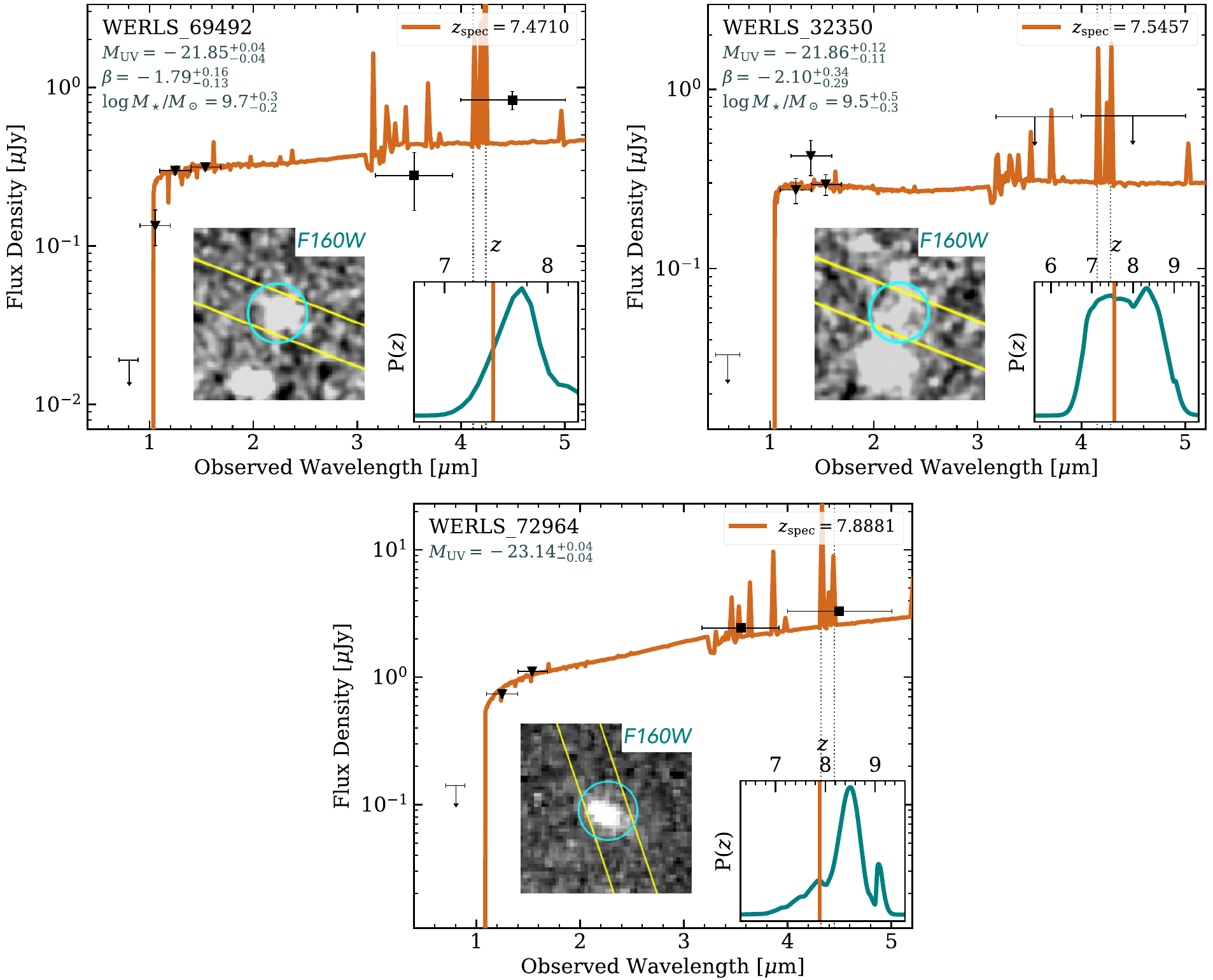}
    \caption{SED fits, image cutout, and photometric redshift PDF for each source in the secure sample. Each primary figure shows the best-fit \texttt{Bagpipes} SED fixed to the spec-$z$ in orange and the photometry in black markers, noted by their instruments wherein circles are ground-based data, triangles are \textit{HST} data, hexagons are \textit{JWST} data, and squares are \textit{Spitzer} data. The wavelength of redshifted \Oiii/\Hb\ is marked with dotted lines. Inset on the left side shows a 3'' cutout (\textit{HST}/F160W), overlaid with the MOSFIRE slit in yellow and the target in cyan. Inset on the right of each primary figure is the \textsc{EAzY} photometric redshift PDF in teal (allowing the redshift to vary) in teal, compared to the spectroscopic redshift solution in orange.}
    \label{fig:securesed}
\end{figure*}

\begin{figure*}[ht!]
    \centering
    \includegraphics[angle=0,trim=0in 0in 0in 0in, clip, width=0.77\textwidth]{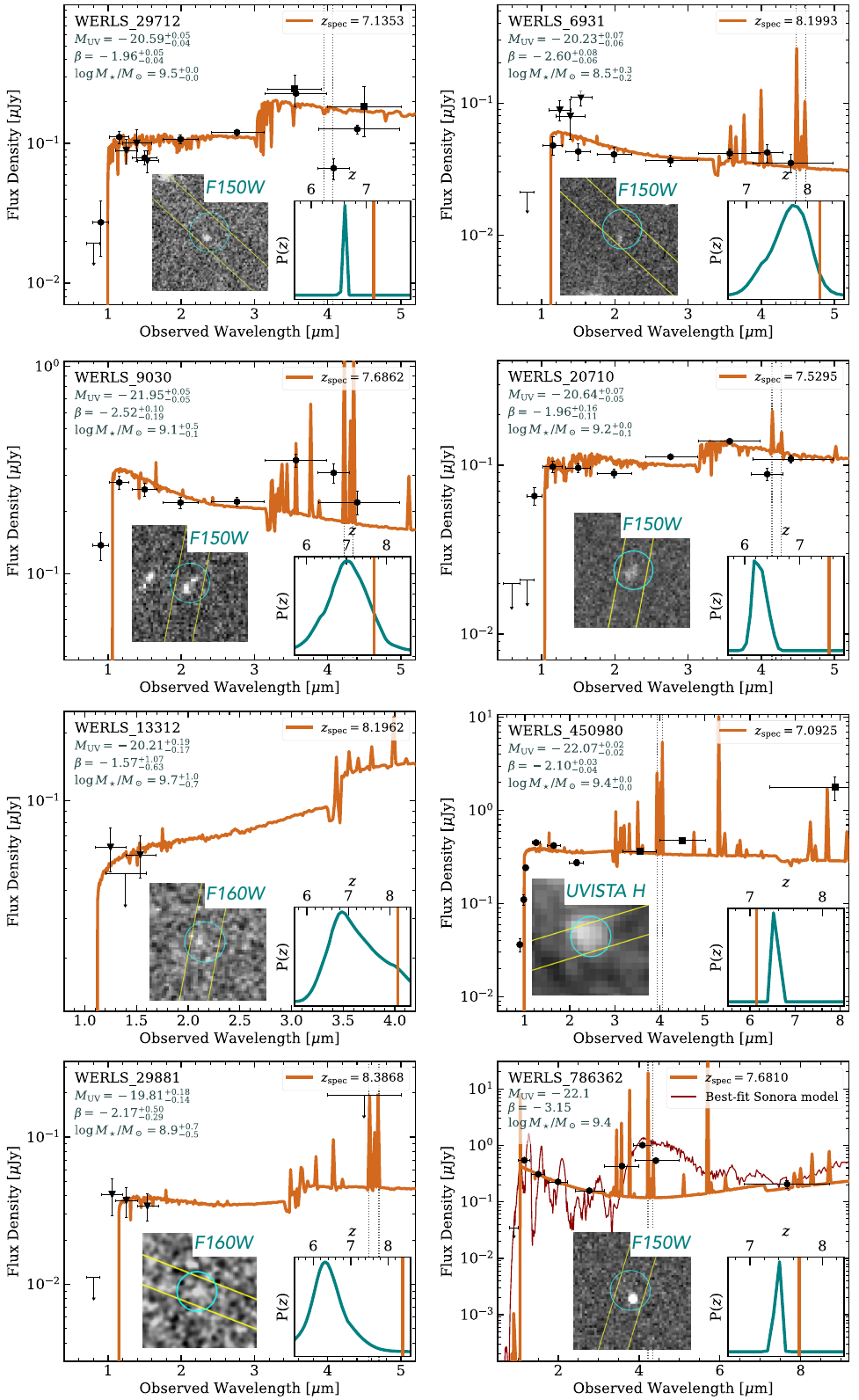}
    \caption{SED fits, image cutout, and photometric redshift PDF for each source in the tentative sample, with format consistent with Figure \ref{fig:securesed}. WERLS\_786362 also shows the best-fit Sonora brown dwarf model in thin red (see \textsection4.2.4 for more details). Here, the cutouts are UltraVISTA/$H$, \textit{HST}/F160W, or \textit{JWST}/F150W as available.}
    \label{fig:tentativesed}
\end{figure*}

We recompute photometric redshift PDFs uniformly for our entire sample of LAEs in order to incorporate the new \textit{JWST} data. We fit each galaxy spectral energy distribution (SED) with \textsc{EAzY} \citep[][]{2008brammer}, which computes linear combinations of pre-defined templates to derive photometric redshift probability distribution functions (PDFs) based on the $\chi^2$ of the templates. Given that our sources are LAEs, we adopt the template set detailed in \citet{2022larson}, with the standard \texttt{tweak\_fsps\_QSF\_12\_v3} set of 12 FSPS \citep{2010conroy} and additional bluer LAE models, designed for high redshift star-forming galaxies like those we target here. Specifically, we use Set 3: Reduced \Lya\ from \citet{2022larson}, which include models with emission lines added but with \Lya\ reduced to 1/10 of that produced by \texttt{CLOUDY} \citep{2017ferland} to emulate a 10\% \Lya\ escape fraction, meant to represent typical $4<z<7$ galaxies. We allow the redshift to vary from $0.01<z<10.0$ with a step size of $\Delta z = 0.01$, and assume no redshift prior in order to derive the photometric redshift PDF. The results of these fits compared to the \Lya-derived spectroscopic redshifts are shown as subpanels in Figure \ref{fig:securesed} (for the secure sample) and in Figure \ref{fig:tentativesed} (for the tentative sample).

While sources were originally selected via photometric constraints derived from SED fitting, with new redshift constraints in hand from spectroscopy, we improve upon these SED characterizations using the Bayesian SED fitting code, \texttt{Bagpipes} \citep{2018carnall}. For each source, we fix the redshift to the \Lya-derived spectroscopic redshift. As we are able to fix the redshifts, we adopt a non-parametric star formation history (SFH) via the \citet{2019leja} continuity SFH model. We adopt 7 age bins, the most recent bin capturing the SFR in the last 10 Myr and a maximum epoch of star-formation at $z=20$. We adopt the bursty continuity prior from \citet{2022tacchella}. We allow the metallicity to vary from $0.001<Z/Z_\odot<2.5$ with a log-uniform prior. We adopt a \citet{2001calzetti} dust attenuation law, and we allow the attenuation to vary from $0.001<\rm{A_V}<3$ with a log-uniform prior. We also include a nebular component; \texttt{Bagpipes} uses the \texttt{CLOUDY} photoionization models \citep{2017ferland} to generate \Hii\ regions, and follows \citet{2017byler} wherein total nebular emission is the sum of emission from \Hii\ regions of different ages. We allow the ionization parameter $\log U$ to vary from -4 to -1, with a Gaussian prior with $\mu,\sigma = (-2, 0.25)$. 

\begin{figure}[h!]
    \centering
    \includegraphics[angle=0,trim=0in 0in 0in 0in, clip, width=0.48\textwidth]{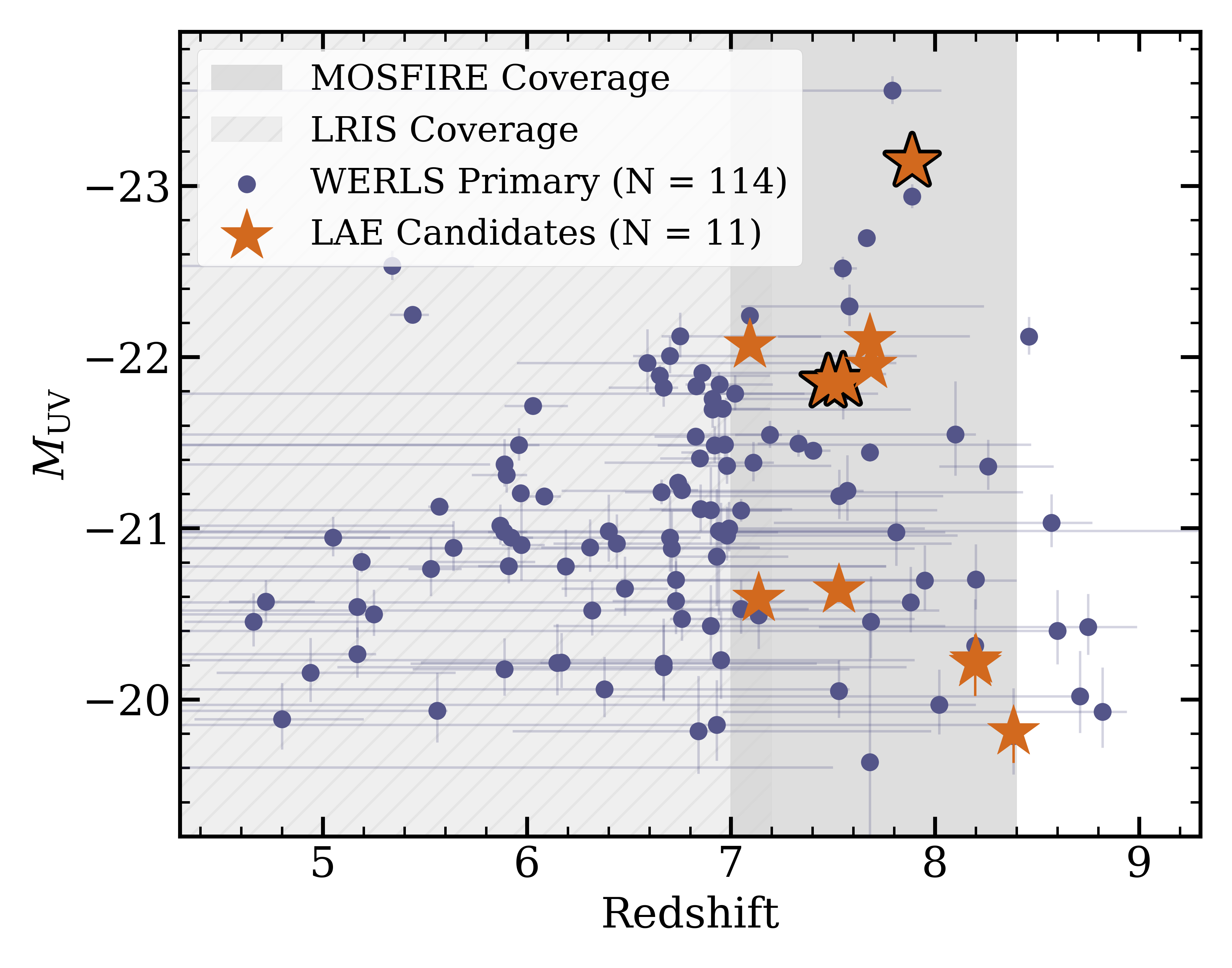}
    \caption{$M_{\text{UV}}$ versus redshift for all galaxies targeted in our MOSFIRE observations. Blue points denote all WERLS Primary targets (here, observed with MOSFIRE but selected for both LRIS and MOSFIRE wavelength ranges) and orange stars show the LAEs reported in this paper (black outlines highlight the secure subsample). Redshifts are photo-$z$'s except for the LAEs, which have \Lya-derived spectroscopic redshifts as reported in this paper. Our targets are selected to be UV-bright --- 91\% of the sample have $M_{\text{UV}} \leq -20$ --- but span a wide range of $M_{\text{UV}}$. The range of redshifted \Lya\ emission accessible with MOSFIRE $Y$-band and LRIS are represented as gray regions.}
    \label{fig:muv}
\end{figure}

From the best-fit \texttt{Bagpipes} model, we calculate the absolute UV magnitude ($M_{\rm UV}$) of each source (listed in Table \ref{tab:LAE}). The distribution of $M_{\rm UV}$ versus best available redshift for the sample is shown in Figure \ref{fig:muv}. Our targets are UV-bright by selection, but span a wide range of UV magnitude, with our spectroscopic sample spanning $-23.14 \leq M_{\rm UV} \leq -19.81$.

We derive physical properties from the best-fit \texttt{Bagpipes} model for each source (note that the following excludes values listed for WERLS\_786362, which are not derived from the best-fit model to the photometry, but instead are taken directly from the generated \texttt{Bagpipes} model). As expected for LAEs, our galaxies are generally star-forming but show a broad range, with star formation rates (SFRs) ranging 0.1-90\,\Msun\, yr$^{-1}$ and a median SFR of 10\,\Msun\, yr$^{-1}$. We estimate a median stellar mass of $\sim2.3\times10^9$\,\Msun\ and ranging $8.5<$ log ($M_\star$/ \Msun) $<9.7$, suggesting our sample represents massive EoR galaxies but not necessarily the most extreme at this epoch. Given their \Lya\ detections and implied ionized photon escape, we expect these galaxies to be fairly blue; this is supported by their measured rest-frame UV slopes ($\beta$), ranging $-2.6<\beta<-1.6$ with a median of $\beta\sim2.1$. The measured properties for the sample are listed in Table \ref{tab:LAE}.

\subsection{Individual Sources}

In the following subsections, we detail the spectroscopic and photometric analysis and properties for each source, and use this to justify their placement in the secure or tentative subsample.

\subsubsection{WERLS\_72964}

WERLS\_72964 is on mask wmmc06 with a SNR = 4.0 \Lya\ emission line corresponding to \zs = 7.8881, and is in our secure sample. It lies within the COSMOS CANDELS coverage, is not in the COSMOS2020 catalog, and lies outside of PRIMER-COSMOS. We fit the target's CANDELS photometry (ID: 72964) and find a photometric redshift consistent with the spectroscopic solution within uncertainty, with \zp $=8.5^{+0.2}_{-0.7}$. Further, the source is detected significantly in IRAC Channel 1 (3.6\um) and Channel 2 (4.5\um), with some photometric excess (0.33\,mag) in Channel 2 where there should be contamination from \Oiii+\Hb nebular emission at both the spectroscopic and photometric redshift solutions. Co-author vetting resulted in a score of 3.4/4 (a maximum score of 4 reflects the scenario wherein all 22 co-authors voted the source as secure).

\subsubsection{WERLS\_69492 a.k.a EGS-zs8-2}

WERLS\_69492 is a known LAE, first reported as EGS-zs8-2 in \citet{2016roberts-borsani}, who detect \Lya\ at 4.7$\sigma$ with MOSFIRE and measure a redshift of \zs = 7.4777; this observation is later supported by \citet{2017stark}, who report a 7.4$\sigma$ \Lya\ detection for the target. It is in our secure sample and is our highest SNR \Lya\ detection, with SNR = 7.7. The target is on mask wmme01 in the EGS field and is covered by CANDELS (ID: 69492) but does not fall in the CEERS \textit{JWST}/NIRCam coverage. We fit the CANDELS photometry and find a photometric redshift of \zp = $7.8^{+0.2}_{-0.4}$, consistent within uncertainty with our \Lya-derived \zs = 7.4763. The \texttt{Bagpipes} SED fit suggests excess from nebular emission, which is supported by the strong photometric excess in IRAC Channel 2 (4.5\um) of 1.19\,mag \citep[corroborated by][]{2016roberts-borsani}. Co-author vetting resulted in a score of 3.8/4.

The target was also observed by CEERS with \textit{JWST}/NIRSpec (MSA ID: 698) with medium resolution in the G140M, G235M, and G395M filters. The CEERS reduction confirms the redshift of the source via multiple strong, rest-frame optical emission lines (including \Oiii+\Hb), with \zs = 7.4710 \citep{2023jung}. We attribute the small $\Delta z = 0.005$ between spectroscopic redshift solutions to the expected velocity offset between \Lya\ and the systemic redshift. Here, \Lya\ emission is redshifted from the rest-frame optical emission with a velocity offset of $\Delta v \sim188$\,km\,s$^{-1}$, consistent within uncertainty with the measured velocity offset in \citet{2023jung} of $142\pm142$\,km\,s$^-1$.

\subsubsection{WERLS\_32350}

WERLS\_32350 is on mask wmme01 with a SNR = 5.6 \Lya\ emission line corresponding to \zs=7.5501 and is in our secure sample. It is selected from EGS CANDELS (ID: 32350) and is not within the CEERS \textit{JWST}/NIRCam coverage. We fit the CANDELS photometry and find a photometric redshift of \zp = $8.3^{+0.2}_{-1.3}$, consistent within uncertainty with our \Lya-derived \zs = 7.5501. The \texttt{Bagpipes} SED fit suggests excess from \Oiii+\Hb\ nebular emission, which is supported by the strong photometric excess in IRAC Channel 2 (4.5\um) of 0.91\,mag. Co-author vetting resulted in a score of 3.0/4.

The target has multiple MOSFIRE $Y$-band observations, through this program and in \citet{2022jung}, who report a \Lya\ detection at $z = 7.7759\pm0.0012$ (ID: z8\_32350), inconsistent with our $z=7.5501$ \Lya\ detection, which was likely not discovered in the \citet{2022jung} automated line search as it partially overlaps with a sky line, and was excluded in their automated search. The target was observed by CEERS with \textit{JWST}/NIRSpec (MSA ID: 689) with medium resolution in the G140M, G235M, and G395M filters. The CEERS reduction confirms the redshift of the source via multiple strong, rest-frame optical emission lines (including \Oiii+\Hb), with \zs $= 7.5457\pm0.0001$ \citep{2023jung}. This is consistent with the WERLS \Lya-derived redshift, with a small $\Delta z$ between this spectroscopic redshift solutions to the expected offset between systemic and \Lya-derived redshifts. The measured velocity offset of \Lya\ is $\Delta v\sim154$\,km\,s$^{-1}$, with \Lya\ emission redshifted from rest-frame optical emission, consistent within uncertainty with the measured velocity offset in \citet{2023jung} of $221\pm109$\,km\,s$^-1$.

\subsubsection{WERLS\_786362}

WERLS\_786362 is on mask wmmc05 and has a SNR = 4.8 \Lya\ emission line corresponding to \zs=7.6810. It lies within the COSMOS CANDELS coverage and within PRIMER-COSMOS. This source was imaged by both NIRCam and MIRI with PRIMER, and has a faint detection in MIRI/F770W. We fit the target's \textit{JWST}+\textit{HST} photometry (CANDELS ID: 786362) and find a best-fit SED with \zp =$7.41^{+0.09}_{-0.12}$, consistent with the spectroscopic solution. Co-author vetting resulted in a score of 3.6/4. 

The photometry for WERLS\_786362 demonstrates a particularly blue slope at short wavelengths in addition to a red slope at long wavelengths (specifically, it is very blue in F115W$-$F200W and red in F277W$-$F444W), which could be indicative of a brown dwarf \citep[e.g.][]{2023hainline,2023burgasser,2023langeroodi}. Its compact morphology, akin to the ``little red dots'' discovered recently in \textit{JWST} images, is consistent with both a high redshift galaxy or a brown dwarf. We fit the photometry to a generated \texttt{Bagpipes} galaxy model with a very blue component with boosted \Lya\ and a red dusty component with boosted \Hb, as well as a brown dwarf with T=1000\,K and log $g=3.0$ from the Sonora models \citep{2021marley}, and find that both models plausibly match the data. Given that our \texttt{Bagpipes} galaxy model is physically extreme, the photometric evidence suggests the brown dwarf case is more likely. However, the brown dwarf solution does not account for the strong emission line we detect with MOSFIRE, which supports the the high redshift galaxy case assuming the line is \Lya. As we lack multi-line confirmation of the source and therefore are unable to distinguish between the brown dwarf and high redshift LAE cases, we place this source in our tentative sample.

\subsubsection{WERLS\_29712}

WERLS\_29712 has a candidate \Lya\ line at SNR = 4.1 with \zs = 7.1353. It is on mask wmmu01 in the UDS field. We fit the target's UDS CANDELS (ID: 29712) and PRIMER-UDS photometry  and find \zp = $6.8^{+0.1}_{-5.5}$, with a small peak in the photometric redshift PDF at $z\sim1-2$, and 9\% of the total PDF at $z<3$. Co-author vetting resulted in a score of 2.9/4. This lower redshift solution is inconsistent with the spectroscopic solution; given this discrepancy along with the faint signal in the 2D spectrum, we place it in the tentative sample.

\subsubsection{WERLS\_6931}

WERLS\_6931 has a candidate \Lya\ line with SNR = 6.0, corresponding to \zs = 8.1993. It is on mask wmmu01 in the UDS field. We fit the target's CANDELS (ID: 6931) and PRIMER-UDS photometry  and find \zp = $6.8^{+0.8}_{-5.3}$, with a peak in the photometric redshift PDF at $z\sim1-2$, and 24\% of the total PDF at $z<3$. Co-author vetting resulted in a score of 2.7/4. While the \zs\ solution falls within the broader high redshift peak of the photometric redshift PDF, given that only 35\% of the redshift PDF is at $z>7$, we place it in the tentative sample.

\subsubsection{WERLS\_9030}

WERLS\_9030 is on mask wmmc02 and has a moderate \Lya\ emission line candidate detection at SNR = 4.4 corresponding to \zs = 7.6862. It lies within the COSMOS CANDELS and PRIMER-COSMOS coverage, and is not in the COSMOS2020 catalog. This source was imaged by both NIRCam and MIRI with PRIMER, and has a faint detection in MIRI/F770W. We fit the target's CANDELS (ID: 9030) and PRIMER photometry and find a photometric redshift solution of \zp = $7.0^{+0.4}_{-5.6}$, consistent with the spectroscopic solution within errors, with a smaller peak in the photometric redshift PDF at $z\sim1-2$, and 22\% of the total PDF at $z<3$. This lower redshift solution is inconsistent with the spectroscopic solution; fixing the SED to the \zs\ returns a solution wherein the F814W flux is partially contaminated by \Lya\ emission, which is consistent with its MOSFIRE spectrum. Co-author vetting resulted in a score of 2.8/4. As the \Lya\ line is fairly faint in the 2D spectrum, we place it in the tentative sample.

\subsubsection{WERLS\_20710}

WERLS\_20710 is on mask wmmc02 and has a moderately detected (SNR = 5.2) \Lya\ emission line candidate corresponding to \zs = 7.5295. It lies within the COSMOS CANDELS coverage and within PRIMER-COSMOS. Co-author vetting resulted in a score of 2.7/4. We fit the target's \textit{JWST}+\textit{HST} photometry (CANDELS ID: 20710) and find a photometric redshift of \zp = $6.5^{+0.08}_{-4.8}$, somewhat inconsistent with the spectroscopic solution; given this we place this source in our tentative sample.

\subsubsection{WERLS\_13312}

WERLS\_13312 is on mask wmmc02 and has a moderately detected (SNR = 5.4) \Lya\ emission line candidate corresponding to \zs = 8.1962. It lies within the COSMOS CANDELS coverage but not in PRIMER-COSMOS or COSMOS-Web, and is not in the COSMOS2020 catalog. We fit the target's CANDELS photometry (ID: 13312) and find a best-fit SED with \zp = $8.1^{+0.1}_{-1.5}$, consistent with the spectroscopic solution. However, the SED appears poorly constrained as the target is undetected in $Y$-band and bluer, and is not detected in IRAC. Co-author vetting resulted in a score of 3.2/4. Though the line was highly ranked by co-authors, given its dearth of secure multi-band photometry and the absence of clear negative signal in the 2D spectrum, we place this source in our tentative sample.

\subsubsection{WERLS\_450980}

WERLS\_450980 is selected from the COSMOS2020 catalog, is on mask wmmc01, and is in our tentative sample. It has candidate \Lya\ emission detected at SNR = 6.7, giving \zs = 7.0925. There is a much fainter positive signal with SNR = 3.9 just blueward of the candidate emission line; this could indicate the signal is the \Oii\ doublet rather than \Lya. Assuming the brighter emission line candidate is \Oii\ with rest frame wavelength $3729\,\rm \AA$, that would place the source at $z=1.639$, and we would expect to find the rest-frame $3727\,\rm \AA$\ emission line $\sim6\, \rm \AA$\ blueward. Here, the fainter signal is nearly twice that separation, at $\sim11\, \rm \AA$\ from our candidate emission line. Therefore, we find the \Lya\ line identification to be more likely. Using the COSMOS2020 photometry (ID: 450980), we find the photometric redshift to be well-matched to the spectroscopic solution, with \zp = $7.3^{+0.1}_{-0.8}$. Co-author vetting resulted in a score of 2.7/4. While the line appears real and the SED results are consistent with our \Lya-derived spectroscopic solution, we include the source in the tentative sample due to the uncertainty in our line identification.

\subsubsection{WERLS\_29881}

WERLS\_29881 is in our tentative sample and is our highest redshift LAE, detected at SNR = 4.6 with \zs = 8.3868. It is on mask wmme01 in the EGS field, but lies outside of the CEERS coverage. We fit the target's CANDELS photometry (ID: 29881) and find a best-fit SED with \zp = $6.4^{+0.4}_{-4.9}$, with a smaller peak in the photometric redshift PDF at $z\sim1$, and 10\% of the total PDF at $z<3$. This is inconsistent with the spectroscopic solution; fixing the SED to the \zs\ returns a solution wherein the \textit{HST}/F814W flux is partially contaminated by \Lya\ emission, which is consistent with its MOSFIRE spectrum. Further, while the source is undetected in IRAC Channel 1 (3.6\um), it is detected in IRAC Channel 2 (4.5\um), where there should be contamination from \Oiii+\Hb\ nebular emission at the spectroscopic redshift. Co-author vetting resulted in a score of 3.3/4. While co-authors ranked this detection highly, given the line detection is at the edge of the MOSFIRE wavelength coverage, it is statistically more likely to be noise, and we place it in the tentative sample.

\subsection{\textit{JWST}/NIRSpec Observations}

Five of our primary targets in the EGS/CEERS field were observed with \textit{JWST}/NIRSpec \citep{2023boker} as part of the CEERS program: WERLS\_69492, WERLS\_32350, WERLS\_35089, WERLS\_45153, and WERLS\_40898. For two of these five sources (WERLS\_69492 and WERLS\_32350), the NIRSpec data secures the WERLS \Lya-derived redshift, anchored by strong nebular emission lines in the near-infrared, namely \Oiii\ and \Hb\ (see \textsection4.2.2. and \textsection4.2.3. for more details). For the other three sources, we did not identify \Lya\ emission in the WERLS MOSFIRE data, explained for each source in more detail below.

MOSFIRE Primary target WERLS\_40898 is on mask wmme03 in the EGS field and had no \Lya\ detection in the WERLS MOSFIRE data. It was observed by CEERS with \textit{JWST}/NIRSpec (MSA ID: 1027) with both PRISM mode and with medium resolution in the G140M, G235M, and G395M filters. As reported in \citet{2023arrabalharo}, the CEERS reduction secures the redshift of the source via multiple strong, rest-frame optical emission lines (including \Oiii+\Hb), with \zs = 7.820$^{+0.001}_{-0.001}$ \citep[see also][]{2022heintz,2023sanders}. The medium resolution spectrum also shows strong \Lya\ emission at $10732.1\,\rm \AA$, and \citet{2023tang} find a fairly large \Lya\ velocity offset from the systemic redshift of $\sim323$\,km\,s$^{-1}$. Based on the strength of the NIRSpec detection, we expect \Lya\ to be detectable by WERLS for this source. However, the spectrum is contaminated by a sky line at the observed \Lya\ wavelength. Additionally, mask wmme03 was less than half complete, with only 8 ABBA sequences taken (out of our goal of 20+ sequences or about 4 hours of total exposure time). 

WERLS\_35089 is a MOSFIRE Primary target on mask wmme01 in the EGS field and had no \Lya\ detection in the WERLS MOSFIRE data, corroborated by the MOSFIRE $Y$-band non-detection in \citet{2022jung}. It was observed by CEERS with \textit{JWST}/NIRSpec (MSA ID: 716) with PRISM mode only. The CEERS reduction shows a single bright emission line, which given photometric redshift priors can be identified securely as \Ha, with \zs = 6.959 (Arrabal Haro et al. in prep); the redshift is also consistent with \Oiii\ falling in the detector gap for these observations. While there is no detectable \Lya\ emission line in the spectrum, the \Lya\ break is detected, which anchors the spectroscopic redshift solution.

WERLS Primary target WERLS\_45143 is on mask wmme01 in the EGS field and had no \Lya\ detection in the WERLS MOSFIRE data. It was observed by CEERS with \textit{JWST}/NIRSpec (MSA ID: 717) with both PRISM mode and with medium resolution in the G140M, G235M, and G395M filters. The CEERS reduction secures the redshift of the source via multiple strong emission lines (including \Oiii+\Hb as well as \Ha), with \zs = 6.934 (Arrabal Haro et al. in prep). The NIRSpec spectrum shows no \Lya\ emission, but the \Lya\ break is detected.

\section{Discussion}

The primary aims of the WERLS experiment in its entirety are to 1) conduct a census of \Lya\ emission in known, luminous EoR galaxy candidates to map ionization bubbles in the IGM on scales larger than their expected size, 2) directly compare the \Lya-inferred location of ionized bubbles to underlying galaxy density maps to be measured via deep \textit{JWST}/NIRCam imaging to directly constrain the environments of LAEs, related to the drivers of reionization, and 3) increase the number of spectroscopically-confirmed bright EoR sources to inform photometric redshift calibration of fainter
EoR galaxies exclusively detected by \textit{JWST}. This paper (in part) addresses the first and third aims, by reporting new \Lya\ detections for bright galaxies at $z\sim7-8$. Here, we discuss the efficacy of this census of \Lya, and look ahead to future efforts to address the latter goals of the WERLS experiment.

\subsection{Observed LAE Yield}

To estimate the expected LAE yield for our observations, we consider the photometric redshift PDFs for the entire subsample of WERLS Primary targets ($N=114$), the portion of the spectra that are blocked by OH sky lines, and the expected neutral fraction at $z\sim7-8$ that would further reduce the number of observable LAEs. To compare these estimates to our observed sample of LAEs ($N=11$), we separate out the MOSFIRE Primary targets ($N=33$), which have photometric redshift solutions that peak strongly at $z\sim7-8$ where \Lya\ is detectable in our data, and the other WERLS Primary (LRIS and MOSFIRE+LRIS) targets ($N=27$ and $N=54$, respectively), which have broader photometric redshift solutions within the EoR and/or are LRIS Primary targets at $z\lesssim7$. For each subset and the total Primary target sample, we find the total photometric redshift cumulative distributions, and use this to quantify the number of targets that would fall in the \Lya\ redshift range detectable for our observations. This quantity is then multiplied by the typical fraction of the spectrum that is contaminated by sky lines, $\sim38\%$, assuming an uncertainty of $\pm10\%$ to account for variation both in the exact wavelength coverage of each spectrum and in the seeing (and therefore, width of the sky lines). Finally, we multiply this number by the fraction of the IGM that is ionized at our target epoch; given that our observations do not directly constrain the ionized fraction, we allow it to vary. Importantly, the \Lya\ fraction as we consider here is not equal to the IGM neutral fraction, as the neutral fraction also depends on \Lya\ velocity offsets and ionized bubble sizes. We do not attempt to constrain the neutral fraction here, and apply a varying average neutral fraction (which should broadly contain the variations due to velocity offsets and ionized bubbles) to serve as an upper limit for our LAE yield estimation.

\begin{figure}[h!]
    \centering
    \includegraphics[angle=0,trim=0in 0in 0in 0in, clip, width=0.48\textwidth]{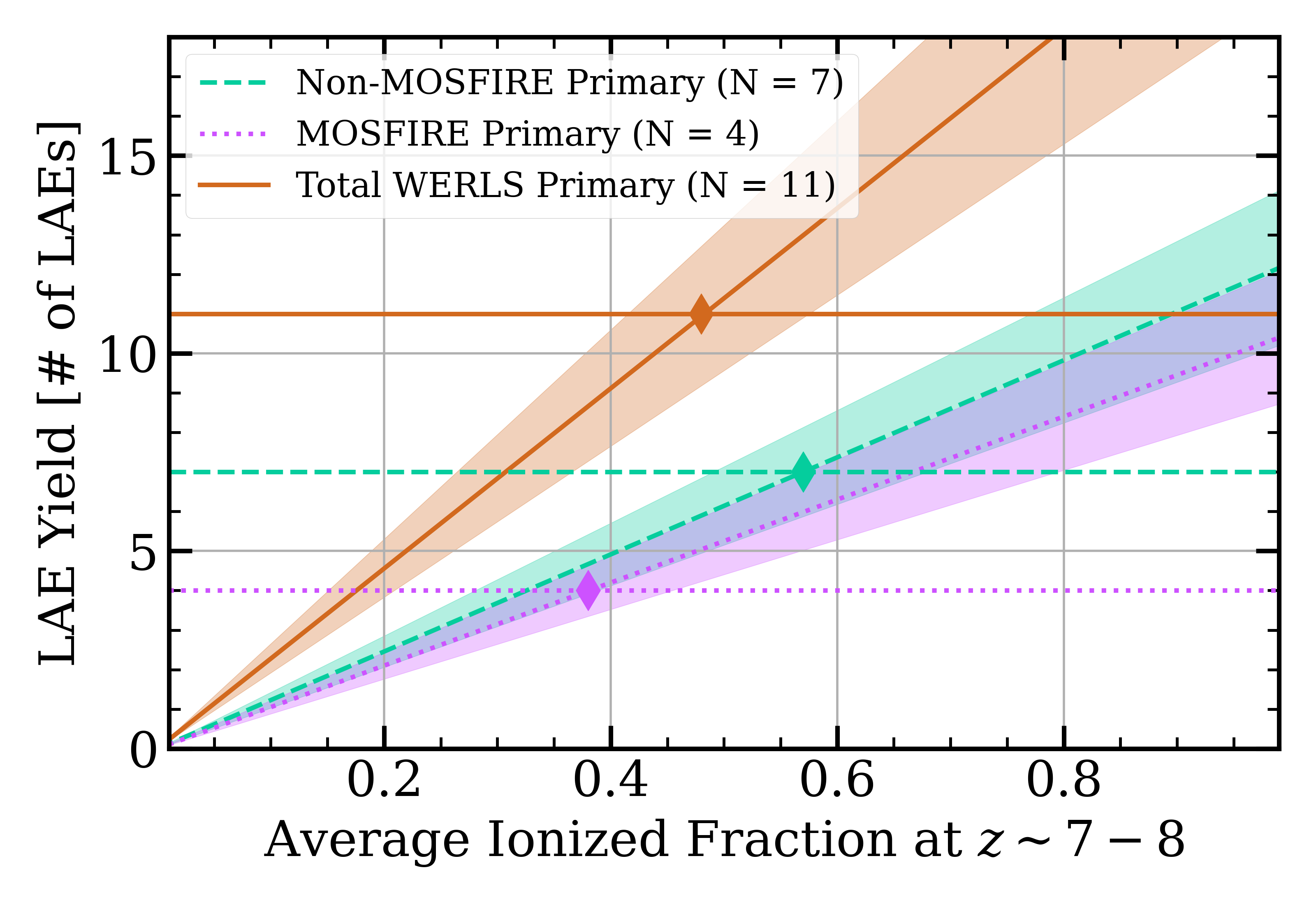}
    \caption{LAE yield for our MOSFIRE observations versus ionized fraction (horizontal lines) compared to the expected detection rate given the photometric redshift PDFs for our targets and portion of the spectrum blocked by sky lines. The MOSFIRE Primary sample (targets most likely at $z\sim7-8$) is shown in purple, the non-MOSFIRE WERLS Primary sample (drawn from a broader redshift range within the EoR) is shown in cyan, and the total WERLS Primary sample is in orange. The uncertainty on sky line contamination is represented by the shaded regions. Our total LAE yield ($N=11$) supports the scenario wherein reionization is at its midpoint at $z\sim7-8$.}
    \label{fig:yield}
\end{figure}

Our expected LAE yield as a function of the average cosmic ionized fraction is shown in Figure \ref{fig:yield}. We detect $N=7$ Non-MOSFIRE Primary targets and $N=4$ MOSFIRE Primary targets, for a total of 11 LAEs. None of our 11 LAEs were LRIS Primary targets, but the subsample did contain MOSFIRE Primary and both MOSFIRE+LRIS Primary targets. Given this, we can estimate the yield simply by taking the number of detected LAEs ($N=11$) over the number of targeted candidates from the two primary categories (MOSFIRE, $N=33$ and MOSFIRE+LRIS, $N=54$), for a $\sim13$\% yield of \Lya\ emission observed from our EoR targets. This is broadly consistent with expectations for a Universe that is half-ionized at $z\sim7-8$ (see Figure \ref{fig:yield}), and demonstrates the relative success of WERLS at detecting \Lya\ in UV-bright EoR galaxies.

\subsection{Possible Ionized Bubble at $z=7.68$}

Observing \Lya\ emission at $z\sim7-8$ is fairly unlikely as the IGM maintains a fairly high neutral fraction at these redshifts. EoR LAEs are thought to be more readily observable if they sit within ionized bubbles, wherein the emitted \Lya\ is protected by an ionized region large enough for the \Lya\ photons to scatter out of resonance with the neutral IGM and remain observable. Similar works reporting EoR \Lya\ detections note potential overdensities near their targets \citep[e.g.][]{2020jung,2021endsley,2022larson}, and suggest these overdensities can support larger (and therefore, more easily detectable) ionized bubbles. The threshold radius for \Lya\ observability has been approximated at $\sim1\,$ physical Mpc \citep{2014dijkstra} before resonant scattering is sufficiently diminished as photons are cosmologically redshifted. The exact scale of this size depends on the \Lya\ velocity offset from the systemic redshift, which would reduce this threshold to a smaller radius.

We identify one potential overdensity in the WERLS MOSFIRE sample, with two targets at $z = 7.68$ in the COSMOS field: WERLS\_786362 at $z = 7.6810$ and WERLS\_9030 at $z = 7.6862$. Given their proximity in redshift of $\Delta z = 0.0052$, the pair has a very small line-of-sight separation of 0.2\,pMpc, less than half the line-of-sight separations of the $z=7.7$ galaxy group in EGS77 \citep[0.7\,pMpc;][]{2020tilvi}. However, the two galaxies are more distant in projection at a separation of 9.6\,arcmin, corresponding to a transverse separation of 2.9\,pMpc. 

Following the methods in \citet{2021endsley} and \citet{2022larson}, we estimate the expected ionized bubble radius that would be produced by each galaxy, assuming no \Lya\ velocity offset as we lack systemic redshifts for these targets (effectively, this serves as an upper limit). From \citet{2000cen} and \citet{2021endsley}, we calculate the ionized bubble radius $R$ produced by each galaxy as

\begin{equation}
    R = \left(\frac{3\, {\dot{N}}_{\rm ion}\, f_{\rm esc}\, t}{4\, \pi\, n_{\rm HI}(z)} \right)^{1/3}.
\end{equation}

Considering first the denominator, the proper volume density of neutral hydrogen ($n_{\rm HI}(z)$) can be derived from cosmological parameters \citep[see Equations 2 \& 3 in][]{2022larson}. We adopt \textit{Planck} measurements of the helium mass fraction, Hubble constant, and baryon density \citep{2016collaboration,2020collaboration}.

Each variable in the numerator represents the ionizing production, output, and transmission from the individual galaxy. The ionizing production is represented by $t$, the duration of the current star formation episode; for our targets this is unconstrained, so we follow \citet{2022larson} and take $t = 20\,$Myr, noting that a longer episode would increase the bubble radius. The transmission of ionizing photons is represented by the escape fraction ($f_{\rm esc}$), which is unconstrained for our targets; we allow this parameter to vary from [0-1], and calculate bubble radius as a function of $f_{\rm esc}$. Lastly, the ionizing photon output from the galaxy is represented by the intrinsic ionizing emissivity, ${\dot{N}}_{\rm ion}$. 

We calculate ${\dot{N}}_{\rm ion}$ from the product of the ionizing photons production efficiency ($\xi_{\rm ion}$) and the specific nonionizing UV luminosity ($\rho_{\rm UV}$). Here, we assume $\xi_{\rm ion} = 25.6$, to be consistent with measurements of $\xi_{\rm ion}$ for both local analogs of EoR galaxies \citep{2021tang} and bright EoR sources \citep{2016stark}, as well as model predictions from \citet{2019finkelstein}. Finally, $\rho_{\rm UV}$ can be estimated from the observed brightness of the galaxy (here, its apparent $H$-band magnitude) and its redshift. 

\begin{figure}[h!]
    \centering
    \includegraphics[angle=0,trim=0in 0in 0in 0in, clip, width=0.48\textwidth]{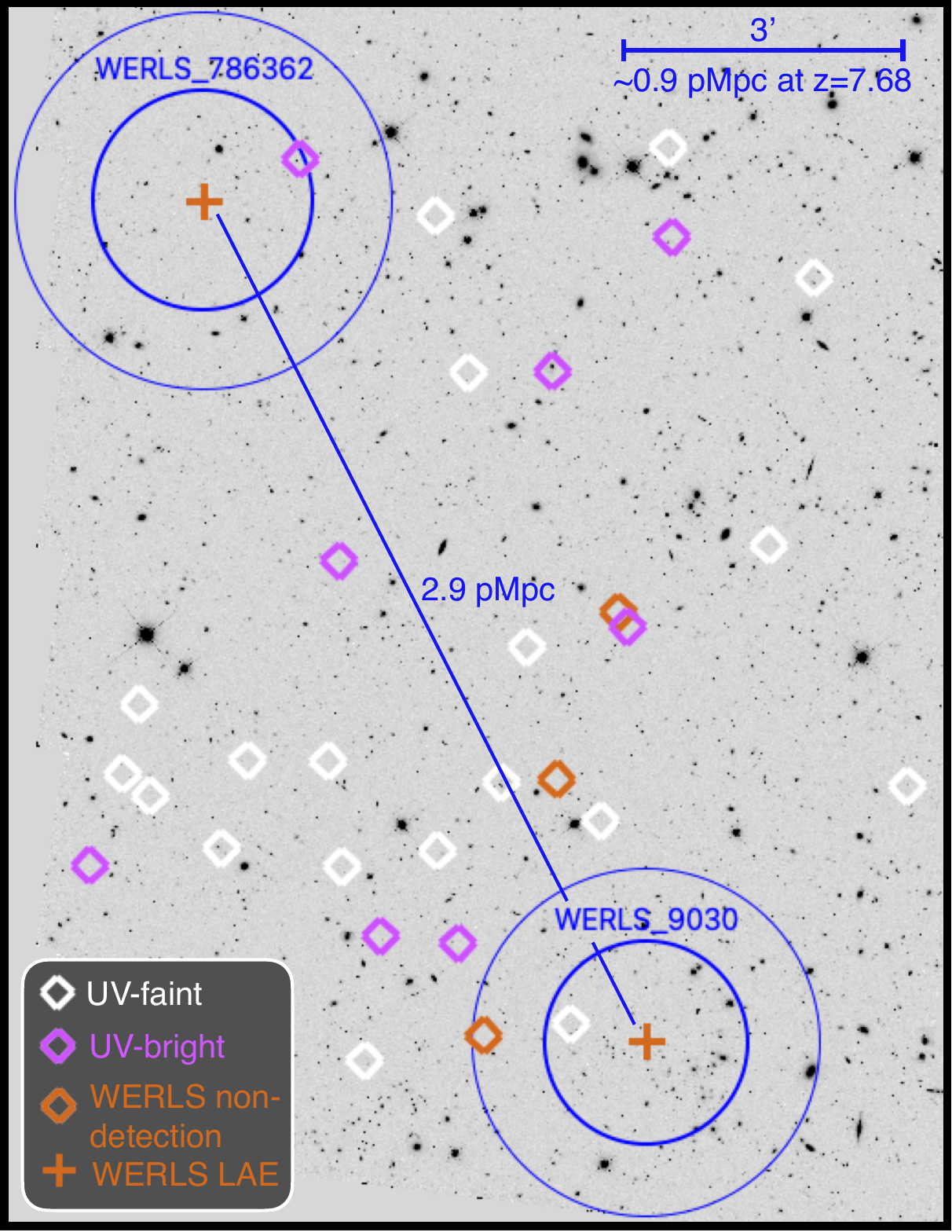}
    \caption{Spatial positions of the two $z=7.68$ LAEs in our sample (orange plus signs) and the approximate ionized bubbles they would each produce for $f_{\rm esc} = 0.2$ (inner blue circles) and for $f_{\rm esc} = 1.0$ (outer blue circles). The two systems bound a spherical region with radius $R\sim1.5$\,pMpc, a plausible scale for a coherent ionized bubble at this redshift if additional sources contribute to the local ionizing photon budget. The positions of sources at similar photometric redshifts are marked as diamonds, with UV-bright ($\Muv\leq-20$) sources in purple, UV-faint ($-20<\Muv\leq-18$) sources in white, and sources targeted with WERLS MOSFIRE but undetected in \Lya\ shown in orange. The entire $\sim9\times12$\,arcmin region shown here lies within the full COSMOS-Web coverage, with nearly every source also within PRIMER-COSMOS. The figure is projected on the CANDELS \textit{HST}/F160W image.}
    \label{fig:bubble}
\end{figure}

By applying Equation 1 individually to WERLS\_786362 and WERLS\_9030, we calculate an ionized bubble radius dependent on the escape fraction (see Figure \ref{fig:bubble}). Taking $f_{\rm esc} = 0.2$ as an example, we find $R = 0.34$\,pMpc for WERLS\_786362 and $R = 0.31$\,pMpc for the slighlty fainter WERLS\_9030. As a physically implausible upper bound, allowing all ionizing photons to escape ($f_{\rm esc} = 1.0$) would produce a $R = 0.58$\,pMpc bubble for WERLS\_786362 and $R = 0.53$\,pMpc for WERLS\_9030. In the $f_{\rm esc} = 0.2$ case, estimated bubble sizes are about $5\times$ smaller than the spherical region defined by the observed separation between the galaxies ($R=1.5$\,pMpc); even in the $f_{\rm esc} = 1.0$ case, the sum of the individual bubble radii ($1.1$\,pMpc) is less than the galaxies' separation. Therefore, we determine that the physical scenario wherein their individual ionized bubbles alone can easily overlap and create a common, larger ionized bubble is unlikely. Note that in the case where the intrinsic \Lya\ emission is much stronger than the observed \Lya\ line, emission would be observable through a smaller bubble with some significant \Lya\ loss.

The galaxies' separation is consistent with the expected sizes of ionized regions at $z\sim7-8$ \citep{2017furlanetto,2018daloisio} as well as observations of large ionized regions \citep{2021endsley,2022endsleya}, and the presence of a large ionized bubble could facilitate the escape of \Lya\ photons. We consider the possibility that the two LAEs are both embedded in a larger ionized structure of $R\gtrsim1.5$\,pMpc. This scenario is possible if a local overdensity exists in the region, wherein the local ionizing photon budget is supported by emission from UV-bright galaxies (that may fall outside the WERLS MOSFIRE coverage), and/or fainter EoR galaxies (below the WERLS target criteria). In order to produce a larger ionized bubble encompassing both $z=7.68$ LAEs in our sample, taking a nominal escape fraction of $f_{\rm esc} = 0.2$, approximately four additional systems of similar ionizing power at similar redshifts would need to be located within a $\sim23$\,arcmin$^2$ area. We search the COSMOS CANDELS catalog for galaxies at $7.58 \leq z_{\rm phot} \leq 7.78$ within a $\sim5\,$arcmin radius circle bounded by the two WERLS LAEs, and find 10 (30) candidates with $\Muv\leq-20$ ($-20<\Muv\leq -18$). Three of the 10 UV-bright sources were targeted with WERLS, and were not detected in \Lya. Figure \ref{fig:bubble} shows the spatial distribution of the LAEs and potential nearby systems. Targeting the remaining candidates --- including sources fainter than the WERLS criteria --- to search for other nearby LAEs at $z\sim7.68$ offers an explicit hypothesis that can be tested with future MOSFIRE observations to determine if the $z=7.68$ LAEs reported here occupy a single, larger ionized bubble.

\subsection{WERLS Synergy with \textit{JWST}}

While the aim of the WERLS Keck spectroscopy is to detect \Lya\ from UV-bright EoR galaxies that likely trace high density peaks, the ultimate goal of the WERLS experiment is to use this census of \Lya\ emission to then map the ionization state of the Universe and better constrain both the sources and process of reionization. This broader goal relies on planned and upcoming \textit{JWST}/NIRCam deep imaging. COSMOS-Web, CEERS, and PRIMER will be able to construct detailed maps of the underlying mass in large scale structure on (5\,Mpc)$^3$ scales by detecting thousands of EoR galaxies at luminosities $10-30\times$ fainter than our UV-bright WERLS targets. By design, the majority of our targets sit within COSMOS-Web, as the survey is large enough to mitigate cosmic variance and to capture reionization on scales larger than its expected patchiness. The smaller but deeper CEERS and PRIMER surveys cover areas comparable to a single ionized bubble in the EoR, but include fainter sources and provide a finer sampling of ionized bubbles; the cosmological context of bubbles found in these deeper programs can then be informed by the larger statistical samples in COSMOS-Web.

This WERLS synergy between \textit{JWST} imaging and Keck spectroscopy to pinpoint beacons of reionization via \Lya\ detection and then map the underlying galaxy density can tell us which galaxies are primarily responsible for reionizing the Universe. Upon completion of both the NIRCam surveys and the full Keck WERLS program, we plan to make this measurement through careful cross-correlation of the two maps (in particular using a two-point correlation function). From these maps, we will test the hypothesis that either massive, intrinsically bright galaxies drove reionization in highly clustered regions or that more common low-mass galaxies drive a more homogeneous reionization process.

WERLS synergy with \textit{JWST} is also powerful from an entirely spectroscopic context. By design, WERLS is a \Lya\ detection experiment; indeed, given the neutral fraction of the IGM halfway through the EoR, we do not expect to detect \Lya\ in most of our EoR targets. Given the lack of other emission lines near the target \Lya\ emission line, there exists some uncertainty in spectroscopic confirmation, for both detections and non-detections. NIRSpec observations can provide, both for LAEs and non-LAEs, unambiguous spectroscopic confirmation via multiple rest-frame optical emission lines.

The five NIRSpec-derived redshifts for both LAEs and non-LAEs in WERLS allows broader exploration of the \Lya-detection experiment goal of WERLS. With MOSFIRE, we obtain deep, high resolution spectra ($R \sim3500$) of \Lya\ from our sources, which serves as a crucial step in the detection experiment. For example, while we detect \Lya\ in our MOSFIRE observation of WERLS\_32350, \citet{2023jung} did not detect \Lya\ from WERLS\_32350 (referred to in their work as z8\_32350) with their medium resolution NIRSpec G140M grating data, which they attribute to the faintness of its \Lya\ emission, below the detection limit. The case of WERLS\_40898 --- a NIRSpec-confirmed LAE that was undetected in WERLS because of the presence of an atmospheric line at the expected wavelength of the line --- demonstrates one challenge of ground-based near-infrared observing and serves as a reminder that care should be taken when interpreting results related to completeness and yield for this survey and others like it; we account for sky line contamination in our LAE yield estimation in \textsection 5.1. Additionally, we are able to securely confirm two WERLS targets (WERLS\_35089 and WERLS\_45153) as non-LAEs via their NIRSpec spectra. These cases demonstrate the synergy between the WERLS/Keck observations targeting \Lya\ with deep, high resolution near-infrared spectra from the ground, and relatively inexpensive multi-line spectroscopic confirmation from \textit{JWST}/NIRSpec; with these two instruments the WERLS \Lya\ detection experiment can be robustly and efficiently conducted.

\section{Summary}

In this paper, we present the first results from the WERLS program, specifically Keck I/MOSFIRE $Y$-band spectroscopic observations for 114 known, UV-bright EoR candidates in COSMOS, EGS, and UDS. We summarize our results as follows.

\begin{enumerate}
\item We spectroscopically identify 11 LAEs from $z\sim7-8$, with 3 secure and 8 tentative LAEs spanning \Lya-derived spectroscopic redshifts of $7.0925 \leq z \leq 8.3868$ and absolute UV magnitudes of $-23.14 < \Muv < -19.81$.

\item We find an observed LAE yield of $\sim13$\%, which is broadly consistent with expectations for a Universe that is half-ionized at $z\sim7-8$, illustrating the success of WERLS at detecting \Lya\ in UV-bright EoR galaxies.

\item We identify one potential overdensity in the WERLS MOSFIRE sample, with two targets at $z = 7.68$ in the COSMOS field that are separated by $2.9$\,pMpc (9.6\,arcmin). Based on their estimated individual ionized bubble radii, the two galaxies could occupy a common ionized bubble if nearby galaxies within a $\sim1.5$\,pMpc volume ($\sim4$ UV-bright galaxies for a nominal $f_{\rm esc} = 0.2$) contribute to the local ionizing photon budget.

\end{enumerate}

The first year of WERLS has demonstrated its efficacy at detecting LAEs near the midpoint of the EoR. Combined with large-scale mass density maps of the field derived from deep \textit{JWST}/NIRCam imaging, future synergistic Keck+\textit{JWST} efforts provide a powerful tool for pinpointing beacons of reionization and mapping the ionization state of the Universe, enabling robust tests regarding the primary drivers and the timeline of reionization.

\section*{Acknowledgements}

The authors wish to recognize and acknowledge the very significant cultural role and reverence that the summit of Maunakea has always had within the indigenous Hawaiian community. We are most fortunate to have the opportunity to conduct observations from this mountain.

O.R.C, C.M.C., and others at UT Austin acknowledge that they work at an institution that sits on indigenous land. The Tonkawa lived in central Texas, and the Comanche and Apache moved through this area. We pay our respects to all the American Indian and Indigenous Peoples and communities who have been or have become a part of these lands and territories in Texas. We are grateful to be able to live, work, collaborate, and learn on this piece of Turtle Island.

We would like to thank the \texttt{PypeIt} team for help in the reduction and customization work for \texttt{PypeIt}/MOSFIRE, in particular Debora Pelliccia. 

This work was supported by PID40/2022A\_N078 (PIs C. Casey \& J. Kartaltepe). Data presented herein were obtained at the W. M. Keck Observatory from telescope time allocated to the National Aeronautics and Space Administration through the agency’s scientific partnership with the California Institute of Technology and the University of California. The Observatory was made possible by the generous financial support of the W. M. Keck Foundation. 

This material is based on work supported by the National Science Foundation Graduate Research Fellowship under grant number DGE 2137420.

This work is partly based on observations made with the NASA/ESA/CSA \textit{JWST}. The data were obtained from the Mikulski Archive for Space Telescopes at the Space Telescope Science Institute, which is operated by the Association of Universities for Research in Astronomy, Inc., under NASA contract NAS 5-03127 for \textit{JWST}. These observations are associated with programs JWST-ERS-01345 and JWST-GO-01837.

C.M.C. thanks the National Science Foundation for support through grants AST-1814034 and AST-2009577 as well as the University of Texas at Austin College of Natural Sciences for support; C.M.C. also acknowledges support from the Research Corporation for Science Advancement from a 2019 Cottrell Scholar Award sponsored by IF/THEN, an initiative of Lyda Hill Philanthropies. 

J.S.K. thanks the National Science Foundation for support through grant AST-2009572. J.S.K. and B.N.V. thank NASA/ California Institute of Technology-JPL for support through grant 1668074. A.B. thanks the Emerson Summer Undergraduate Research Fellowship through the College of Science at the Rochester Institute of Technology.

The authors would like to thank our Keck Observatory Support Astronomers, Jim Lyke and Chien-Hsiu Lee, for their assistance during our observing runs. Special thanks to all of our Observing Assistants for their work driving the telescope during our observations: Matthew (Matt) Wahl, Marita (Rita) Morris, Tony Ridenour, Carolyn Jordan, and John Pelletier.

\software{\texttt{astropy} \citep{2013astropycollaboration,2018astropycollaboration,2022astropycollaboration}, \texttt{numpy} \citep{2020harris}, \textsc{EAzY} \citep{2008brammer}, \textsc{LePHARE} \citep{2011arnouts}, \texttt{Jupyter} \citep{2016kluyver}, \texttt{matplotlib} \citep{2023caswell}, \texttt{scipy} \citep{2020virtanen}, \textsc{SourceExtractor} \citep{1996bertin}, STScI JWST Calibration Pipeline \citep[\href{jwst-pipeline.readthedocs.io}{jwst-pipeline.readthedocs.io;}][]{2023rigby}, \textsc{MAGMA} (\href{https://www2.keck.hawaii.edu/inst/mosfire/magma.html}{www2.keck.hawaii.edu/inst/mosfire/magma.html}), \texttt{PypeIt} \citep[][]{2020prochaskaa}, \textsc{MosfireDRP} (\href{https://keck-datareductionpipelines.github.io/MosfireDRP/}{keck-datareductionpipelines.github.io/MosfireDRP}), \texttt{Bagpipes} \citep{2018carnall}, \texttt{photutils} \citep{2023bradley}, \texttt{pypher} \citep{2016boucaud}, \texttt{grizli} \citep{2023grizli}}

\facility{Keck:I (MOSFIRE)}

\section*{Data Availability}

The target list and spectroscopic results presented herein are available in machine-readable format; see Appendix and Tables \ref{tab:positions} \& \ref{tab:appendix}. Reduced MOSFIRE spectra will be provided upon request.

\bibliographystyle{aasjournal}
\bibliography{werls}

\begin{thebibliography}{}
\expandafter\ifx\csname natexlab\endcsname\relax\def\natexlab#1{#1}\fi
\providecommand{\url}[1]{\href{#1}{#1}}
\providecommand{\dodoi}[1]{doi:~\href{http://doi.org/#1}{\nolinkurl{#1}}}
\providecommand{\doeprint}[1]{\href{http://ascl.net/#1}{\nolinkurl{http://ascl.net/#1}}}
\providecommand{\doarXiv}[1]{\href{https://arxiv.org/abs/#1}{\nolinkurl{https://arxiv.org/abs/#1}}}

\bibitem[{Arnouts {et~al.}(1999)Arnouts, Cristiani, Moscardini, Matarrese,
  Lucchin, Fontana, \& Giallongo}]{1999arnouts}
Arnouts, S., Cristiani, S., Moscardini, L., {et~al.} 1999, Monthly Notices of
  the Royal Astronomical Society, 310, 540,
  \dodoi{10.1046/j.1365-8711.1999.02978.x}

\bibitem[{Arnouts \& Ilbert(2011)}]{2011arnouts}
Arnouts, S., \& Ilbert, O. 2011, Astrophysics Source Code Library,
  ascl:1108.009

\bibitem[{Arons \& McCray(1970)}]{1970arons}
Arons, J., \& McCray, R. 1970, Astrophysical Letters, 5, 123

\bibitem[{Arrabal~Haro {et~al.}(2023)Arrabal~Haro, Dickinson, Finkelstein,
  Fujimoto, Fern{\'a}ndez, Kartaltepe, Jung, Cole, Burgarella, Chworowsky,
  Hutchison, Morales, Papovich, Simons, Amor{\'i}n, Backhaus, Bagley,
  Bisigello, Calabr{\`o}, Castellano, Cleri, Dav{\'e}, Dekel, Ferguson,
  Fontana, Gawiser, Giavalisco, Harish, Hathi, Hirschmann, Holwerda,
  {Huertas-Company}, Koekemoer, Larson, Lucas, Mobasher,
  {P{\'e}rez-Gonz{\'a}lez}, Pirzkal, Rose, Santini, Trump, {de la Vega}, Wang,
  Weiner, Wilkins, Yang, Yung, \& Zavala}]{2023arrabalharo}
Arrabal~Haro, P., Dickinson, M., Finkelstein, S.~L., {et~al.} 2023, The
  Astrophysical Journal, 951, L22, \dodoi{10.3847/2041-8213/acdd54}

\bibitem[{Ashby {et~al.}(2015)Ashby, Willner, Fazio, Dunlop, Egami, Faber,
  Ferguson, Grogin, Hora, Huang, Koekemoer, Labb{\'e}, \& Wang}]{2015ashby}
Ashby, M. L.~N., Willner, S.~P., Fazio, G.~G., {et~al.} 2015, The Astrophysical
  Journal Supplement Series, 218, 33, \dodoi{10.1088/0067-0049/218/2/33}

\bibitem[{{Astropy Collaboration} {et~al.}(2013){Astropy Collaboration},
  Robitaille, Tollerud, Greenfield, Droettboom, Bray, Aldcroft, Davis,
  Ginsburg, {Price-Whelan}, Kerzendorf, Conley, Crighton, Barbary, Muna,
  Ferguson, Grollier, Parikh, Nair, Unther, Deil, Woillez, Conseil, Kramer,
  Turner, Singer, Fox, Weaver, Zabalza, Edwards, Azalee~Bostroem, Burke, Casey,
  Crawford, Dencheva, Ely, Jenness, Labrie, Lim, Pierfederici, Pontzen, Ptak,
  Refsdal, Servillat, \& Streicher}]{2013astropycollaboration}
{Astropy Collaboration}, Robitaille, T.~P., Tollerud, E.~J., {et~al.} 2013,
  Astronomy and Astrophysics, 558, A33, \dodoi{10.1051/0004-6361/201322068}

\bibitem[{{Astropy Collaboration} {et~al.}(2018){Astropy Collaboration},
  {Price-Whelan}, Sip{\H o}cz, G{\"u}nther, Lim, Crawford, Conseil, Shupe,
  Craig, Dencheva, Ginsburg, VanderPlas, Bradley, {P{\'e}rez-Su{\'a}rez}, {de
  Val-Borro}, Aldcroft, Cruz, Robitaille, Tollerud, Ardelean, Babej, Bach,
  Bachetti, Bakanov, Bamford, Barentsen, Barmby, Baumbach, Berry, Biscani,
  Boquien, Bostroem, Bouma, Brammer, Bray, Breytenbach, Buddelmeijer, Burke,
  Calderone, Cano~Rodr{\'i}guez, Cara, Cardoso, Cheedella, Copin, Corrales,
  Crichton, D'Avella, Deil, Depagne, Dietrich, Donath, Droettboom, Earl, Erben,
  Fabbro, Ferreira, Finethy, Fox, Garrison, Gibbons, Goldstein, Gommers, Greco,
  Greenfield, Groener, Grollier, Hagen, Hirst, Homeier, Horton, Hosseinzadeh,
  Hu, Hunkeler, Ivezi{\'c}, Jain, Jenness, Kanarek, Kendrew, Kern, Kerzendorf,
  Khvalko, King, Kirkby, Kulkarni, Kumar, Lee, Lenz, Littlefair, Ma, Macleod,
  Mastropietro, McCully, Montagnac, Morris, Mueller, Mumford, Muna, Murphy,
  Nelson, Nguyen, Ninan, N{\"o}the, Ogaz, Oh, Parejko, Parley, Pascual, Patil,
  Patil, Plunkett, Prochaska, Rastogi, Reddy~Janga, Sabater, Sakurikar,
  Seifert, Sherbert, {Sherwood-Taylor}, Shih, Sick, Silbiger, Singanamalla,
  Singer, Sladen, Sooley, Sornarajah, Streicher, Teuben, Thomas, Tremblay,
  Turner, Terr{\'o}n, {van Kerkwijk}, {de la Vega}, Watkins, Weaver, Whitmore,
  Woillez, Zabalza, \& {Astropy Contributors}}]{2018astropycollaboration}
{Astropy Collaboration}, {Price-Whelan}, A.~M., Sip{\H o}cz, B.~M., {et~al.}
  2018, The Astronomical Journal, 156, 123, \dodoi{10.3847/1538-3881/aabc4f}

\bibitem[{{Astropy Collaboration} {et~al.}(2022){Astropy Collaboration},
  {Price-Whelan}, Lim, Earl, Starkman, Bradley, Shupe, Patil, Corrales,
  Brasseur, N{\"o}the, Donath, Tollerud, Morris, Ginsburg, Vaher, Weaver,
  Tocknell, Jamieson, {van Kerkwijk}, Robitaille, Merry, Bachetti, G{\"u}nther,
  Aldcroft, {Alvarado-Montes}, Archibald, B{\'o}di, Bapat, Barentsen,
  Baz{\'a}n, Biswas, Boquien, Burke, Cara, Cara, Conroy, Conseil, Craig, Cross,
  Cruz, D'Eugenio, Dencheva, Devillepoix, Dietrich, Eigenbrot, Erben, Ferreira,
  {Foreman-Mackey}, Fox, Freij, Garg, Geda, Glattly, Gondhalekar, Gordon,
  Grant, Greenfield, Groener, Guest, Gurovich, Handberg, Hart,
  {Hatfield-Dodds}, Homeier, Hosseinzadeh, Jenness, Jones, Joseph, Kalmbach,
  Karamehmetoglu, Ka{\l}uszy{\'n}ski, Kelley, Kern, Kerzendorf, Koch, Kulumani,
  Lee, Ly, Ma, MacBride, Maljaars, Muna, Murphy, Norman, O'Steen, Oman,
  Pacifici, Pascual, {Pascual-Granado}, Patil, Perren, Pickering, Rastogi,
  Roulston, Ryan, Rykoff, Sabater, Sakurikar, Salgado, Sanghi, Saunders,
  Savchenko, Schwardt, {Seifert-Eckert}, Shih, Jain, Shukla, Sick, Simpson,
  Singanamalla, Singer, Singhal, Sinha, Sip{\H o}cz, Spitler, Stansby,
  Streicher, {\v S}umak, Swinbank, Taranu, Tewary, Tremblay, {de Val-Borro},
  Van~Kooten, Vasovi{\'c}, Verma, {de Miranda Cardoso}, Williams, Wilson,
  Winkel, {Wood-Vasey}, Xue, Yoachim, Zhang, Zonca, \& {Astropy Project
  Contributors}}]{2022astropycollaboration}
{Astropy Collaboration}, {Price-Whelan}, A.~M., Lim, P.~L., {et~al.} 2022, The
  Astrophysical Journal, 935, 167, \dodoi{10.3847/1538-4357/ac7c74}

\bibitem[{Becker {et~al.}(2015)Becker, Bolton, Madau, Pettini, {Ryan-Weber}, \&
  Venemans}]{2015becker}
Becker, G.~D., Bolton, J.~S., Madau, P., {et~al.} 2015, Monthly Notices of the
  Royal Astronomical Society, 447, 3402, \dodoi{10.1093/mnras/stu2646}

\bibitem[{Bertin \& Arnouts(1996)}]{1996bertin}
Bertin, E., \& Arnouts, S. 1996, Astronomy and Astrophysics Supplement, v.117,
  p.393-404, 117, 393, \dodoi{10.1051/aas:1996164}

\bibitem[{B{\"o}ker {et~al.}(2023)B{\"o}ker, Beck, Birkmann, Giardino, Keyes,
  Kumari, Muzerolle, Rawle, Zeidler, {Abul-Huda}, {Alves de Oliveira}, Arribas,
  Bechtold, Bhatawdekar, Bonaventura, Bunker, Cameron, Carniani, Charlot,
  Curti, Espinoza, Ferruit, Franx, Jakobsen, Karakla, {L{\'o}pez-Caniego},
  L{\"u}tzgendorf, Maiolino, Manjavacas, Marston, Moseley, Ogle, Perna,
  {Pe{\~n}a-Guerrero}, Pirzkal, Plesha, Proffitt, Rauscher, Rix, {Rodr{\'i}guez
  del Pino}, Rustamkulov, Sabbi, Sing, Sirianni, {te Plate}, {\'U}beda,
  Wahlgren, Wislowski, Wu, \& Willott}]{2023boker}
B{\"o}ker, T., Beck, T.~L., Birkmann, S.~M., {et~al.} 2023, Publications of the
  Astronomical Society of the Pacific, 135, 038001,
  \dodoi{10.1088/1538-3873/acb846}

\bibitem[{Bolan {et~al.}(2021)Bolan, Lemaux, Mason, Brada{\v c}, Treu, Strait,
  Pelliccia, Pentericci, \& Malkan}]{2021bolan}
Bolan, P., Lemaux, B.~C., Mason, C., {et~al.} 2021, Inferring the {{IGM Neutral
  Fraction}} at z \textasciitilde{} 6-8 with {{Low-Luminosity Lyman Break
  Galaxies}}

\bibitem[{Boucaud {et~al.}(2016)Boucaud, Bocchio, Abergel, Orieux, Dole, \&
  {Hadj-Youcef}}]{2016boucaud}
Boucaud, A., Bocchio, M., Abergel, A., {et~al.} 2016, Astronomy \&
  Astrophysics, 596, A63, \dodoi{10.1051/0004-6361/201629080}

\bibitem[{Bouwens {et~al.}(2022)Bouwens, Smit, Schouws, Stefanon, Bowler,
  Endsley, Gonzalez, Inami, Stark, Oesch, Hodge, Aravena, {da Cunha}, Dayal,
  de~Looze, Ferrara, Fudamoto, Graziani, Li, Nanayakkara, Pallottini,
  Schneider, Sommovigo, Topping, {van der Werf}, Algera, Barrufet, Hygate,
  Labb{\'e}, Riechers, \& Witstok}]{2022bouwens}
Bouwens, R.~J., Smit, R., Schouws, S., {et~al.} 2022, The Astrophysical
  Journal, 931, 160, \dodoi{10.3847/1538-4357/ac5a4a}

\bibitem[{Bradley {et~al.}(2023)Bradley, Sip{\H o}cz, Robitaille, Tollerud,
  Vin{\'i}cius, Deil, Barbary, Wilson, Busko, Donath, G{\"u}nther, Cara, Lim,
  Me{\ss}linger, Conseil, Bostroem, Droettboom, Bray, Bratholm, Jamieson,
  Ginsburg, Barentsen, Craig, Morris, Perrin, Rathi, Pascual, Perren, Georgiev,
  \& Kerzendorf}]{2023bradley}
Bradley, L., Sip{\H o}cz, B., Robitaille, T., {et~al.} 2023, Astropy/Photutils:
  1.9.0, Zenodo, \dodoi{10.5281/zenodo.8248020}

\bibitem[{{Brammer}(2023)}]{2023grizli}
{Brammer}, G. 2023, {grizli}, 1.8.2, Zenodo,  Zenodo,
  \dodoi{10.5281/zenodo.7712834}

\bibitem[{Brammer {et~al.}(2008)Brammer, {van Dokkum}, \& Coppi}]{2008brammer}
Brammer, G.~B., {van Dokkum}, P.~G., \& Coppi, P. 2008, The Astrophysical
  Journal, 686, 1503, \dodoi{10.1086/591786}

\bibitem[{Bremer \& Dayal(2023)}]{2023bremer}
Bremer, J., \& Dayal, P. 2023, Monthly Notices of the Royal Astronomical
  Society, 524, 118, \dodoi{10.1093/mnras/stad1844}

\bibitem[{Bunker {et~al.}(2010)Bunker, Wilkins, Ellis, Stark, Lorenzoni, Chiu,
  Lacy, Jarvis, \& Hickey}]{2010bunker}
Bunker, A.~J., Wilkins, S., Ellis, R.~S., {et~al.} 2010, Monthly Notices of the
  Royal Astronomical Society, 409, 855,
  \dodoi{10.1111/j.1365-2966.2010.17350.x}

\bibitem[{Bunker {et~al.}(2023)Bunker, Saxena, Cameron, Willott, {Curtis-Lake},
  Jakobsen, Carniani, Smit, Maiolino, Witstok, Curti, D'Eugenio, Jones,
  Ferruit, Arribas, Charlot, Chevallard, Giardino, {de Graaff}, Looser,
  Luetzgendorf, Maseda, Rawle, Rix, Rodriguez Del~Pino, Alberts, Egami,
  Eisenstein, Endsley, Hainline, Hausen, Johnson, Rieke, Rieke, Robertson,
  Shivaei, Stark, Sun, Tacchella, Tang, Williams, Willmer, Baker, Baum,
  Bhatawdekar, Bowler, Boyett, Chen, Circosta, Helton, Ji, Lyu, Nelson,
  Parlanti, Perna, Sandles, Scholtz, Suess, Topping, Uebler, Wallace, \&
  Whitler}]{2023bunker}
Bunker, A.~J., Saxena, A., Cameron, A.~J., {et~al.} 2023, {{JADES NIRSpec
  Spectroscopy}} of {{GN-z11}}: {{Lyman-}}\$\textbackslash alpha\$ Emission and
  Possible Enhanced Nitrogen Abundance in a \$z=10.60\$ Luminous Galaxy,
  \dodoi{10.48550/arXiv.2302.07256}

\bibitem[{{Burgasser} {et~al.}(2023){Burgasser}, {Gerasimov}, {Bezanson},
  {Labbe}, {Brammer}, {Cutler}, {Furtak}, {Greene}, {Leja}, {Pan}, {Price},
  {Wang}, {Weaver}, {Whitaker}, {Fujimoto}, {Kokorev}, {Dayal}, {Nanayakkara},
  {Williams}, \& {Zitrin}}]{2023burgasser}
{Burgasser}, A.~J., {Gerasimov}, R., {Bezanson}, R., {et~al.} 2023, arXiv
  e-prints, arXiv:2308.12107, \dodoi{10.48550/arXiv.2308.12107}

\bibitem[{Bushouse {et~al.}(2022)Bushouse, Eisenhamer, Dencheva, Davies,
  Greenfield, Morrison, Hodge, Simon, Grumm, Droettboom, Slavich, Sosey, Pauly,
  Miller, Jedrzejewski, Hack, Davis, Crawford, Law, Gordon, Regan, Cara,
  MacDonald, Bradley, Shanahan, Jamieson, Teodoro, \& Williams}]{2022bushouse}
Bushouse, H., Eisenhamer, J., Dencheva, N., {et~al.} 2022, Zenodo,
  \dodoi{10.5281/zenodo.7325378}

\bibitem[{Byler {et~al.}(2017)Byler, Dalcanton, Conroy, \& Johnson}]{2017byler}
Byler, N., Dalcanton, J.~J., Conroy, C., \& Johnson, B.~D. 2017, The
  Astrophysical Journal, 840, 44, \dodoi{10.3847/1538-4357/aa6c66}

\bibitem[{Calzetti(2001)}]{2001calzetti}
Calzetti, D. 2001, Publications of the Astronomical Society of the Pacific,
  113, 1449, \dodoi{10.1086/324269}

\bibitem[{Cameron {et~al.}(2023)Cameron, Saxena, Bunker, D'Eugenio, Carniani,
  Maiolino, {Curtis-Lake}, Ferruit, Jakobsen, Arribas, Bonaventura, Charlot,
  Chevallard, Curti, Looser, Maseda, Rawle, Rodr{\'i}guez Del~Pino, Smit,
  {\"U}bler, Willott, Witstok, Egami, Eisenstein, Johnson, Hainline, Rieke,
  Robertson, Stark, Tacchella, Williams, Willmer, Bhatawdekar, Bowler, Boyett,
  Circosta, Helton, Jones, Kumari, Ji, Nelson, Parlanti, Sandles, Scholtz, \&
  Sun}]{2023cameron}
Cameron, A.~J., Saxena, A., Bunker, A.~J., {et~al.} 2023, {{JADES}}:
  {{Probing}} Interstellar Medium Conditions at \$z\textbackslash sim5.5-9.5\$
  with Ultra-Deep {{JWST}}/{{NIRSpec}} Spectroscopy,
  \dodoi{10.48550/arXiv.2302.04298}

\bibitem[{Carnall {et~al.}(2018)Carnall, McLure, Dunlop, \&
  Dav{\'e}}]{2018carnall}
Carnall, A.~C., McLure, R.~J., Dunlop, J.~S., \& Dav{\'e}, R. 2018, Monthly
  Notices of the Royal Astronomical Society, 480, 4379,
  \dodoi{10.1093/mnras/sty2169}

\bibitem[{Casey {et~al.}(2022)Casey, Kartaltepe, Drakos, Franco, Harish,
  Paquereau, Ilbert, Rose, Cox, Nightingale, Robertson, Silverman, Koekemoer,
  Massey, McCracken, Rhodes, Akins, Amvrosiadis, {Arango-Toro}, Bagley,
  Bongiorno, Capak, Champagne, Chartab, Chavez~Ortiz, Chworowsky, Cooke,
  Cooper, Darvish, Ding, Faisst, Finkelstein, Fujimoto, Gentile, Gillman,
  Gould, Gozaliasl, Hayward, He, Hemmati, Hirschmann, Jahnke, Jin, Khostovan,
  Kokorev, Lambrides, Laigle, Larson, Leung, Liu, Liaudat, Long, Magdis,
  Mahler, Mainieri, Manning, Maraston, Martin, McCleary, McKinney, McPartland,
  Mobasher, Pattnaik, Renzini, Rich, Sanders, Sattari, Scognamiglio, Scoville,
  Sheth, Shuntov, Sparre, Suzuki, Talia, Toft, Trakhtenbrot, Urry, Valentino,
  Vanderhoof, Vardoulaki, Weaver, Whitaker, Wilkins, Yang, \&
  Zavala}]{2022casey}
Casey, C.~M., Kartaltepe, J.~S., Drakos, N.~E., {et~al.} 2022, {{COSMOS-Web}}:
  {{An Overview}} of the {{JWST Cosmic Origins Survey}},
  \dodoi{10.48550/arXiv.2211.07865}

\bibitem[{Castellano {et~al.}(2016)Castellano, Dayal, Pentericci, Fontana,
  Hutter, Brammer, Merlin, Grazian, Pilo, Amorin, Cristiani, Dickinson,
  Ferrara, Gallerani, Giallongo, Giavalisco, Guaita, Koekemoer, Maiolino,
  Paris, Santini, Vallini, Vanzella, \& Wagg}]{2016castellano}
Castellano, M., Dayal, P., Pentericci, L., {et~al.} 2016, The Astrophysical
  Journal, 818, L3, \dodoi{10.3847/2041-8205/818/1/L3}

\bibitem[{Caswell {et~al.}(2023)Caswell, de~Andrade, Lee, Droettboom, Hoffmann,
  Klymak, Hunter, Firing, Stansby, Varoquaux, Nielsen, Gustafsson, Root, May,
  Elson, Sepp{\"a}nen, Lee, Dale, Sunden, {hannah}, McDougall, Straw, Hobson,
  Lucas, Gohlke, Vincent, Yu, Ma, Silvester, \& Moad}]{2023caswell}
Caswell, T.~A., de~Andrade, E.~S., Lee, A., {et~al.} 2023,
  Matplotlib/Matplotlib: {{REL}}: V3.7.2, Zenodo,
  \dodoi{10.5281/zenodo.8118151}

\bibitem[{Cen \& Haiman(2000)}]{2000cen}
Cen, R., \& Haiman, Z. 2000, The Astrophysical Journal, 542, L75,
  \dodoi{10.1086/312937}

\bibitem[{Chabrier(2003)}]{2003chabrier}
Chabrier, G. 2003, Publications of the Astronomical Society of the Pacific,
  115, 763, \dodoi{10.1086/376392}

\bibitem[{Conroy {et~al.}(2010)Conroy, White, \& Gunn}]{2010conroy}
Conroy, C., White, M., \& Gunn, J.~E. 2010, The Astrophysical Journal, 708, 58,
  \dodoi{10.1088/0004-637X/708/1/58}

\bibitem[{{Curtis-Lake} {et~al.}(2023){Curtis-Lake}, Carniani, Cameron,
  Charlot, Jakobsen, Maiolino, Bunker, Witstok, Smit, Chevallard, Willott,
  Ferruit, Arribas, Bonaventura, Curti, D'Eugenio, Franx, Giardino, Looser,
  L{\"u}tzgendorf, Maseda, Rawle, Rix, {Rodr{\'i}guez del Pino}, {\"U}bler,
  Sirianni, Dressler, Egami, Eisenstein, Endsley, Hainline, Hausen, Johnson,
  Rieke, Robertson, Shivaei, Stark, Tacchella, Williams, Willmer, Bhatawdekar,
  Bowler, Boyett, Chen, {de Graaff}, Helton, Hviding, Jones, Kumari, Lyu,
  Nelson, Perna, Sandles, Saxena, Suess, Sun, Topping, Wallace, \&
  Whitler}]{2023curtis-lake}
{Curtis-Lake}, E., Carniani, S., Cameron, A., {et~al.} 2023, Nature Astronomy,
  7, 622, \dodoi{10.1038/s41550-023-01918-w}

\bibitem[{D'Aloisio {et~al.}(2018)D'Aloisio, McQuinn, Davies, \&
  Furlanetto}]{2018daloisio}
D'Aloisio, A., McQuinn, M., Davies, F.~B., \& Furlanetto, S.~R. 2018, Monthly
  Notices of the Royal Astronomical Society, 473, 560,
  \dodoi{10.1093/mnras/stx2341}

\bibitem[{Dijkstra {et~al.}(2014)Dijkstra, Wyithe, Haiman, Mesinger, \&
  Pentericci}]{2014dijkstra}
Dijkstra, M., Wyithe, S., Haiman, Z., Mesinger, A., \& Pentericci, L. 2014,
  Monthly Notices of the Royal Astronomical Society, 440, 3309,
  \dodoi{10.1093/mnras/stu531}

\bibitem[{Endsley \& Stark(2022)}]{2022endsleya}
Endsley, R., \& Stark, D.~P. 2022, Monthly Notices of the Royal Astronomical
  Society, 511, 6042, \dodoi{10.1093/mnras/stac524}

\bibitem[{Endsley {et~al.}(2021)Endsley, Stark, Charlot, Chevallard, Robertson,
  Bouwens, \& Stefanon}]{2021endsley}
Endsley, R., Stark, D.~P., Charlot, S., {et~al.} 2021, Monthly Notices of the
  Royal Astronomical Society, 502, 6044, \dodoi{10.1093/mnras/stab432}

\bibitem[{{Euclid Collaboration} {et~al.}(2022){Euclid Collaboration}, Moneti,
  McCracken, Shuntov, Kauffmann, Capak, Davidzon, Ilbert, Scarlata, Toft,
  Weaver, Chary, Cuby, Faisst, Masters, McPartland, Mobasher, Sanders,
  Scaramella, Stern, Szapudi, Teplitz, Zalesky, Amara, Auricchio, Bodendorf,
  Bonino, Branchini, {Brau-Nogue}, Brescia, Brinchmann, Capobianco, Carbone,
  Carretero, Castander, Castellano, Cavuoti, Cimatti, Cledassou, Congedo,
  Conselice, Conversi, Copin, Corcione, Costille, Cropper, Da~Silva,
  Degaudenzi, Douspis, Dubath, Duncan, Dupac, Dusini, Farrens, Ferriol,
  Fosalba, Frailis, Franceschi, Fumana, Garilli, Gillis, Giocoli, Granett,
  Grazian, Grupp, Haugan, Hoekstra, Holmes, Hormuth, Hudelot, Jahnke, Kermiche,
  Kiessling, Kilbinger, Kitching, Kohley, K{\"u}mmel, Kunz, {Kurki-Suonio},
  Ligori, Lilje, Lloro, Maiorano, Mansutti, Marggraf, Markovic, Marulli,
  Massey, Maurogordato, Meneghetti, Merlin, Meylan, Moresco, Moscardini,
  Munari, Niemi, Padilla, Paltani, Pasian, Pedersen, Pires, Poncet, Popa,
  Pozzetti, Raison, Rebolo, Rhodes, Rix, Roncarelli, Rossetti, Saglia,
  Schneider, Secroun, Seidel, Serrano, Sirignano, Sirri, Stanco,
  {Tallada-Cresp{\'i}}, Taylor, Tereno, {Toledo-Moreo}, Torradeflot, Wang,
  Welikala, Weller, Zamorani, Zoubian, Andreon, Bardelli, Camera,
  {Graci{\'a}-Carpio}, Medinaceli, Mei, Polenta, Romelli, Sureau, Tenti,
  Vassallo, Zacchei, Zucca, Baccigalupi, {Balaguera-Antol{\'i}nez}, Bernardeau,
  Biviano, Bolzonella, Bozzo, Burigana, Cabanac, Cappi, Carvalho, Casas,
  Castignani, {Colodro-Conde}, Coupon, Courtois, Di~Ferdinando, Farina,
  Finelli, {Flose-Reimberg}, Fotopoulou, Galeotta, Ganga, {Garcia-Bellido},
  Gaztanaga, Gozaliasl, Hook, Joachimi, Kansal, Keihanen, Kirkpatrick,
  Lindholm, Mainetti, Maino, Maoli, Martinelli, Martinet, Maturi, Metcalf,
  Morgante, Morisset, Nucita, Patrizii, Potter, Renzi, Riccio, S{\'a}nchez,
  Sapone, Schirmer, Schultheis, Scottez, Sefusatti, Teyssier, Tubio, Tutusaus,
  Valiviita, Viel, \& Hildebrandt}]{2022euclidcollaboration}
{Euclid Collaboration}, Moneti, A., McCracken, H.~J., {et~al.} 2022, Astronomy
  and Astrophysics, 658, A126, \dodoi{10.1051/0004-6361/202142361}

\bibitem[{Faisst(2016)}]{2016faisst}
Faisst, A.~L. 2016, The Astrophysical Journal, 829, 99,
  \dodoi{10.3847/0004-637X/829/2/99}

\bibitem[{Faisst {et~al.}(2014)Faisst, Capak, Carollo, Scarlata, \&
  Scoville}]{2014faisst}
Faisst, A.~L., Capak, P., Carollo, C.~M., Scarlata, C., \& Scoville, N. 2014,
  The Astrophysical Journal, 788, 87, \dodoi{10.1088/0004-637X/788/1/87}

\bibitem[{Ferland {et~al.}(2017)Ferland, Chatzikos, Guzm{\'a}n, Lykins, {van
  Hoof}, Williams, Abel, Badnell, Keenan, Porter, \& Stancil}]{2017ferland}
Ferland, G.~J., Chatzikos, M., Guzm{\'a}n, F., {et~al.} 2017, Revista Mexicana
  de Astronomia y Astrofisica, 53, 385, \dodoi{10.48550/arXiv.1705.10877}

\bibitem[{Finkelstein(2016)}]{2016finkelstein}
Finkelstein, S.~L. 2016, Publications of the Astronomical Society of Australia,
  33, e037, \dodoi{10.1017/pasa.2016.26}

\bibitem[{Finkelstein {et~al.}(2010)Finkelstein, Papovich, Giavalisco, Reddy,
  Ferguson, Koekemoer, \& Dickinson}]{2010finkelstein}
Finkelstein, S.~L., Papovich, C., Giavalisco, M., {et~al.} 2010, The
  Astrophysical Journal, 719, 1250, \dodoi{10.1088/0004-637X/719/2/1250}

\bibitem[{Finkelstein {et~al.}(2012)Finkelstein, Papovich, Ryan, Pawlik,
  Dickinson, Ferguson, Finlator, Koekemoer, Giavalisco, Cooray, Dunlop, Faber,
  Grogin, Kocevski, \& Newman}]{2012finkelstein}
Finkelstein, S.~L., Papovich, C., Ryan, R.~E., {et~al.} 2012, The Astrophysical
  Journal, 758, 93, \dodoi{10.1088/0004-637X/758/2/93}

\bibitem[{Finkelstein {et~al.}(2013)Finkelstein, Papovich, Dickinson, Song,
  Tilvi, Koekemoer, Finkelstein, Mobasher, Ferguson, Giavalisco, Reddy, Ashby,
  Dekel, Fazio, Fontana, Grogin, Huang, Kocevski, Rafelski, Weiner, \&
  Willner}]{2013finkelstein}
Finkelstein, S.~L., Papovich, C., Dickinson, M., {et~al.} 2013, Nature, 502,
  524, \dodoi{10.1038/nature12657}

\bibitem[{Finkelstein {et~al.}(2019)Finkelstein, D'Aloisio, Paardekooper, Ryan,
  Behroozi, Finlator, Livermore, Upton~Sanderbeck, Dalla~Vecchia, \&
  Khochfar}]{2019finkelstein}
Finkelstein, S.~L., D'Aloisio, A., Paardekooper, J.-P., {et~al.} 2019, The
  Astrophysical Journal, 879, 36, \dodoi{10.3847/1538-4357/ab1ea8}

\bibitem[{Finkelstein {et~al.}(2022)Finkelstein, Bagley, Song, Larson,
  Papovich, Dickinson, Finkelstein, Koekemoer, Pirzkal, Somerville, Yung,
  Behroozi, Ferguson, Giavalisco, Grogin, Hathi, Hutchison, Jung, Kocevski,
  Kawinwanichakij, {Rojas-Ruiz}, Ryan, Snyder, \& Tacchella}]{2022finkelstein}
Finkelstein, S.~L., Bagley, M., Song, M., {et~al.} 2022, The Astrophysical
  Journal, 928, 52, \dodoi{10.3847/1538-4357/ac3aed}

\bibitem[{Flury {et~al.}(2022)Flury, Jaskot, Ferguson, Worseck, Makan,
  Chisholm, {Saldana-Lopez}, Schaerer, McCandliss, Wang, Ford, Heckman, Ji,
  Giavalisco, Amorin, Atek, Blaizot, Borthakur, Carr, Castellano, Cristiani,
  De~Barros, Dickinson, Finkelstein, Fleming, Fontanot, Garel, Grazian, Hayes,
  Henry, Mauerhofer, Micheva, Oey, Ostlin, Papovich, Pentericci, Ravindranath,
  Rosdahl, Rutkowski, Santini, Scarlata, Teplitz, Thuan, Trebitsch, Vanzella,
  Verhamme, \& Xu}]{2022flury}
Flury, S.~R., Jaskot, A.~E., Ferguson, H.~C., {et~al.} 2022, The Astrophysical
  Journal Supplement Series, 260, 1, \dodoi{10.3847/1538-4365/ac5331}

\bibitem[{Fujimoto {et~al.}(2023)Fujimoto, Arrabal~Haro, Dickinson,
  Finkelstein, Kartaltepe, Larson, Burgarella, Bagley, Behroozi, Chworowsky,
  Hirschmann, Trump, Wilkins, Yung, Koekemoer, Papovich, Pirzkal, Ferguson,
  Fontana, Grogin, Grazian, Kewley, Kocevski, Lotz, Pentericci, Ravindranath,
  Somerville, Wilkins, Amor{\'i}n, Backhaus, Calabr{\`o}, Casey, Cooper,
  Fern{\'a}ndez, Franco, Giavalisco, Hathi, Harish, Hutchison, Iyer, Jung,
  Lucas, \& Zavala}]{2023fujimoto}
Fujimoto, S., Arrabal~Haro, P., Dickinson, M., {et~al.} 2023, The Astrophysical
  Journal, 949, L25, \dodoi{10.3847/2041-8213/acd2d9}

\bibitem[{Furlanetto {et~al.}(2017)Furlanetto, Mirocha, Mebane, \&
  Sun}]{2017furlanetto}
Furlanetto, S.~R., Mirocha, J., Mebane, R.~H., \& Sun, G. 2017, Monthly Notices
  of the Royal Astronomical Society, 472, 1576, \dodoi{10.1093/mnras/stx2132}

\bibitem[{{Gaia Collaboration} {et~al.}(2023){Gaia Collaboration}, Vallenari,
  Brown, Prusti, {de Bruijne}, Arenou, Babusiaux, Biermann, Creevey, Ducourant,
  Evans, Eyer, Guerra, Hutton, Jordi, Klioner, Lammers, Lindegren, Luri,
  Mignard, Panem, Pourbaix, Randich, Sartoretti, Soubiran, Tanga, Walton,
  {Bailer-Jones}, Bastian, Drimmel, Jansen, Katz, Lattanzi, {van Leeuwen},
  Bakker, Cacciari, Casta{\~n}eda, De~Angeli, Fabricius, Fouesneau, Fr{\'e}mat,
  Galluccio, Guerrier, Heiter, Masana, Messineo, Mowlavi, Nicolas,
  Nienartowicz, Pailler, Panuzzo, Riclet, Roux, Seabroke, Sordo, Th{\'e}venin,
  {Gracia-Abril}, Portell, Teyssier, Altmann, Andrae, Audard, {Bellas-Velidis},
  Benson, Berthier, Blomme, Burgess, Busonero, Busso, C{\'a}novas, Carry,
  Cellino, Cheek, Clementini, Damerdji, Davidson, {de Teodoro},
  Nu{\~n}ez~Campos, Delchambre, Dell'Oro, Esquej,
  {Fern{\'a}ndez-Hern{\'a}ndez}, Fraile, Garabato, {Garc{\'i}a-Lario}, Gosset,
  Haigron, Halbwachs, Hambly, Harrison, Hern{\'a}ndez, Hestroffer, Hodgkin,
  Holl, Jan{\ss}en, {Jevardat de Fombelle}, Jordan, {Krone-Martins}, Lanzafame,
  L{\"o}ffler, Marchal, Marrese, Moitinho, Muinonen, Osborne, Pancino, Pauwels,
  {Recio-Blanco}, Reyl{\'e}, Riello, Rimoldini, Roegiers, Rybizki, Sarro,
  Siopis, Smith, Sozzetti, Utrilla, {van Leeuwen}, Abbas, {\'A}brah{\'a}m,
  Abreu~Aramburu, Aerts, Aguado, Ajaj, {Aldea-Montero}, Altavilla, {\'A}lvarez,
  Alves, Anders, Anderson, Anglada~Varela, Antoja, Baines, Baker,
  {Balaguer-N{\'u}{\~n}ez}, Balbinot, Balog, Barache, Barbato, Barros, Barstow,
  Bartolom{\'e}, Bassilana, Bauchet, Becciani, Bellazzini, Berihuete, Bernet,
  Bertone, Bianchi, Binnenfeld, {Blanco-Cuaresma}, Blazere, Boch, Bombrun,
  Bossini, Bouquillon, Bragaglia, Bramante, Breedt, Bressan, Brouillet,
  Brugaletta, Bucciarelli, Burlacu, Butkevich, Buzzi, Caffau, Cancelliere,
  {Cantat-Gaudin}, Carballo, Carlucci, Carnerero, Carrasco, Casamiquela,
  Castellani, {Castro-Ginard}, Chaoul, Charlot, Chemin, Chiaramida, Chiavassa,
  Chornay, Comoretto, Contursi, Cooper, Cornez, Cowell, Crifo, Cropper, Crosta,
  Crowley, Dafonte, Dapergolas, David, David, {de Laverny}, De~Luise, De~March,
  De~Ridder, {de Souza}, {de Torres}, {del Peloso}, {del Pozo}, Delbo, Delgado,
  Delisle, Demouchy, Dharmawardena, Di~Matteo, Diakite, Diener, Distefano,
  Dolding, Edvardsson, Enke, Fabre, Fabrizio, Faigler, Fedorets, Fernique,
  Fienga, Figueras, Fournier, Fouron, Fragkoudi, Gai, {Garcia-Gutierrez},
  {Garcia-Reinaldos}, {Garc{\'i}a-Torres}, Garofalo, Gavel, Gavras, Gerlach,
  Geyer, Giacobbe, Gilmore, Girona, Giuffrida, Gomel, Gomez,
  {Gonz{\'a}lez-N{\'u}{\~n}ez}, {Gonz{\'a}lez-Santamar{\'i}a},
  {Gonz{\'a}lez-Vidal}, Granvik, Guillout, Guiraud,
  {Guti{\'e}rrez-S{\'a}nchez}, Guy, Hatzidimitriou, Hauser, Haywood, Helmer,
  Helmi, Sarmiento, Hidalgo, Hilger, H{\l}adczuk, Hobbs, Holland, Huckle,
  Jardine, Jasniewicz, {Jean-Antoine Piccolo}, {Jim{\'e}nez-Arranz}, Jorissen,
  Juaristi~Campillo, Julbe, Karbevska, Kervella, Khanna, Kontizas, Kordopatis,
  Korn, K{\'o}sp{\'a}l, {Kostrzewa-Rutkowska}, Kruszy{\'n}ska, Kun, Laizeau,
  Lambert, Lanza, Lasne, Le~Campion, Lebreton, Lebzelter, Leccia, Leclerc,
  {Lecoeur-Taibi}, Liao, Licata, Lindstr{\o}m, Lister, Livanou, Lobel, Lorca,
  Loup, Madrero~Pardo, Magdaleno~Romeo, Managau, Mann, Manteiga, Marchant,
  Marconi, Marcos, Marcos~Santos, Mar{\'i}n~Pina, Marinoni, Marocco, Marshall,
  Martin~Polo, {Mart{\'i}n-Fleitas}, Marton, Mary, Masip, Massari,
  {Mastrobuono-Battisti}, Mazeh, McMillan, Messina, Michalik, Millar, Mints,
  Molina, Molinaro, Moln{\'a}r, Monari, Mongui{\'o}, Montegriffo, Montero, Mor,
  Mora, Morbidelli, Morel, Morris, Muraveva, Murphy, Musella, Nagy, Noval,
  Oca{\~n}a, Ogden, Ordenovic, Osinde, Pagani, Pagano, Palaversa, Palicio,
  {Pallas-Quintela}, Panahi, {Payne-Wardenaar}, Pe{\~n}alosa~Esteller,
  Penttil{\"a}, Pichon, Piersimoni, Pineau, Plachy, Plum, Poggio, Pr{\v s}a,
  Pulone, Racero, Ragaini, Rainer, Raiteri, Rambaux, Ramos, {Ramos-Lerate},
  Re~Fiorentin, Regibo, Richards, Rios~Diaz, Ripepi, Riva, Rix, Rixon,
  Robichon, Robin, Robin, Roelens, Rogues, Rohrbasser, {Romero-G{\'o}mez},
  Rowell, Royer, Ruz~Mieres, Rybicki, Sadowski, S{\'a}ez~N{\'u}{\~n}ez,
  Sagrist{\`a}~Sell{\'e}s, Sahlmann, Salguero, Samaras, Sanchez~Gimenez, Sanna,
  Santove{\~n}a, Sarasso, Schultheis, Sciacca, Segol, Segovia, S{\'e}gransan,
  Semeux, Shahaf, Siddiqui, Siebert, Siltala, Silvelo, Slezak, Slezak, Smart,
  Snaith, Solano, Solitro, Souami, Souchay, Spagna, Spina, Spoto, Steele,
  Steidelm{\"u}ller, Stephenson, S{\"u}veges, Surdej, Szabados, {Szegedi-Elek},
  Taris, Taylor, Teixeira, Tolomei, Tonello, Torra, Torra, Torralba~Elipe,
  Trabucchi, Tsounis, Turon, Ulla, Unger, Vaillant, {van Dillen}, {van Reeven},
  Vanel, Vecchiato, Viala, Vicente, Voutsinas, Weiler, Wevers, Wyrzykowski,
  Yoldas, Yvard, Zhao, Zorec, Zucker, \& Zwitter}]{2023gaiacollaboration}
{Gaia Collaboration}, Vallenari, A., Brown, A. G.~A., {et~al.} 2023, Astronomy
  and Astrophysics, 674, A1, \dodoi{10.1051/0004-6361/202243940}

\bibitem[{Grogin {et~al.}(2011)Grogin, Kocevski, Faber, Ferguson, Koekemoer,
  Riess, Acquaviva, Alexander, Almaini, Ashby, Barden, Bell, Bournaud, Brown,
  Caputi, Casertano, Cassata, Castellano, Challis, Chary, Cheung, Cirasuolo,
  Conselice, Roshan~Cooray, Croton, Daddi, Dahlen, Dav{\'e}, {de Mello}, Dekel,
  Dickinson, Dolch, Donley, Dunlop, Dutton, Elbaz, Fazio, Filippenko,
  Finkelstein, Fontana, Gardner, Garnavich, Gawiser, Giavalisco, Grazian, Guo,
  Hathi, H{\"a}ussler, Hopkins, Huang, Huang, Jha, Kartaltepe, Kirshner, Koo,
  Lai, Lee, Li, Lotz, Lucas, Madau, McCarthy, McGrath, McIntosh, McLure,
  Mobasher, Moustakas, Mozena, Nandra, Newman, Niemi, Noeske, Papovich,
  Pentericci, Pope, Primack, Rajan, Ravindranath, Reddy, Renzini, Rix, Robaina,
  Rodney, Rosario, Rosati, Salimbeni, Scarlata, Siana, Simard, Smidt,
  Somerville, Spinrad, Straughn, Strolger, Telford, Teplitz, Trump, {van der
  Wel}, Villforth, Wechsler, Weiner, Wiklind, Wild, Wilson, Wuyts, Yan, \&
  Yun}]{2011grogin}
Grogin, N.~A., Kocevski, D.~D., Faber, S.~M., {et~al.} 2011, The Astrophysical
  Journal Supplement Series, 197, 35, \dodoi{10.1088/0067-0049/197/2/35}

\bibitem[{Hainline {et~al.}(2023)Hainline, Helton, Johnson, Sun, Topping,
  Leisenring, Baker, Eisenstein, Hausen, Hviding, Lyu, Robertson, Tacchella,
  Williams, Willmer, \& Roellig}]{2023hainline}
Hainline, K.~N., Helton, J.~M., Johnson, B.~D., {et~al.} 2023, Brown {{Dwarf
  Candidates}} in the {{JADES}} and {{CEERS Extragalactic Surveys}},
  \dodoi{10.48550/arXiv.2309.03250}

\bibitem[{Harris {et~al.}(2020)Harris, Millman, {van der Walt}, Gommers,
  Virtanen, Cournapeau, Wieser, Taylor, Berg, Smith, Kern, Picus, Hoyer, {van
  Kerkwijk}, Brett, Haldane, {del R{\'i}o}, Wiebe, Peterson,
  {G{\'e}rard-Marchant}, Sheppard, Reddy, Weckesser, Abbasi, Gohlke, \&
  Oliphant}]{2020harris}
Harris, C.~R., Millman, K.~J., {van der Walt}, S.~J., {et~al.} 2020, Nature,
  585, 357, \dodoi{10.1038/s41586-020-2649-2}

\bibitem[{Hashimoto {et~al.}(2018)Hashimoto, Laporte, Mawatari, Ellis, Inoue,
  Zackrisson, {Roberts-Borsani}, Zheng, Tamura, Bauer, Fletcher, Harikane,
  Hatsukade, Hayatsu, Matsuda, Matsuo, Okamoto, Ouchi, Pell{\'o}, Rydberg,
  Shimizu, Taniguchi, Umehata, \& Yoshida}]{2018hashimoto}
Hashimoto, T., Laporte, N., Mawatari, K., {et~al.} 2018, Nature, 557, 392,
  \dodoi{10.1038/s41586-018-0117-z}

\bibitem[{Heintz {et~al.}(2022)Heintz, Brammer, {Gim{\'e}nez-Arteaga}, Strait,
  Lagos, Vijayan, Matthee, Watson, Mason, Hutter, Toft, Fynbo, \&
  Oesch}]{2022heintz}
Heintz, K.~E., Brammer, G.~B., {Gim{\'e}nez-Arteaga}, C., {et~al.} 2022,
  Dilution of Chemical Enrichment in Galaxies 600 {{Myr}} after the {{Big
  Bang}}, \dodoi{10.48550/arXiv.2212.02890}

\bibitem[{Hoag {et~al.}(2017)Hoag, Brada{\v c}, Trenti, Treu, Schmidt, Huang,
  Lemaux, He, Bernard, Abramson, Mason, Morishita, Pentericci, \&
  Schrabback}]{2017hoag}
Hoag, A., Brada{\v c}, M., Trenti, M., {et~al.} 2017, Nature Astronomy, 1,
  0091, \dodoi{10.1038/s41550-017-0091}

\bibitem[{Hoag {et~al.}(2018)Hoag, Brada{\v c}, Brammer, Huang, Treu, Mason,
  Castellano, Di~Criscienzo, Jones, Kelly, Pentericci, Ryan, Schmidt, \&
  Trenti}]{2018hoag}
Hoag, A., Brada{\v c}, M., Brammer, G., {et~al.} 2018, The Astrophysical
  Journal, 854, 39, \dodoi{10.3847/1538-4357/aaa9c2}

\bibitem[{Hoag {et~al.}(2019)Hoag, Brada{\v c}, Huang, Mason, Treu, Schmidt,
  Trenti, Strait, Lemaux, Finney, \& Paddock}]{2019hoag}
Hoag, A., Brada{\v c}, M., Huang, K., {et~al.} 2019, The Astrophysical Journal,
  878, 12, \dodoi{10.3847/1538-4357/ab1de7}

\bibitem[{Horne(1986)}]{1986horne}
Horne, K. 1986, Publications of the Astronomical Society of the Pacific, 98,
  609, \dodoi{10.1086/131801}

\bibitem[{Hu {et~al.}(2019)Hu, Wang, Zheng, Malhotra, Rhoads, Infante,
  Barrientos, Yang, Jiang, Kang, Perez, Wold, Hibon, Jiang, Khostovan, Valdes,
  Walker, Galaz, Coughlin, Harish, Kong, Pharo, \& Zheng}]{2019hu}
Hu, W., Wang, J., Zheng, Z.-Y., {et~al.} 2019, The Astrophysical Journal, 886,
  90, \dodoi{10.3847/1538-4357/ab4cf4}

\bibitem[{Hutchison {et~al.}(2020)Hutchison, Walawender, \&
  Kwok}]{2020hutchison}
Hutchison, T.~A., Walawender, J., \& Kwok, S.~H. 2020, 11447, 114476A,
  \dodoi{10.1117/12.2562864}

\bibitem[{Ilbert {et~al.}(2006)Ilbert, Arnouts, McCracken, Bolzonella, Bertin,
  Le~F{\`e}vre, Mellier, Zamorani, Pell{\`o}, Iovino, Tresse, Le~Brun, Bottini,
  Garilli, Maccagni, Picat, Scaramella, Scodeggio, Vettolani, Zanichelli,
  Adami, Bardelli, Cappi, Charlot, Ciliegi, Contini, Cucciati, Foucaud,
  Franzetti, Gavignaud, Guzzo, Marano, Marinoni, Mazure, Meneux, Merighi,
  Paltani, Pollo, Pozzetti, Radovich, Zucca, Bondi, Bongiorno, Busarello, {de
  La Torre}, Gregorini, Lamareille, Mathez, Merluzzi, Ripepi, Rizzo, \&
  Vergani}]{2006ilbert}
Ilbert, O., Arnouts, S., McCracken, H.~J., {et~al.} 2006, Astronomy and
  Astrophysics, Volume 457, Issue 3, October III 2006, pp.841-856, 457, 841,
  \dodoi{10.1051/0004-6361:20065138}

\bibitem[{Inoue {et~al.}(2016)Inoue, Tamura, Matsuo, Mawatari, Shimizu,
  Shibuya, Ota, Yoshida, Zackrisson, Kashikawa, Kohno, Umehata, Hatsukade, Iye,
  Matsuda, Okamoto, \& Yamaguchi}]{2016inoue}
Inoue, A.~K., Tamura, Y., Matsuo, H., {et~al.} 2016, Science, 352, 1559,
  \dodoi{10.1126/science.aaf0714}

\bibitem[{Izotov {et~al.}(2016)Izotov, Schaerer, Thuan, Worseck, Guseva,
  Orlitov{\'a}, \& Verhamme}]{2016izotov}
Izotov, Y.~I., Schaerer, D., Thuan, T.~X., {et~al.} 2016, Monthly Notices of
  the Royal Astronomical Society, 461, 3683, \dodoi{10.1093/mnras/stw1205}

\bibitem[{Izotov {et~al.}(2018)Izotov, Schaerer, Worseck, Guseva, Thuan,
  Verhamme, Orlitov{\'a}, \& Fricke}]{2018izotov}
Izotov, Y.~I., Schaerer, D., Worseck, G., {et~al.} 2018, Monthly Notices of the
  Royal Astronomical Society, 474, 4514, \dodoi{10.1093/mnras/stx3115}

\bibitem[{Jung {et~al.}(2019)Jung, Finkelstein, Dickinson, Hutchison, Larson,
  Papovich, Pentericci, Song, Ferguson, Guo, Malhotra, Mobasher, Rhoads, Tilvi,
  \& Wold}]{2019jung}
Jung, I., Finkelstein, S.~L., Dickinson, M., {et~al.} 2019, The Astrophysical
  Journal, 877, 146, \dodoi{10.3847/1538-4357/ab1bde}

\bibitem[{Jung {et~al.}(2020)Jung, Finkelstein, Dickinson, Hutchison, Larson,
  Papovich, Pentericci, Straughn, Guo, Malhotra, Rhoads, Song, Tilvi, \&
  Wold}]{2020jung}
---. 2020, The Astrophysical Journal, 904, 144,
  \dodoi{10.3847/1538-4357/abbd44}

\bibitem[{Jung {et~al.}(2022)Jung, Finkelstein, Larson, Hutchison, Straughn,
  Bagley, Castellano, Cleri, Cooper, Dickinson, Ferguson, Holwerda, Kartaltepe,
  Kim, Koekemoer, Papovich, Park, Pentericci, {Perez-Gonzalez}, Song,
  Tacchella, Weiner, Willmer, \& Zavala}]{2022jung}
Jung, I., Finkelstein, S.~L., Larson, R.~L., {et~al.} 2022, New \$z {$>$} 7\$
  {{Lyman-alpha Emitters}} in {{EGS}}: {{Evidence}} of an {{Extended Ionized
  Structure}} at \$z \textbackslash sim 7.7\$,
  \dodoi{10.48550/arXiv.2212.09850}

\bibitem[{Jung {et~al.}(2023)Jung, Finkelstein, Arrabal~Haro, Dickinson,
  Ferguson, Hutchison, Kartaltepe, Larson, Simons, Papovich, Park, Pentericci,
  Trump, Amorin, Backhaus, Casey, Cheng, Cleri, Cooper, Cooper, Gardner,
  Gawiser, Grazian, Hathi, Hirschmann, Koekemoer, Lucas, Mobasher,
  Ravindranath, Straughn, Yung, \& {de la Vega}}]{2023jung}
Jung, I., Finkelstein, S.~L., Arrabal~Haro, P., {et~al.} 2023, {{CEERS}}:
  {{Diversity}} of {{Lyman-Alpha Emitters}} during the {{Epoch}} of
  {{Reionization}}, \dodoi{10.48550/arXiv.2304.05385}

\bibitem[{Kakiichi {et~al.}(2016)Kakiichi, Dijkstra, Ciardi, \&
  Graziani}]{2016kakiichi}
Kakiichi, K., Dijkstra, M., Ciardi, B., \& Graziani, L. 2016, Monthly Notices
  of the Royal Astronomical Society, 463, 4019, \dodoi{10.1093/mnras/stw2193}

\bibitem[{Kannan {et~al.}(2022)Kannan, Garaldi, Smith, Pakmor, Springel,
  Vogelsberger, \& Hernquist}]{2022kannan}
Kannan, R., Garaldi, E., Smith, A., {et~al.} 2022, Monthly Notices of the Royal
  Astronomical Society, 511, 4005, \dodoi{10.1093/mnras/stab3710}

\bibitem[{Kluyver {et~al.}(2016)Kluyver, {Ragan-Kelley}, P{\'e}rez, Granger,
  Bussonnier, Frederic, Kelley, Hamrick, Grout, Corlay, Ivanov, Avila, Abdalla,
  Willing, \& {Jupyter Development Team}}]{2016kluyver}
Kluyver, T., {Ragan-Kelley}, B., P{\'e}rez, F., {et~al.} 2016, Jupyter
  {{Notebooks}}\textemdash a Publishing Format for Reproducible Computational
  Workflows, 87--90, \dodoi{10.3233/978-1-61499-649-1-87}

\bibitem[{Koekemoer {et~al.}(2011)Koekemoer, Faber, Ferguson, Grogin, Kocevski,
  Koo, Lai, Lotz, Lucas, McGrath, Ogaz, Rajan, Riess, Rodney, Strolger,
  Casertano, Castellano, Dahlen, Dickinson, Dolch, Fontana, Giavalisco,
  Grazian, Guo, Hathi, Huang, {van der Wel}, Yan, Acquaviva, Alexander,
  Almaini, Ashby, Barden, Bell, Bournaud, Brown, Caputi, Cassata, Challis,
  Chary, Cheung, Cirasuolo, Conselice, Roshan~Cooray, Croton, Daddi, Dav{\'e},
  {de Mello}, {de Ravel}, Dekel, Donley, Dunlop, Dutton, Elbaz, Fazio,
  Filippenko, Finkelstein, Frazer, Gardner, Garnavich, Gawiser, Gruetzbauch,
  Hartley, H{\"a}ussler, Herrington, Hopkins, Huang, Jha, Johnson, Kartaltepe,
  Khostovan, Kirshner, Lani, Lee, Li, Madau, McCarthy, McIntosh, McLure,
  McPartland, Mobasher, Moreira, Mortlock, Moustakas, Mozena, Nandra, Newman,
  Nielsen, Niemi, Noeske, Papovich, Pentericci, Pope, Primack, Ravindranath,
  Reddy, Renzini, Rix, Robaina, Rosario, Rosati, Salimbeni, Scarlata, Siana,
  Simard, Smidt, Snyder, Somerville, Spinrad, Straughn, Telford, Teplitz,
  Trump, Vargas, Villforth, Wagner, Wandro, Wechsler, Weiner, Wiklind, Wild,
  Wilson, Wuyts, \& Yun}]{2011koekemoer}
Koekemoer, A.~M., Faber, S.~M., Ferguson, H.~C., {et~al.} 2011, The
  Astrophysical Journal Supplement Series, 197, 36,
  \dodoi{10.1088/0067-0049/197/2/36}

\bibitem[{Kriek {et~al.}(2015)Kriek, Shapley, Reddy, Siana, Coil, Mobasher,
  Freeman, {de Groot}, Price, Sanders, Shivaei, Brammer, Momcheva, Skelton,
  {van Dokkum}, Whitaker, Aird, Azadi, Kassis, Bullock, Conroy, Dav{\'e},
  Kere{\v s}, \& Krumholz}]{2015kriek}
Kriek, M., Shapley, A.~E., Reddy, N.~A., {et~al.} 2015, The Astrophysical
  Journal Supplement Series, 218, 15, \dodoi{10.1088/0067-0049/218/2/15}

\bibitem[{Laigle {et~al.}(2016)Laigle, McCracken, Ilbert, Hsieh, Davidzon,
  Capak, Hasinger, Silverman, Pichon, Coupon, Aussel, Le~Borgne, Caputi,
  Cassata, Chang, Civano, Dunlop, Fynbo, Kartaltepe, Koekemoer, Le~F{\`e}vre,
  Le~Floc'h, Leauthaud, Lilly, Lin, Marchesi, {Milvang-Jensen}, Salvato,
  Sanders, Scoville, Smolcic, Stockmann, Taniguchi, Tasca, Toft, Vaccari, \&
  Zabl}]{2016laigle}
Laigle, C., McCracken, H.~J., Ilbert, O., {et~al.} 2016, The Astrophysical
  Journal Supplement Series, 224, 24, \dodoi{10.3847/0067-0049/224/2/24}

\bibitem[{Langeroodi {et~al.}(2023)Langeroodi, Hjorth, \&
  Zhang}]{2023langeroodi}
Langeroodi, D., Hjorth, J., \& Zhang, Z. 2023, Little {{Red Dots}} or {{Brown
  Dwarfs}}? {{NIRSpec Discovery}} of {{Three Brown Dwarfs Masquerading}} as
  {{NIRCam-Selected Highly-Reddened AGNs}}, \dodoi{10.48550/arXiv.2308.10900}

\bibitem[{Larson {et~al.}(2022)Larson, Finkelstein, Hutchison, Papovich,
  Bagley, Dickinson, {Rojas-Ruiz}, Ferguson, Jung, Giavalisco, Grazian,
  Pentericci, \& Tacchella}]{2022larson}
Larson, R.~L., Finkelstein, S.~L., Hutchison, T.~A., {et~al.} 2022, The
  Astrophysical Journal, 930, 104, \dodoi{10.3847/1538-4357/ac5dbd}

\bibitem[{{Leja} {et~al.}(2019){Leja}, {Carnall}, {Johnson}, {Conroy}, \&
  {Speagle}}]{2019leja}
{Leja}, J., {Carnall}, A.~C., {Johnson}, B.~D., {Conroy}, C., \& {Speagle},
  J.~S. 2019, \apj, 876, 3, \dodoi{10.3847/1538-4357/ab133c}

\bibitem[{Lilly {et~al.}(2009)Lilly, Le~Brun, Maier, Mainieri, Mignoli,
  Scodeggio, Zamorani, Carollo, Contini, Kneib, Le~F{\`e}vre, Renzini,
  Bardelli, Bolzonella, Bongiorno, Caputi, Coppa, Cucciati, {de la Torre}, {de
  Ravel}, Franzetti, Garilli, Iovino, Kampczyk, Kovac, Knobel, Lamareille,
  Le~Borgne, Pello, Peng, {P{\'e}rez-Montero}, Ricciardelli, Silverman, Tanaka,
  Tasca, Tresse, Vergani, Zucca, Ilbert, Salvato, Oesch, Abbas, Bottini, Capak,
  Cappi, Cassata, Cimatti, Elvis, Fumana, Guzzo, Hasinger, Koekemoer,
  Leauthaud, Maccagni, Marinoni, McCracken, Memeo, Meneux, Porciani, Pozzetti,
  Sanders, Scaramella, Scarlata, Scoville, Shopbell, \& Taniguchi}]{2009lilly}
Lilly, S.~J., Le~Brun, V., Maier, C., {et~al.} 2009, The Astrophysical Journal
  Supplement Series, 184, 218, \dodoi{10.1088/0067-0049/184/2/218}

\bibitem[{Magee {et~al.}(2023)Magee, Casey, Cooper, Melendez, Fong,
  Karltaltepe, Long, Stawinski, Champagne, Cooper, Faisst, Maraston, \&
  Collaboration}]{2023magee}
Magee, J., Casey, C.~M., Cooper, O.~R., {et~al.} 2023, Research Notes of the
  AAS, 7, 110, \dodoi{10.3847/2515-5172/acd9a5}

\bibitem[{{Marley} {et~al.}(2021){Marley}, {Saumon}, {Visscher}, {Lupu},
  {Freedman}, {Morley}, {Fortney}, {Seay}, {Smith}, {Teal}, \&
  {Wang}}]{2021marley}
{Marley}, M.~S., {Saumon}, D., {Visscher}, C., {et~al.} 2021, \apj, 920, 85,
  \dodoi{10.3847/1538-4357/ac141d}

\bibitem[{{Marques-Chaves} {et~al.}(2021){Marques-Chaves}, Schaerer,
  {\'A}lvarez-M{\'a}rquez, Colina, {Dessauges-Zavadsky}, {P{\'e}rez-Fournon},
  {Saldana-Lopez}, \& Verhamme}]{2021marques-chaves}
{Marques-Chaves}, R., Schaerer, D., {\'A}lvarez-M{\'a}rquez, J., {et~al.} 2021,
  Monthly Notices of the Royal Astronomical Society, 507, 524,
  \dodoi{10.1093/mnras/stab2187}

\bibitem[{Mason {et~al.}(2015)Mason, Trenti, \& Treu}]{2015mason}
Mason, C.~A., Trenti, M., \& Treu, T. 2015, The Astrophysical Journal, 813, 21,
  \dodoi{10.1088/0004-637X/813/1/21}

\bibitem[{Mason {et~al.}(2019)Mason, Fontana, Treu, Schmidt, Hoag, Abramson,
  Amorin, Brada{\v c}, Guaita, Jones, Henry, Malkan, Pentericci, Trenti, \&
  Vanzella}]{2019mason}
Mason, C.~A., Fontana, A., Treu, T., {et~al.} 2019, Monthly Notices of the
  Royal Astronomical Society, 485, 3947, \dodoi{10.1093/mnras/stz632}

\bibitem[{McCracken {et~al.}(2012)McCracken, {Milvang-Jensen}, Dunlop, Franx,
  Fynbo, Le~F{\`e}vre, Holt, Caputi, Goranova, Buitrago, Emerson, Freudling,
  Hudelot, {L{\'o}pez-Sanjuan}, Magnard, Mellier, M{\o}ller, Nilsson,
  Sutherland, Tasca, \& Zabl}]{2012mccrackena}
McCracken, H.~J., {Milvang-Jensen}, B., Dunlop, J., {et~al.} 2012, Astronomy
  \&amp; Astrophysics, Volume 544, id.A156,
  {$<$}NUMPAGES{$>$}11{$<$}/NUMPAGES{$>$} pp., 544, A156,
  \dodoi{10.1051/0004-6361/201219507}

\bibitem[{McElwain {et~al.}(2023)McElwain, Feinberg, Perrin, Clampin, Mountain,
  Lallo, Lajoie, Kimble, Bowers, Stark, Acton, Atkinson, Barinek, Barto,
  Basinger, Beck, Bergkoetter, Bluth, Boucarut, Brady, Brooks, Brown, Byard,
  Carey, Carrasquilla, Chae, Chaney, Chayer, Chonis, Cohen, Cole, Comeau, Coon,
  Coppock, Coyle, Dean, Dziak, Eisenhower, Flagey, Franck, Gallagher, Gilman,
  Glassman, Green, Grieco, Haase, Hadjimichael, Hagopian, Hahn, Hartig, Havey,
  Hayden, Hellekson, Hicks, Holfeltz, Howard, Huguet, Jahne, Johnson, Johnston,
  Jurling, Kegley, Kennard, {Keski-Kuha}, Knight, Kulp, Levi, Levine, Lightsey,
  Luetgens, Mather, Matthews, McKay, Mehalick, Mel{\'e}ndez, Mosier, Murphy,
  Nelan, Niedner, Nol, Ohara, Ohl, Olczak, Osborne, Park, Perrygo, Pueyo,
  Redding, Regan, Reynolds, Rifelli, Rigby, Sabatke, Saif, Scorse, Seo, Shi,
  Sigrist, Smith, Smith, Smith, Sohn, Stahl, Telfer, Terlecki, Texter, Buren,
  Campen, Vila, Voyton, Waldman, Walker, Weiser, Wells, West, Whitman, Wolf, \&
  Zielinski}]{2023mcelwain}
McElwain, M.~W., Feinberg, L.~D., Perrin, M.~D., {et~al.} 2023, Publications of
  the Astronomical Society of the Pacific, 135, 058001,
  \dodoi{10.1088/1538-3873/acada0}

\bibitem[{McLean {et~al.}(2012)McLean, Steidel, Epps, Konidaris, Matthews,
  Adkins, Aliado, Brims, Canfield, Cromer, Fucik, Kulas, Mace, Magnone,
  Rodriguez, Rudie, Trainor, Wang, Weber, \& Weiss}]{2012mclean}
McLean, I.~S., Steidel, C.~C., Epps, H.~W., {et~al.} 2012, 8446, 84460J,
  \dodoi{10.1117/12.924794}

\bibitem[{McLure {et~al.}(2010)McLure, Dunlop, Cirasuolo, Koekemoer, Sabbi,
  Stark, Targett, \& Ellis}]{2010mclure}
McLure, R.~J., Dunlop, J.~S., Cirasuolo, M., {et~al.} 2010, Monthly Notices of
  the Royal Astronomical Society, 403, 960,
  \dodoi{10.1111/j.1365-2966.2009.16176.x}

\bibitem[{Menzel {et~al.}(2023)Menzel, Davis, Parrish, Lawrence, Stewart,
  Cooper, Irish, Mosier, Levine, Pitman, Walsh, Maghami, Thomson, Wooldridge,
  Boucarut, Feinberg, Turner, Kalia, \& Bowers}]{2023menzel}
Menzel, M., Davis, M., Parrish, K., {et~al.} 2023, Publications of the
  Astronomical Society of the Pacific, 135, 058002,
  \dodoi{10.1088/1538-3873/acbb9f}

\bibitem[{Naidu {et~al.}(2020)Naidu, Tacchella, Mason, Bose, Oesch, \&
  Conroy}]{2020naidu}
Naidu, R.~P., Tacchella, S., Mason, C.~A., {et~al.} 2020, The Astrophysical
  Journal, 892, 109, \dodoi{10.3847/1538-4357/ab7cc9}

\bibitem[{Oesch {et~al.}(2012)Oesch, Bouwens, Illingworth, Gonzalez, Trenti,
  {van Dokkum}, Franx, Labb{\'e}, Carollo, \& Magee}]{2012oesch}
Oesch, P.~A., Bouwens, R.~J., Illingworth, G.~D., {et~al.} 2012, The
  Astrophysical Journal, 759, 135, \dodoi{10.1088/0004-637X/759/2/135}

\bibitem[{Oesch {et~al.}(2015)Oesch, {van Dokkum}, Illingworth, Bouwens,
  Momcheva, Holden, {Roberts-Borsani}, Smit, Franx, Labb{\'e}, Gonz{\'a}lez, \&
  Magee}]{2015oesch}
Oesch, P.~A., {van Dokkum}, P.~G., Illingworth, G.~D., {et~al.} 2015, The
  Astrophysical Journal, 804, L30, \dodoi{10.1088/2041-8205/804/2/L30}

\bibitem[{Oesch {et~al.}(2023)Oesch, Brammer, Naidu, Bouwens, Chisholm,
  Illingworth, Matthee, Nelson, Qin, Reddy, Shapley, Shivaei, {van Dokkum},
  Weibel, Whitaker, Wuyts, {Covelo-Paz}, Endsley, Fudamoto, Giovinazzo,
  {Herard-Demanche}, Kerutt, Kramarenko, Labbe, Leonova, Lin, Magee,
  Marchesini, Maseda, Mason, Matharu, Meyer, Neufeld, Prieto~Lyon, Schaerer,
  Sharma, Shuntov, Smit, Stefanon, Wyithe, \& Xiao}]{2023oesch}
Oesch, P.~A., Brammer, G., Naidu, R.~P., {et~al.} 2023, The {{JWST FRESCO
  Survey}}: {{Legacy NIRCam}}/{{Grism Spectroscopy}} and {{Imaging}} in the Two
  {{GOODS Fields}}, \dodoi{10.48550/arXiv.2304.02026}

\bibitem[{Oke \& Gunn(1983)}]{1983oke}
Oke, J.~B., \& Gunn, J.~E. 1983, The Astrophysical Journal, 266, 713,
  \dodoi{10.1086/160817}

\bibitem[{Oke {et~al.}(1995)Oke, Cohen, Carr, Cromer, Dingizian, Harris,
  Labrecque, Lucinio, Schaal, Epps, \& Miller}]{1995oke}
Oke, J.~B., Cohen, J.~G., Carr, M., {et~al.} 1995, Publications of the
  Astronomical Society of the Pacific, 107, 375, \dodoi{10.1086/133562}

\bibitem[{Ono {et~al.}(2012)Ono, Ouchi, Mobasher, Dickinson, Penner, Shimasaku,
  Weiner, Kartaltepe, Nakajima, Nayyeri, Stern, Kashikawa, \&
  Spinrad}]{2012ono}
Ono, Y., Ouchi, M., Mobasher, B., {et~al.} 2012, The Astrophysical Journal,
  744, 83, \dodoi{10.1088/0004-637X/744/2/83}

\bibitem[{Ouchi {et~al.}(2020)Ouchi, Ono, \& Shibuya}]{2020ouchi}
Ouchi, M., Ono, Y., \& Shibuya, T. 2020, Annual Review of Astronomy and
  Astrophysics, vol. 58, p.617-659, 58, 617,
  \dodoi{10.1146/annurev-astro-032620-021859}

\bibitem[{Paardekooper {et~al.}(2015)Paardekooper, Khochfar, \&
  Dalla~Vecchia}]{2015paardekooper}
Paardekooper, J.-P., Khochfar, S., \& Dalla~Vecchia, C. 2015, Monthly Notices
  of the Royal Astronomical Society, 451, 2544, \dodoi{10.1093/mnras/stv1114}

\bibitem[{Pentericci {et~al.}(2018)Pentericci, Vanzella, Castellano, Fontana,
  De~Barros, Grazian, Marchi, Bradac, Conselice, Cristiani, Dickinson,
  Finkelstein, Giallongo, Guaita, Koekemoer, Maiolino, Santini, \&
  Tilvi}]{2018pentericci}
Pentericci, L., Vanzella, E., Castellano, M., {et~al.} 2018, Astronomy \&amp;
  Astrophysics, Volume 619, id.A147, {$<$}NUMPAGES{$>$}16{$<$}/NUMPAGES{$>$}
  pp., 619, A147, \dodoi{10.1051/0004-6361/201732465}

\bibitem[{{Planck Collaboration} {et~al.}(2016){Planck Collaboration}, Ade,
  Aghanim, Arnaud, Ashdown, Aumont, Baccigalupi, Banday, Barreiro, Bartlett,
  Bartolo, Battaner, Battye, Benabed, Beno{\^i}t, {Benoit-L{\'e}vy}, Bernard,
  Bersanelli, Bielewicz, Bock, Bonaldi, Bonavera, Bond, Borrill, Bouchet,
  Boulanger, Bucher, Burigana, Butler, Calabrese, Cardoso, Catalano, Challinor,
  Chamballu, Chary, Chiang, Chluba, Christensen, Church, Clements, Colombi,
  Colombo, Combet, Coulais, Crill, Curto, Cuttaia, Danese, Davies, Davis, {de
  Bernardis}, {de Rosa}, {de Zotti}, Delabrouille, D{\'e}sert, Di~Valentino,
  Dickinson, Diego, Dolag, Dole, Donzelli, Dor{\'e}, Douspis, Ducout, Dunkley,
  Dupac, Efstathiou, Elsner, En{\ss}lin, Eriksen, Farhang, Fergusson, Finelli,
  Forni, Frailis, Fraisse, Franceschi, Frejsel, Galeotta, Galli, Ganga,
  Gauthier, Gerbino, Ghosh, Giard, {Giraud-H{\'e}raud}, Giusarma, Gjerl{\o}w,
  {Gonz{\'a}lez-Nuevo}, G{\'o}rski, Gratton, Gregorio, Gruppuso, Gudmundsson,
  Hamann, Hansen, Hanson, Harrison, Helou, {Henrot-Versill{\'e}},
  {Hern{\'a}ndez-Monteagudo}, Herranz, Hildebrandt, Hivon, Hobson, Holmes,
  Hornstrup, Hovest, Huang, Huffenberger, Hurier, Jaffe, Jaffe, Jones, Juvela,
  Keih{\"a}nen, Keskitalo, Kisner, Kneissl, Knoche, Knox, Kunz, {Kurki-Suonio},
  Lagache, L{\"a}hteenm{\"a}ki, Lamarre, Lasenby, Lattanzi, Lawrence, Leahy,
  Leonardi, Lesgourgues, Levrier, Lewis, Liguori, Lilje, {Linden-V{\o}rnle},
  {L{\'o}pez-Caniego}, Lubin, {Mac{\'i}as-P{\'e}rez}, Maggio, Maino, Mandolesi,
  Mangilli, Marchini, Maris, Martin, Martinelli, {Mart{\'i}nez-Gonz{\'a}lez},
  Masi, Matarrese, McGehee, Meinhold, Melchiorri, Melin, Mendes, Mennella,
  Migliaccio, Millea, Mitra, {Miville-Desch{\^e}nes}, Moneti, Montier,
  Morgante, Mortlock, Moss, Munshi, Murphy, Naselsky, Nati, Natoli,
  Netterfield, {N{\o}rgaard-Nielsen}, Noviello, Novikov, Novikov, Oxborrow,
  Paci, Pagano, Pajot, Paladini, Paoletti, Partridge, Pasian, Patanchon,
  Pearson, Perdereau, Perotto, Perrotta, Pettorino, Piacentini, Piat,
  Pierpaoli, Pietrobon, Plaszczynski, Pointecouteau, Polenta, Popa, Pratt,
  Pr{\'e}zeau, Prunet, Puget, Rachen, Reach, Rebolo, Reinecke, Remazeilles,
  Renault, Renzi, Ristorcelli, Rocha, Rosset, Rossetti, Roudier, {Rouill{\'e}
  d'Orfeuil}, {Rowan-Robinson}, {Rubi{\~n}o-Mart{\'i}n}, Rusholme, Said,
  Salvatelli, Salvati, Sandri, Santos, Savelainen, Savini, Scott, Seiffert,
  Serra, Shellard, Spencer, Spinelli, Stolyarov, Stompor, Sudiwala, Sunyaev,
  Sutton, {Suur-Uski}, Sygnet, Tauber, Terenzi, Toffolatti, Tomasi, Tristram,
  Trombetti, Tucci, Tuovinen, T{\"u}rler, Umana, Valenziano, Valiviita,
  Van~Tent, Vielva, Villa, Wade, Wandelt, Wehus, White, White, Wilkinson, Yvon,
  Zacchei, \& Zonca}]{2016collaboration}
{Planck Collaboration}, Ade, P. a.~R., Aghanim, N., {et~al.} 2016, Astronomy
  \&amp; Astrophysics, Volume 594, id.A13,
  {$<$}NUMPAGES{$>$}63{$<$}/NUMPAGES{$>$} pp., 594, A13,
  \dodoi{10.1051/0004-6361/201525830}

\bibitem[{{Planck Collaboration} {et~al.}(2020){Planck Collaboration}, Aghanim,
  Akrami, Ashdown, Aumont, Baccigalupi, Ballardini, Banday, Barreiro, Bartolo,
  Basak, Battye, Benabed, Bernard, Bersanelli, Bielewicz, Bock, Bond, Borrill,
  Bouchet, Boulanger, Bucher, Burigana, Butler, Calabrese, Cardoso, Carron,
  Challinor, Chiang, Chluba, Colombo, Combet, Contreras, Crill, Cuttaia, {de
  Bernardis}, {de Zotti}, Delabrouille, Delouis, Di~Valentino, Diego, Dor{\'e},
  Douspis, Ducout, Dupac, Dusini, Efstathiou, Elsner, En{\ss}lin, Eriksen,
  Fantaye, Farhang, Fergusson, {Fernandez-Cobos}, Finelli, Forastieri, Frailis,
  Fraisse, Franceschi, Frolov, Galeotta, Galli, Ganga, {G{\'e}nova-Santos},
  Gerbino, Ghosh, {Gonz{\'a}lez-Nuevo}, G{\'o}rski, Gratton, Gruppuso,
  Gudmundsson, Hamann, Handley, Hansen, Herranz, Hildebrandt, Hivon, Huang,
  Jaffe, Jones, Karakci, Keih{\"a}nen, Keskitalo, Kiiveri, Kim, Kisner, Knox,
  Krachmalnicoff, Kunz, {Kurki-Suonio}, Lagache, Lamarre, Lasenby, Lattanzi,
  Lawrence, Le~Jeune, Lemos, Lesgourgues, Levrier, Lewis, Liguori, Lilje,
  Lilley, Lindholm, {L{\'o}pez-Caniego}, Lubin, Ma, {Mac{\'i}as-P{\'e}rez},
  Maggio, Maino, Mandolesi, Mangilli, {Marcos-Caballero}, Maris, Martin,
  Martinelli, {Mart{\'i}nez-Gonz{\'a}lez}, Matarrese, Mauri, McEwen, Meinhold,
  Melchiorri, Mennella, Migliaccio, Millea, Mitra, {Miville-Desch{\^e}nes},
  Molinari, Montier, Morgante, Moss, Natoli, {N{\o}rgaard-Nielsen}, Pagano,
  Paoletti, Partridge, Patanchon, Peiris, Perrotta, Pettorino, Piacentini,
  Polastri, Polenta, Puget, Rachen, Reinecke, Remazeilles, Renzi, Rocha,
  Rosset, Roudier, {Rubi{\~n}o-Mart{\'i}n}, {Ruiz-Granados}, Salvati, Sandri,
  Savelainen, Scott, Shellard, Sirignano, Sirri, Spencer, Sunyaev, {Suur-Uski},
  Tauber, Tavagnacco, Tenti, Toffolatti, Tomasi, Trombetti, Valenziano,
  Valiviita, Van~Tent, Vibert, Vielva, Villa, Vittorio, Wandelt, Wehus, White,
  White, Zacchei, \& Zonca}]{2020collaboration}
{Planck Collaboration}, Aghanim, N., Akrami, Y., {et~al.} 2020, Astronomy
  \&amp; Astrophysics, Volume 641, id.A6,
  {$<$}NUMPAGES{$>$}67{$<$}/NUMPAGES{$>$} pp., 641, A6,
  \dodoi{10.1051/0004-6361/201833910}

\bibitem[{Prochaska {et~al.}(2020)Prochaska, Hennawi, Cooke, Westfall, Wang,
  {EmAstro}, {Tiffanyhsyu}, Wasserman, Villaume, {Marijana777}, Schindler,
  Young, Simha, Wilde, Tejos, Isbell, Fl{\"o}rs, Sandford, Vasovi{\'c}, Betts,
  \& Holden}]{2020prochaskaa}
Prochaska, J.~X., Hennawi, J., Cooke, R., {et~al.} 2020, Zenodo,
  \dodoi{10.5281/zenodo.3743493}

\bibitem[{Rieke {et~al.}(2023)Rieke, Kelly, Misselt, Stansberry, Boyer, Beatty,
  Egami, Florian, Greene, Hainline, Leisenring, Roellig, Schlawin, Sun, Tinnin,
  Williams, Willmer, Wilson, Clark, Rohrbach, Brooks, Canipe, Correnti,
  DiFelice, Gennaro, Girard, Hartig, Hilbert, Koekemoer, Nikolov, Pirzkal,
  Rest, Robberto, Sunnquist, Telfer, Wu, Ferry, Lewis, Baum, Beichman, Doyon,
  Dressler, Eisenstein, Ferrarese, Hodapp, Horner, Jaffe, Johnstone, Krist,
  Martin, McCarthy, Meyer, Rieke, Trauger, \& Young}]{2023rieke}
Rieke, M.~J., Kelly, D.~M., Misselt, K., {et~al.} 2023, Publications of the
  Astronomical Society of the Pacific, 135, 028001,
  \dodoi{10.1088/1538-3873/acac53}

\bibitem[{Rigby {et~al.}(2023)Rigby, Perrin, McElwain, Kimble, Friedman, Lallo,
  Doyon, Feinberg, Ferruit, Glasse, Rieke, Rieke, Wright, Willott, Colon,
  Milam, Neff, Stark, Valenti, Abell, Abney, {Abul-Huda}, Acton, Adams, Adler,
  Aguilar, Ahmed, Albert, Alberts, Aldridge, Allen, Altenburg,
  {\'A}lvarez-M{\'a}rquez, {Alves de Oliveira}, Andersen, Anderson, Anderson,
  Argyriou, Armstrong, Arribas, Artigau, Arvai, Atkinson, Bacon, Bair, Banks,
  Barrientes, Barringer, Bartosik, Bast, Baudoz, Beatty, Bechtold, Beck,
  Bergeron, Bergkoetter, Bhatawdekar, Birkmann, Blazek, Blome, Boccaletti,
  B{\"o}ker, Boia, Bonaventura, Bond, Bosley, Boucarut, Bourque, Bouwman,
  Bower, Bowers, Boyer, Bradley, Brady, Braun, Breda, Bresnahan, Bright, Britt,
  Bromenschenkel, Brooks, Brooks, Brown, Brown, Brown, Bunker, Burger,
  Bushouse, Cale, Cameron, Cameron, Canipe, Caplinger, Caputo, Cara, Carey,
  Carniani, Carrasquilla, Carruthers, Case, Catherine, Chance, Chapman,
  Charlot, Charlow, Chayer, Chen, Cherinka, Chichester, Chilton, Chonis,
  Clampin, Clark, Clark, Coe, Coleman, Comber, Comeau, Connolly, Cooper,
  Cooper, Coppock, Correnti, Cossou, Coulais, Coyle, Cracraft, Curti, Cuturic,
  Davis, Davis, Dean, DeLisa, {deMeester}, Dencheva, Dencheva, DePasquale,
  Deschenes, Hunor~Detre, Diaz, Dicken, DiFelice, Dillman, Dixon, Doggett,
  Donaldson, Douglas, DuPrie, Dupuis, Durning, Easmin, Eck, Edeani, Egami,
  Ehrenwinkler, Eisenhamer, Eisenhower, Elie, Elliott, Elliott, Ellis,
  Engesser, Espinoza, Etienne, Etxaluze, Falini, Feeney, Ferry, Filippazzo,
  Fincham, Fix, Flagey, Florian, Flynn, Fontanella, Ford, Forshay, Fox, Franz,
  Fu, Fullerton, Galkin, Galyer, Garc{\'i}a~Mar{\'i}n, Gardner, Gardner,
  Garland, Garrett, Gasman, Gaspar, Gaudreau, Gauthier, Geers, Geithner,
  Gennaro, Giardino, Girard, Giuliano, Glassmire, Glauser, Glazer, Godfrey,
  Golimowski, Gollnitz, Gong, Gonzaga, Gordon, Gordon, Goudfrooij, Greene,
  Greenhouse, Grimaldi, Groebner, Grundy, Guillard, Gutman, Ha, Haderlein,
  Hagedorn, Hainline, Haley, Hami, Hamilton, Hammel, Hansen, Harkins, Harr,
  Hart, Hart, Hartig, Hashimoto, Haskins, Hathaway, Havey, Hayden, Hecht,
  {Heller-Boyer}, Henriques, Henry, Hermann, Hernandez, Hesman, Hicks, Hilbert,
  Hines, Hoffman, Holfeltz, Holler, Hoppa, Hott, Howard, Howard, Hunter,
  Hunter, Hurst, Husemann, Hustak, Ilinca~Ignat, Illingworth, Irish, Jackson,
  Jahromi, Jakobsen, James, James, Januszewski, Jenkins, Jirdeh, Johnson,
  Johnson, Jones, Jones, Jones, Jones, Jordan, Jordan, Jurczyk, Jurling,
  Kaleida, Kalmanson, Kammerer, Kang, Kao, Karakla, Kavanagh, Kelly, Kendrew,
  Kennedy, Kenny, {Keski-kuha}, Keyes, Kidwell, Kinzel, Kirk, Kirkpatrick,
  Kirshenblat, Klaassen, Knapp, Knight, Knollenberg, Koehler, Koekemoer,
  Kovacs, Kulp, Kumari, Kyprianou, La~Massa, Labador, Labiano, Lagage, Lajoie,
  Lallo, Lam, Lamb, Lambros, Lampenfield, Langston, Larson, Law, Lawrence, Lee,
  Leisenring, Lepo, Leveille, Levenson, Levine, Levy, Lewis, Lewis, Libralato,
  Lightsey, Link, Liu, Lo, Lockwood, Logue, Long, Long, Loomis,
  {Lopez-Caniego}, Lorenzo~Alvarez, {Love-Pruitt}, Lucy, Luetzgendorf, Maghami,
  Maiolino, Major, Malla, Malumuth, Manjavacas, Mannfolk, Marrione, Marston,
  Martel, Maschmann, Masci, Masciarelli, Maszkiewicz, Mather, McKenzie, McLean,
  McMaster, Melbourne, Mel{\'e}ndez, Menzel, Merz, Meyett, Meza, Miskey,
  Misselt, Moller, Morrison, Morse, Moseley, Mosier, Mountain, Mueckay,
  Mueller, Mullally, Murphy, Murray, Murray, Mustelier, Muzerolle, Mycroft,
  Myers, Myrick, Nanavati, Nance, Nayak, Naylor, Nelan, Nickson, Nielson,
  {Nieto-Santisteban}, Nikolov, {Noriega-Crespo}, O'Shaughnessy, O'Sullivan,
  Ochs, Ogle, Oleszczuk, Olmsted, Osborne, Ottens, Owens, Pacifici, Pagan,
  Page, Park, Parrish, Patapis, Paul, Pauly, Pavlovsky, Pedder, Peek,
  {Pena-Guerrero}, Penanen, Perez, Perna, Perriello, Phillips, Pietraszkiewicz,
  Pinaud, Pirzkal, Pitman, Piwowar, Platais, Player, Plesha, Pollizi, Polster,
  Pontoppidan, Porterfield, Proffitt, Pueyo, Pulliam, Quirt, Quispe~Neira,
  Ramos~Alarcon, Ramsay, Rapp, Rapp, Rauscher, Ravindranath, Rawle, Regan,
  Reichard, Reis, Ressler, Rest, Reynolds, Rhue, Richon, Rickman, Ridgaway,
  Ritchie, Rix, Robberto, Robinson, Robinson, Robinson, Rock, Rodriguez,
  Rodriguez Del~Pino, Roellig, Rohrbach, Roman, Romelfanger, Rose, Roteliuk,
  Roth, Rothwell, Rowlands, Roy, Royer, Royle, Rui, Rumler, Runnels, Russ,
  Rustamkulov, Ryden, Ryer, Sabata, Sabatke, Sabbi, Samuelson, Sapp,
  Sappington, Sargent, Sauer, Scheithauer, Schlawin, Schlitz, Schmitz,
  Schneider, Schreiber, Schulze, Schwab, Scott, Sembach, Shanahan, Shaughnessy,
  Shaw, Shawger, Shay, Sheehan, Shen, Sherman, Shiao, Shih, Shivaei,
  Sienkiewicz, Sing, Sirianni, Sivaramakrishnan, Skipper, Sloan, Slocum,
  Slowinski, Smith, Smith, Smith, Smith, Snyder, Soh, Sohn, Soto, Spencer,
  Stallcup, Stansberry, Starr, Starr, Stewart, Stiavelli, Straughn, Strickland,
  Stys, Summers, Sun, Sunnquist, Swade, Swam, Swaters, Swoish, Taylor, Taylor,
  Te~Plate, Tea, Teague, Telfer, Temim, Thatte, Thompson, Thompson, Thomson,
  Tikkanen, Tippet, Todd, Toolan, Tran, Trejo, Truong, Tsukamoto, Tustain,
  Tyra, Ubeda, Underwood, Uzzo, Van~Campen, Vandal, Vandenbussche, Vila, Volk,
  Wahlgren, Waldman, Walker, Wander, Warfield, Warner, Wasiak, Watkins, Weaver,
  Weilert, Weiser, Weiss, Weissman, Welty, West, Wheate, Wheatley, Wheeler,
  White, Whiteaker, Whitehouse, Whiteleather, Whitman, Williams, Willmer,
  Willoughby, Wilson, Wirth, Wislowski, Wolf, Wolfe, Wolff, Workman, Wright,
  Wu, Wu, Wymer, Yates, Yeager, Yeates, Yerger, Yoon, Young, Yu, Zak, Zeidler,
  Zhou, Zielinski, Zincke, \& Zonak}]{2023rigby}
Rigby, J., Perrin, M., McElwain, M., {et~al.} 2023, Publications of the
  Astronomical Society of the Pacific, 135, 048001,
  \dodoi{10.1088/1538-3873/acb293}

\bibitem[{{Roberts-Borsani} {et~al.}(2022){Roberts-Borsani}, Treu, Mason,
  Ellis, Laporte, Schmidt, Brada{\v c}, Fontana, Morishita, \&
  Santini}]{2022roberts-borsani}
{Roberts-Borsani}, G., Treu, T., Mason, C., {et~al.} 2022, Nature and
  {{Nurture}}? {{Comparing Ly}}\$\textbackslash alpha\$ {{Detections}} in {{UV
  Bright}} and {{Fainter}} [{{O III}}]+{{H}}\$\textbackslash beta\$
  {{Emitters}} at \$z\textbackslash sim8\$ {{With Keck}}/{{MOSFIRE}}

\bibitem[{{Roberts-Borsani} {et~al.}(2016){Roberts-Borsani}, Bouwens, Oesch,
  Labbe, Smit, Illingworth, {van Dokkum}, Holden, Gonzalez, Stefanon, Holwerda,
  \& Wilkins}]{2016roberts-borsani}
{Roberts-Borsani}, G.~W., Bouwens, R.~J., Oesch, P.~A., {et~al.} 2016, The
  Astrophysical Journal, 823, 143, \dodoi{10.3847/0004-637X/823/2/143}

\bibitem[{Robertson {et~al.}(2013)Robertson, Furlanetto, Schneider, Charlot,
  Ellis, Stark, McLure, Dunlop, Koekemoer, Schenker, Ouchi, Ono, {Curtis-Lake},
  Rogers, Bowler, \& Cirasuolo}]{2013robertson}
Robertson, B.~E., Furlanetto, S.~R., Schneider, E., {et~al.} 2013, The
  Astrophysical Journal, 768, 71, \dodoi{10.1088/0004-637X/768/1/71}

\bibitem[{Rockosi {et~al.}(2010)Rockosi, Stover, Kibrick, Lockwood, Peck,
  Cowley, Bolte, Adkins, Alcott, Allen, Brown, Cabak, Deich, Hilyard, Kassis,
  Lanclos, Lewis, Pfister, Phillips, Robinson, Saylor, Thompson, Ward, Wei, \&
  Wright}]{2010rockosi}
Rockosi, C., Stover, R., Kibrick, R., {et~al.} 2010, 7735, 77350R,
  \dodoi{10.1117/12.856818}

\bibitem[{Saha {et~al.}(2020)Saha, Tandon, Simmonds, Verhamme, Paswan,
  Schaerer, Rutkowski, Borgohain, Elmegreen, Inoue, Combes, Elmegreen, \&
  Paalvast}]{2020saha}
Saha, K., Tandon, S.~N., Simmonds, C., {et~al.} 2020, Nature Astronomy, 4,
  1185, \dodoi{10.1038/s41550-020-1173-5}

\bibitem[{Sanders {et~al.}(2023)Sanders, Shapley, Topping, Reddy, \&
  Brammer}]{2023sanders}
Sanders, R.~L., Shapley, A.~E., Topping, M.~W., Reddy, N.~A., \& Brammer, G.~B.
  2023, Direct {{T}}\_e-Based {{Metallicities}} of Z=2-9 {{Galaxies}} with
  {{JWST}}/{{NIRSpec}}: {{Empirical Metallicity Calibrations Applicable}} from
  {{Reionization}} to {{Cosmic Noon}}, \dodoi{10.48550/arXiv.2303.08149}

\bibitem[{Schouws {et~al.}(2022)Schouws, Stefanon, Bouwens, Smit, Hodge,
  Labb{\'e}, Algera, Boogaard, Carniani, Fudamoto, Holwerda, Illingworth,
  Maiolino, Maseda, Oesch, \& {van der Werf}}]{2022schouws}
Schouws, S., Stefanon, M., Bouwens, R., {et~al.} 2022, The Astrophysical
  Journal, 928, 31, \dodoi{10.3847/1538-4357/ac4605}

\bibitem[{Shibuya {et~al.}(2012)Shibuya, Kashikawa, Ota, Iye, Ouchi, Furusawa,
  Shimasaku, \& Hattori}]{2012shibuya}
Shibuya, T., Kashikawa, N., Ota, K., {et~al.} 2012, The Astrophysical Journal,
  752, 114, \dodoi{10.1088/0004-637X/752/2/114}

\bibitem[{Smit {et~al.}(2015)Smit, Bouwens, Franx, Oesch, Ashby, Willner,
  Labb{\'e}, Holwerda, Fazio, \& Huang}]{2015smit}
Smit, R., Bouwens, R.~J., Franx, M., {et~al.} 2015, The Astrophysical Journal,
  801, 122, \dodoi{10.1088/0004-637X/801/2/122}

\bibitem[{Smit {et~al.}(2018)Smit, Bouwens, Carniani, Oesch, Labb{\'e},
  Illingworth, {van der Werf}, Bradley, Gonzalez, Hodge, Holwerda, Maiolino, \&
  Zheng}]{2018smit}
Smit, R., Bouwens, R.~J., Carniani, S., {et~al.} 2018, Nature, 553, 178,
  \dodoi{10.1038/nature24631}

\bibitem[{Song {et~al.}(2016)Song, Finkelstein, Livermore, Capak, Dickinson, \&
  Fontana}]{2016song}
Song, M., Finkelstein, S.~L., Livermore, R.~C., {et~al.} 2016, The
  Astrophysical Journal, 826, 113, \dodoi{10.3847/0004-637X/826/2/113}

\bibitem[{Stark(2016)}]{2016stark}
Stark, D.~P. 2016, Annual Review of Astronomy and Astrophysics, vol. 54,
  p.761-803, 54, 761, \dodoi{10.1146/annurev-astro-081915-023417}

\bibitem[{Stark {et~al.}(2017)Stark, Ellis, Charlot, Chevallard, Tang, Belli,
  Zitrin, Mainali, Gutkin, {Vidal-Garc{\'i}a}, Bouwens, \& Oesch}]{2017stark}
Stark, D.~P., Ellis, R.~S., Charlot, S., {et~al.} 2017, Monthly Notices of the
  Royal Astronomical Society, 464, 469, \dodoi{10.1093/mnras/stw2233}

\bibitem[{{Tacchella} {et~al.}(2022){Tacchella}, {Finkelstein}, {Bagley},
  {Dickinson}, {Ferguson}, {Giavalisco}, {Graziani}, {Grogin}, {Hathi},
  {Hutchison}, {Jung}, {Koekemoer}, {Larson}, {Papovich}, {Pirzkal},
  {Rojas-Ruiz}, {Song}, {Schneider}, {Somerville}, {Wilkins}, \&
  {Yung}}]{2022tacchella}
{Tacchella}, S., {Finkelstein}, S.~L., {Bagley}, M., {et~al.} 2022, \apj, 927,
  170, \dodoi{10.3847/1538-4357/ac4cad}

\bibitem[{Tang {et~al.}(2021)Tang, Stark, Chevallard, Charlot, Endsley, \&
  Congiu}]{2021tang}
Tang, M., Stark, D.~P., Chevallard, J., {et~al.} 2021, Monthly Notices of the
  Royal Astronomical Society, 503, 4105, \dodoi{10.1093/mnras/stab705}

\bibitem[{Tang {et~al.}(2023)Tang, Stark, Chen, Mason, Topping, Endsley,
  Senchyna, Plat, Lu, Whitler, Robertson, \& Charlot}]{2023tang}
Tang, M., Stark, D.~P., Chen, Z., {et~al.} 2023, {{JWST}}/{{NIRSpec
  Spectroscopy}} of \$z=7-9\$ {{Star Forming Galaxies}} with {{CEERS}}: {{New
  Insight}} into {{Bright Ly}}\$\textbackslash alpha\$ {{Emitters}} in
  {{Ionized Bubbles}}, \dodoi{10.48550/arXiv.2301.07072}

\bibitem[{Tilvi {et~al.}(2020)Tilvi, Malhotra, Rhoads, Coughlin, Zheng,
  Finkelstein, Veilleux, Mobasher, Wang, Probst, Swaters, Hibon, Joshi, Zabl,
  Jiang, Pharo, \& Yang}]{2020tilvi}
Tilvi, V., Malhotra, S., Rhoads, J.~E., {et~al.} 2020, The Astrophysical
  Journal, 891, L10, \dodoi{10.3847/2041-8213/ab75ec}

\bibitem[{Treu {et~al.}(2013)Treu, Schmidt, Trenti, Bradley, \&
  Stiavelli}]{2013treu}
Treu, T., Schmidt, K.~B., Trenti, M., Bradley, L.~D., \& Stiavelli, M. 2013,
  The Astrophysical Journal, 775, L29, \dodoi{10.1088/2041-8205/775/1/L29}

\bibitem[{Virtanen {et~al.}(2020)Virtanen, Gommers, Oliphant, Haberland, Reddy,
  Cournapeau, Burovski, Peterson, Weckesser, Bright, {van der Walt}, Brett,
  Wilson, Millman, Mayorov, Nelson, Jones, Kern, Larson, Carey, Polat, Feng,
  Moore, VanderPlas, Laxalde, Perktold, Cimrman, Henriksen, Quintero, Harris,
  Archibald, Ribeiro, Pedregosa, \& {van Mulbregt}}]{2020virtanen}
Virtanen, P., Gommers, R., Oliphant, T.~E., {et~al.} 2020, Nature Methods, 17,
  261, \dodoi{10.1038/s41592-019-0686-2}

\bibitem[{Weaver {et~al.}(2019)Weaver, Toft, Davidzon, Capak, \&
  McCracken}]{2019weaver}
Weaver, J., Toft, S., Davidzon, I., Capak, P., \& McCracken, H. 2019, The
  {{Farmer}}: {{Improved}} Model-Based Photometry for the next Generation of
  Galaxy Surveys, 9, \dodoi{10.5281/zenodo.3554205}

\bibitem[{Weaver {et~al.}(2022)Weaver, Kauffmann, Ilbert, McCracken, Moneti,
  Toft, Brammer, Shuntov, Davidzon, Hsieh, Laigle, Anastasiou, Jespersen,
  Vinther, Capak, Casey, McPartland, {Milvang-Jensen}, Mobasher, Sanders,
  Zalesky, Arnouts, Aussel, Dunlop, Faisst, Franx, Furtak, Fynbo, Gould, Greve,
  Gwyn, Kartaltepe, Kashino, Koekemoer, Kokorev, Le~F{\`e}vre, Lilly, Masters,
  Magdis, Mehta, Peng, Riechers, Salvato, Sawicki, Scarlata, Scoville, Shirley,
  Silverman, Sneppen, Smol{\u c}i{\'c}, Steinhardt, Stern, Tanaka, Taniguchi,
  Teplitz, Vaccari, Wang, \& Zamorani}]{2022weaver}
Weaver, J.~R., Kauffmann, O.~B., Ilbert, O., {et~al.} 2022, The Astrophysical
  Journal Supplement Series, 258, 11, \dodoi{10.3847/1538-4365/ac3078}

\bibitem[{Whitler {et~al.}(2023)Whitler, Stark, Endsley, Chen, Mason, Topping,
  \& Charlot}]{2023whitler}
Whitler, L., Stark, D.~P., Endsley, R., {et~al.} 2023, Insight from
  {{JWST}}/{{NIRCam}} into Galaxy Overdensities around Bright
  {{Ly}}\$\textbackslash alpha\$ Emitters during Reionization: Implications for
  Ionized Bubbles at \$z \textbackslash sim 9\$,
  \dodoi{10.48550/arXiv.2305.16670}

\bibitem[{Yan {et~al.}(2012)Yan, Finkelstein, Huang, Ryan, Ferguson, Koekemoer,
  Grogin, Dickinson, Newman, Somerville, Dav{\'e}, Faber, Papovich, Guo,
  Giavalisco, Lee, Reddy, Cooray, Siana, Hathi, Fazio, Ashby, Weiner, Lucas,
  Dekel, Pentericci, Conselice, Kocevski, \& Lai}]{2012yan}
Yan, H., Finkelstein, S.~L., Huang, K.-H., {et~al.} 2012, The Astrophysical
  Journal, 761, 177, \dodoi{10.1088/0004-637X/761/2/177}

\bibitem[{Zheng {et~al.}(2011)Zheng, Cen, Weinberg, Trac, \&
  {Miralda-Escud{\'e}}}]{2011zheng}
Zheng, Z., Cen, R., Weinberg, D., Trac, H., \& {Miralda-Escud{\'e}}, J. 2011,
  The Astrophysical Journal, 739, 62, \dodoi{10.1088/0004-637X/739/2/62}

\bibitem[{Zitrin {et~al.}(2015)Zitrin, Ellis, Belli, \& Stark}]{2015zitrin}
Zitrin, A., Ellis, R.~S., Belli, S., \& Stark, D.~P. 2015, The Astrophysical
  Journal, 805, L7, \dodoi{10.1088/2041-8205/805/1/L7}

\end{thebibliography}

\appendix

\section{Spectroscopic Catalog}

Here we provide the full target list, any emission line detections, and any new spectroscopic redshift measurements for the Keck/MOSFIRE observations presented in this paper. Spectroscopic confirmation of EoR LAEs (our primary targets) are discussed in Section \textsection4.2 of the main text; here, we also include the majority of our spectroscopic confirmations, which were lower-redshift filler targets. 

Filler targets are selected at specific redshifts with emission lines accessible with MOSFIRE $Y$-band; \Ha-emitters at $0.5<z<0.7$, \Oii-emitters at $1.6<z<2.0$, and \Ciii-emitters at $4.1<z<4.9$. Stars (for alignment and for flux calibration) were also placed on each mask; these are taken from Gaia DR3 and registered to the same reference astrometry as our source catalogs. Any serendipitously observed sources (with naming convention SERENDIP-X with X being an arbitrary number) are also included in the catalog. Positions of the serendipitous detections are carefully reconstructed using deep imaging and relative offset of other sources on the mask, accurate to $\sim0.5$\,arcsec. Table \ref{tab:positions} shows the first few lines of the data table with the source ID, slitmask, and position of all observed targets, which includes primary targets, filler targets, stars, and serendipitous sources. The full target list is provided in machine-readable form online.

\begin{deluxetable}{lccc}[h]
\setlength{\tabcolsep}{0.1in}
\tablecaption{WERLS/MOSFIRE 2022A Target List \label{tab:positions}} 
\tablehead{
\colhead{ID} &  \colhead{Mask} & \colhead{RA} &\colhead{Dec} }
\startdata
M\_WP551495 & wmmc01 & 150.19379 & 2.17021 \\
        B\_WP450980 & wmmc01 & 150.18992 & 2.16204 \\ 
        star\_75 & wmmc01 & 150.19183 & 2.18124 \\ 
        L\_WP693466 & wmmc01 & 150.18879 & 2.18037 \\ 
        SERENDIP-1 & wmmc01 & 150.18644 & 2.18126 \\ 
        c2020\_pz\_0.664\_22.8\_826089 & wmmc01 & 150.18738 & 2.19157 \\ 
        L\_WP930991 & wmmc01 & 150.18792 & 2.20266 \\ 
        c2020\_pz\_1.65\_22.6\_640174 & wmmc01 & 150.17921 & 2.17518 \\ 
        c2020\_pz\_4.25\_27.1\_702146 & wmmc01 & 150.17733 & 2.18104 \\ 
        ... & ... & ... & ... 
\enddata
\tablecomments{Table \ref{tab:positions} is published in its entirety in machine-readable format. A portion is shown here for guidance regarding its form and content.}
\end{deluxetable}

A total of 330 targets were observed, with 114 primary targets, 15 stars, 35 serendipitous sources, and 166 filler targets. The spectroscopic yield for filler targets is $\sim34$\%, including all 56 spectroscopic redshift identifications, whether tentative (39/56) or secure (17/56). We note observed line wavelengths for serendipitous sources, but in all cases except when multiple lines enabled line species identification \citep[e.g.][]{2023magee}, we do not list line or redshift identifications given their lack of photometric redshift priors in the WERLS target catalog. 

Importantly, quality flags indicating confidence class of spectroscopic confirmation are also included in the table. We follow the convention in the zCOSMOS catalog as described in \citet{2009lilly}, wherein 0 = no redshift measurement attempted, 1 = an insecure redshift, 2 = a likely redshift about which there is some doubt, 3 = a very secure redshift, 4 = a very secure redshift with an exhibition-quality spectrum, 9 = a securely detected single line with prior information that enables redshift identification. We add to this convention scheme the confidence class -1 to indicate continuum detection without line detection, which was only relevant for the stars and for a handful of serendipitous sources. All primary targets in the paper are classified with confidence class of 2 for tentative LAEs or 9 for secure LAEs; these are also indicated in the Notes column by ``tentative\_in\_paper" or ``secure\_in\_paper", respectively. Candidate LAEs that were included in the early vetting process (as described in Section \textsection3.2) but ultimately excluded from this paper are classified with quality flags of 1, and noted by ``early\_vetting" in the notes. For fillers, only sources with confidence classes of 3 or 4 should be considered secure redshifts, as these are the only multi-line detections. Measurements for all other filler targets should be considered tentative. The first few lines of the spectroscopic data table are shown in Table \ref{tab:appendix}, with the full table provided in machine-readable form online.

\begin{deluxetable}{lcccccccc}[h]
\setlength{\tabcolsep}{0.1in}
\tablecaption{WERLS/MOSFIRE 2022A Spectroscopic Catalog \label{tab:appendix}} 
\tablehead{
\colhead{ID} &  \colhead{Redshift} & \colhead{$\lambda_{\rm obs}$} &\colhead{Line Species} & \colhead{Confidence Class\tablenotemark{a}} & \colhead{Notes}  \\
\colhead{} & \colhead{} & \colhead{$\rm \AA$} & \colhead{} & \colhead{} & \colhead{}}
\startdata
        M\_WP551495 & -- & -- & -- & 0 & ~ \\ 
        B\_WP450980 & 7.0925 & 9837 & Ly$\alpha$ & 2 & tentative\_in\_paper \\ 
        star\_75 & -- & -- & -- & -1 & ~ \\ 
        L\_WP693466 & -- & -- & -- & 0 & ~ \\ 
        SERENDIP-1 & -- & 10945 & -- & 0 & ~ \\ 
        c2020\_pz\_0.664\_22.8\_826089 & 0.667 & 10936 & H$\alpha$ & 4 & ~ \\ 
        L\_WP930991 & -- & -- & -- & 0 & ~ \\ 
        c2020\_pz\_1.65\_22.6\_640174 & 1.641 & 9848 & \Oii & 4 & ~ \\ 
        c2020\_pz\_4.25\_27.1\_702146 & -- & -- & -- & 0 & ~ \\ 
        ... & ... & ... & ... & ... & ... \\ 
\enddata
\tablenotetext{}{
$^{a}$\footnotesize Quality flag indicating confidence class of spectroscopic confirmation, where 0 = no redshift measurement attempted, 1 = an insecure redshift, 2 = a likely redshift about which there is some doubt, 3 = a very secure redshift, 4 = a very secure redshift with an exhibition-quality spectrum, 9 = a securely detected single line with prior information that enables redshift identification, and -1 = continuum detection without line detection.}
\tablecomments{Table \ref{tab:appendix} is published in its entirety in machine-readable format. A portion is shown here for guidance regarding its form and content.}
\end{deluxetable}

\end{document}